\documentclass[12pt]{iopart}
\usepackage{epsfig}
\usepackage{cite}
\usepackage[english]{babel}

\newcommand{\eq}[1]{\begin{eqnarray} #1 \end{eqnarray}}
\newcommand{\nn}{\nonumber\\[0.2cm]}
\newcommand{\nno}{\nonumber}

\begin{document}

\review{Fluctuations and Correlations in Nucleus-Nucleus Collisions within Transport Models}

\author{V.P.~Konchakovski$^{1,2}$, M.I.~Gorenstein$^{1,3}$,
      \\E.L.~Bratkovskaya$^{2}$ and  W.~Greiner$^{3}$}

\address{$^1$Bogolyubov Institute for Theoretical Physics, Kiev, Ukraine}
\address{$^2$Institute for Theoretical Physics, University of Frankfurt, Germany}
\address{$^3$Frankfurt Institute for Advanced Studies, University of Frankfurt, Germany}

\begin{abstract}
Particle number fluctuations and correlations in nucleus-nucleus
collisions at SPS and RHIC energies are studied within microscopic
transport approaches. In this review we focus on the Hadron-String-Dynamics
(HSD) and Ultra-relativistic-Quantum-Molecular-Dynamics (UrQMD) models
The obtained results are compared with the available experimental
data as well as with the statistical models and the model of
independent sources. In particular the role of the
experimental centrality selection and acceptance is discussed in detail
for a variety of experimental fluctuations and correlation observables
with the aim to extract information on the critical point in the $(T,\mu_B)$
plane of strongly interacting matter.
\end{abstract}

\maketitle
\tableofcontents
\markboth{}{}
\newpage

\section{Introduction}
\label{ch:intro}

\subsection{Relativistic Heavy-Ion Collisions}

Hot and dense nuclear matter can be generated in the laboratory in
a wide range of temperatures and densities by colliding atomic
nuclei at high energies. In the collision zone the matter is
heated and compressed for a very short period of time. If the
energy pumped into the formed fireball is sufficiently large the
quark-gluon substructure of nucleons comes into play. At moderate
temperatures, nucleons are excited to short-lived states (baryonic
resonances) which decay by the emission of mesons. At higher
temperatures, also baryon-antibaryon pairs are created. This
mixture of baryons, antibaryons and mesons, all strongly
interacting particles, is generally called hadronic matter, or
baryonic matter if baryons prevail. At even higher temperatures or
densities the hadrons melt and the constituents, the quarks and
gluons, form a new phase, the Quark-Gluon Plasma (QGP).

High-energy heavy-ion collision experiments provide the unique
possibility to create and investigate these extreme states of matter.
The study of matter at extremely high temperature and baryon
density, where the hadronic matter transforms to a QGP is the aim of a
variety of experiments at current and future facilities:  NA38, NA49,
NA50, NA60 and NA61/SHINE at the Super-Proton-Synchrotron
(SPS)~\cite{Ab99,Ba99,Al05a,Ar06,Ga06}; PHENIX, STAR, PHOBOS and BRAHMS
at the Relativistic-Heavy-Ion-Collider
(RHIC)~\cite{Ar05,Ba05a,Ad05b,Ad05c}; ALICE at the
Large-Hadron-Collider (LHC)~\cite{Aa08}; CBM and PANDA at the Facility
for Antiproton and Ion Research (FAIR)~\cite{Pe06}; MPD at the
Nuclotron-based Ion Collider fAcility (NICA)~\cite{SiSoTo06}.

Relativistic nucleus-nucleus (A+A) collisions have been studied so
far at beam energies from 0.1 to 2 AGeV at the SIS
(SchwerIonen-Synchrotron), from 2 to 11.6 AGeV at the AGS
(Alternating Gradient Synchrotron) and from 20 to 160 AGeV
at the SPS \cite{Af02,Al04a}. While part of these programs are
closed now, the heavy-ion research has been extended at RHIC with
Au+Au collisions at  c.m. nucleon pair energies $\sqrt{s_{NN}}$
from $\sim 20$ to 200 GeV (equivalent energies in a fixed target
experiment: 0.2 to 21.3 ATeV). In the near future, further
insight into the physics of matter at even more extreme conditions
will be gained at the LHC, which will reach center-of-mass
energies of the few TeV scale. Apart from LHC, the SPS successor
SHINE operates at CERN in order to scan the 10A-158AGeV
energy range with light and intermediate mass nuclei~\cite{Ga06}.
At FAIR, which is expected to start operation in 2015, collisions
of gold nuclei from  5 AGeV up to 35 AGeV will be
studied.
At NICA it is planned to start the experimental program of
colliding Au and/or U ions as well as polarized light nuclei at
energies up to of 5 AGeV in 2013 (an upgrade to
$\sqrt{s_{NN}}~=~9$~GeV is foreseen~\cite{To07a}).

At very high beam energies -- as available at RHIC and LHC -- the
research programs concentrate on the study of the properties of
deconfined QCD matter at very high temperatures and almost zero
net baryon densities, whereas at moderate beam energies (SPS, FAIR
and NICA) experiments focus on the search for structures in the
QCD phase diagram such as the critical endpoint, the predicted
first order phase transition between hadronic and partonic matter,
and the chiral phase transition. The critical endpoint and the
first order phase transition are expected to occur at finite
baryon chemical potential and moderate temperatures.

\begin{figure}[t]
\centerline{\epsfig{file=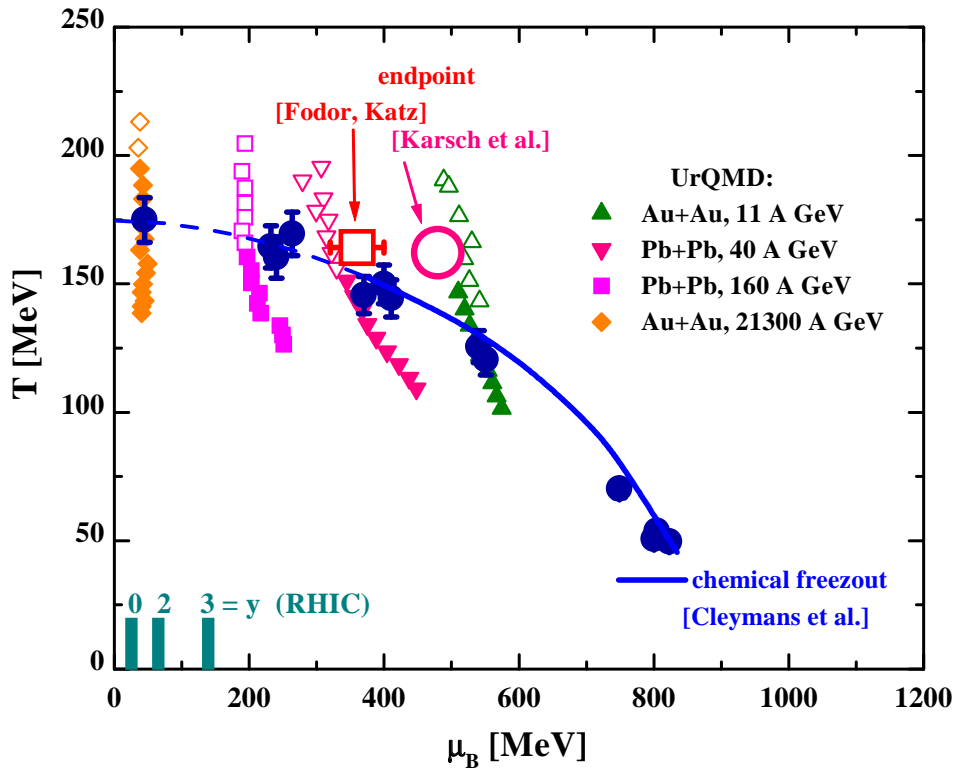,width=0.8\textwidth}} \caption{
The phase diagram with the critical end point at $\mu_B \approx
400 \mbox{ MeV}, T \approx 150 \mbox{ MeV} $ as predicted by
Lattice QCD. In addition, the time evolution in the $T-\mu$-plane
of a central cell in UrQMD calculations \cite{Br99,Br02} is
depicted for different bombarding energies.
At RHIC (see insert at the $\mu_B$ scale) large $\mu_B$ can be
accessible in the fragmentation region only. The figure is taken
from Ref.~\protect{\cite{Br04}}.} \label{fig:phasedia}
\end{figure}

Particle yields or ratios measured at different beam energies and
analyzed with the statistical model provide sets of thermal
parameters, temperature $T$ and baryo-chemical potential
$\mu_B$, which establish a``line of chemical freeze-out"
\cite{BrReSt03,ClRe98,ClRe99}. The results of the fits to
experimental data are shown in a phase diagram of hadronic and
quark-gluon matter in Fig.~\ref{fig:phasedia} \cite{AnBrSt06} with
full circles.

Lattice QCD results \cite{FoKa04,Ka04}
show a crossing, but no first order phase transition to the QGP
for vanishing or small chemical potentials $\mu_B$, i.e. at the
conditions accessible at central rapidities at RHIC full energies.
A first order phase transition is assumed
at high baryochemical potentials or baryon densities. This
corresponds to SPS energies and in the fragmentation region of
RHIC \cite{AnKoMc80,DaGySu85}. The critical baryochemical
potential is predicted \cite{FoKa04,Ka04} to be
$\mu_B^c = 400\pm50$~MeV and the critical temperature
$T_c = 140 \div 160$ MeV.

The thermodynamic parameters $T$ and $\mu_B$ --
as extracted from the microscopic UrQMD approach in the central overlap
region of Pb+Pb or Au+Au collisions \cite{Br04} --
are shown in Fig.~\ref{fig:phasedia}, where the full dots with error
bars denote the chemical freeze-out parameters
determined from fits to the experimental yields
\cite{ClRe99}.
The triangular and quadratic symbols (time-ordered in vertical
sequence) stand for temperatures $T$ and chemical potentials
$\mu_B$ extracted from UrQMD transport calculations in central
Au+Au (Pb+Pb) collisions at RHIC (21.3 ATeV),  SPS (160, 40
AGeV) and AGS (11 AGeV) \cite{Br99,Br02} as a
function of the reaction time (separated by 1 fm/c steps from top
to bottom). The open symbols denote non-equilibrium configurations
and correspond to $T$ parameters extracted from the transverse
momentum distributions, whereas the full symbols denote
configurations in approximate pressure equilibrium in longitudinal
and transverse direction. The $T$ and $\mu_B$ values found in the
transport model suggest that local thermal and chemical
equilibrium may exist in the interesting regions of the phase
diagram: the line of the first order phase transition, the critical
point, and the region of a crossover. Note, however, that most transport
models in their present formulation (such as UrQMD) have no
explicit QGP phase.

\subsection{Fluctuations and Correlations in High Energy Collisions}

Fluctuations and correlations are important characteristics of any
physical system. They  provide essential information about the
effective degrees of freedom.  In general, one can distinguish between
several classes of fluctuations.  There are ``dynamical'' fluctuations
and correlations reflecting the underlying dynamics of the system.
There are also ``trivial'' fluctuations induced by the measurement
process itself, such as finite number statistics, etc. They should be
clarified and subtracted in order to access the dynamical fluctuations
which tell as about the properties of the system as well as
sucseptibilities.

The event-by-event fluctuations in high energy A+A collisions  are
expected to be closely related to the transitions between
different phases of QCD matter. By measuring the fluctuations one
should observe anomalies from the onset of deconfinement and
dynamical instabilities when the expanding system goes through the
1-st order transition line between the QGP and the hadron gas. It
is predicted \cite{GaGoMr04} that the onset of deconfinement
should lead to a non-monotonous behavior in the energy dependence
of multiplicity fluctuations, the so-called ``shark fin''.
Furthermore, the QCD critical point may be signaled by a
characteristic pattern in enhanced fluctuations \cite{StRaSh99}.

\begin{figure}[t!]
\centerline{\epsfig{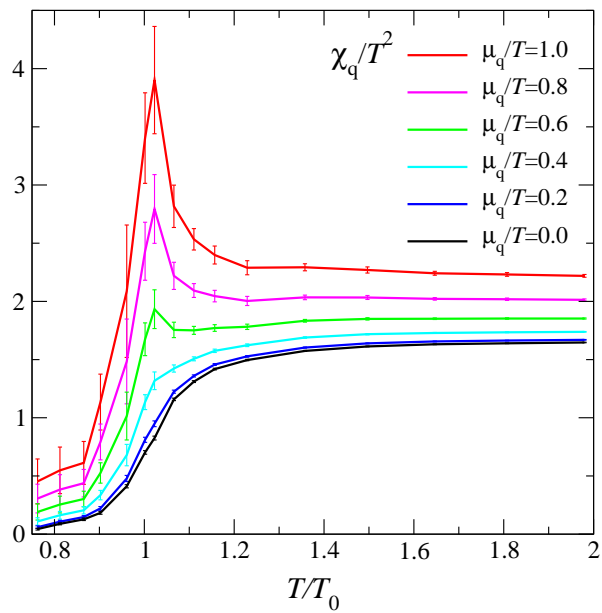}}
\caption{$\chi_q/T^2$ as a function of $T/T_{0}$ for various $\mu_q/T$
according to Ref.~\protect\cite{Al03}.
\label{fig:chis}}
\end{figure}

Indeed, the Lattice QCD calculations in Ref. \cite{Al03} indicate a sharp
increase of the susceptibilities at the critical point
for non-zero chemical potentials. This is illustrated in
Fig. \ref{fig:chis} which shows the dimensionless quark number
susceptibility $\chi_q/T^2$ as a function of $T/T_{0}$ for various
$\mu_q/T$ where the quark number $(q)$ susceptibility is
defined for the system in thermal equilibrium (for a grand-canonical
ensemble) as follows~\cite{Al03}:
\eq{\label{eq:chisq} \frac{\chi_q}{T^2} = \left(
\frac{\partial}{\partial (\mu_{\rm u}/T)}
+\frac{\partial}{\partial (\mu_{\rm d}/T)} \right) \frac{n_{\rm u}
+ n_{\rm d}}{T^3} } with $n_{\rm q}$ denoting the quark number
density. The peak, which develops in $\chi_q$ with increasing
$\mu_q$, is a sign that fluctuations in the baryon density are
growing when approaching the critical endpoint in the $(\mu,T)$.
Thus A+A collisions in the SPS energy region are expected to be a
suitable tool for a search of signatures for the critical point.

An ambitious experimental program for the search of the QCD critical
point has been started by the NA61 Collaboration  at the CERN SPS
\cite{Ga06,La07}. The program includes a variation in the  atomic
number $A$ of the colliding nuclei as well as an energy scan. This
allows to explore the phase diagram in the $T-\mu_B$ plane near the
critical point. One expects to `locate' the position of the critical
point by studying its `fluctuation signals'. High statistics
multiplicity fluctuation data will be taken for p+p, C+C, S+S, In+In,
and Pb+Pb collisions at bombarding energies of $E_{lab}$=10, 20, 30,
40, 80, and  158~AGeV.

The QCD critical point is expected to be experimentally seen as a
non-monotonic dependence of the multiplicity fluctuations, i.e. a
specific combination of atomic number $A$ and bombarding energy
$E_{lab}$ could move the chemical freeze-out of the system close to the
critical point and show a `spike' in the multiplicity fluctuations.
Since the available microscopic transport approaches 
Hadron-String-Dynamics (HSD) and 
Ultra-relativistic-Quantum-Molecular-Dynamics (UrQMD), 
which operate from lower SIS to top RHIC energies, do not include
explicitly a phase transition from a hadronic to a partonic phase, we
can not make a clear suggestion for the  location of the critical point
-- this is beyond the scope of  hadron-string models.  However, this
study should be helpful in the interpretation of the upcoming
experimental data since it will allow to subtract simple dynamical and
geometrical effects from the expected QGP signal.  The deviations of
the future experimental data from the HSD and UrQMD predictions may be
considered as an indication for the critical point signals.

We will deal in particular with particle number event-by-event fluctuations.
They are quantified by the ratio of the variance of the
multiplicity distribution to its mean value, the scaled variance.
Let's introduce the corresponding  notations:
The deviation $\Delta N_A$ from the average number $\langle
N_A\rangle$ of the particle species $A$ is defined by $N_A=\langle
N_A\rangle +\Delta N_A$, while the covariance for species $A$ and
$B$ is
\eq{\label{cov}
\Delta \left(N_A,N_B\right)~\equiv~\langle \Delta N_A \Delta
N_B\rangle~=~\langle N_A N_B\rangle ~-~\langle N_A\rangle \langle
N_B\rangle~,
}
the variance,
\eq{\label{Var}
Var(N_A) ~\equiv~\Delta\left(N_A,N_A\right)~=~\langle \left(\Delta
N_A\right)^2\rangle~=~ \langle N_A^2\rangle -\langle
N_A\rangle^2~,
}
the scaled variance,
\eq{\label{omega}
\omega_A ~\equiv~\frac{Var(N_A)}
{\langle
N_A\rangle}~=~\frac{\langle \left(\Delta
N_A\right)^2\rangle}{\langle N_A\rangle}~=~ \frac{\langle
N_A^2\rangle -\langle N_A\rangle^2}{\langle N_A\rangle}~,
}
and the correlation coefficient,
\eq{\label{corcoef}
\rho_{AB}~\equiv~\frac{\langle\Delta N_A~\Delta
N_B\rangle}{\left[\langle\left(\Delta
N_A\right)^2\rangle~\langle\left(\Delta
N_B\right)^2\rangle\right]^{1/2}}~.
}
Note that an absence of correlations, $\rho_{AB}=0$, implies
$P(N_A,N_B)=P_A(N_A)\times P_B(N_B)$, and $\omega=1$ for the
Poisson multiplicity distribution,
%
$P(N)~=~\overline{N}^N~\exp\left(-~\overline{N}\right)/N!$~.
%
%


\subsection{Outline of the Review}

The Sec. \ref{ch:HSD} provides the basic concept of transport approaches.
The centrality (Sec.~\ref{ch:mult}), energy (Sec.~\ref{ch:ppAA})
and system size dependence (Sec.~\ref{ch:NA61}) of event-by-event
fluctuations in the particle multiplicity are studied in A+A
collisions at the SPS CERN energies.
The results on particle number fluctuations and long-range
forward-backward multiplicity correlations at RHIC BNL energies are
discussed in Sec.~\ref{ch:corr}. Electric charge and baryonic
number fluctuations are considered in Sec.~\ref{ch:QB} while
fluctuations of the particle number ratios are presented in
Sec.~\ref{ch:ratio}.
A summary in Sec.~\ref{ch:summ} concludes the review.

The transport model calculations are performed within the HSD and
UrQMD approaches. The results of both models qualitatively agree to each
other. In some cases -- to illustrate the systematic uncertainties of
the transport model approaches for a specific observable -- we will
present the results of both models. Otherwise the results of only
HSD simulations are presented for brevity.

\section{Concept of Transport Approaches}
\label{ch:HSD}

Two independent transport models that employ hadronic and string
degrees of freedom, i.e. HSD \cite{EhCa96,CaBr99} and UrQMD
\cite{Ba98,Bl99}, have been used for studying the fluctuations in
$p+p$ and $A+A$ collisions. Both approaches take into account the
formation and multiple re-scattering of hadrons in A+A reactions
and thus dynamically describe the generation of pressure in the
hadronic expansion phase. Though quite different in the numerical
realization, both models are based on the same concepts and
dynamical degrees of freedom: strings, quarks, diquarks ($q,
\bar{q}, qq, \bar{q}\bar{q}$) as well as hadronic degrees of
freedom in the confined phase. We stress again that both
approaches do not include an explicit phase transition to a
quark-gluon plasma.

\subsection{The covariant off-shell transport approach  }

The novel HSD model (extended  for off-shell dynamics
\cite{BrCa08}) is based on the solution of generalized covariant
off-shell transport equations for hadronic degrees of freedom (for
the details of the covariant off-shell transport theory we refer
the reader to the review \cite{Ca09} and references therein). As
demonstrated in \cite{BrCa08} the off-shell dynamics is
particularly important for the transport description of the
propagation and interaction of resonances with broad spectral
functions which change their properties in the hot and dense
medium (e.g. for the vector mesons).

The coupled set of transport equations for the phase-space distributions
$N_h(x,p)$ $(x=(t,\vec r), ~ p=(\varepsilon,\vec p))$
of fermion $h$ \cite{EhCa96,CaBr99,We92} with a spectral function
$A_h(x,p)$ (using the Green function decomposition
$i {G}^{<}_h(x,p)=N_h(x,p) A_h(x,p)$)
is formally written as
\eq{
&\phantom{a}\hspace*{-2cm}
 \left\{ \left( \Pi_\mu-\Pi_\nu\partial_\mu^p U_h^\nu
  - m_h^*\partial^p_\mu U_h^S \right)\partial_x^\mu
  + \left( \Pi_\nu \partial^x_\mu U^\nu_h + m^*_h
    \partial^x_\mu U^S_h\right) \partial^\mu_p \right\}
    N_h(x,p) ~ A_h(x,p)                          \nn
&\phantom{a}\hspace*{-2cm}
  -\{ i{\Sigma}^{<},  Re~{G}^{R} \}
 = \sum_{h_2 h_3 h_4} Tr_2 Tr_3 Tr_4 ~
   [T^{\dagger} T]_{12\rightarrow 34}
   \delta^4 (\Pi +\Pi_2-\Pi_3-\Pi_4 ) \label{Ehg24} \\
&\phantom{a}\hspace*{-2cm}
 \hspace*{15em}  \times  A_{h}(x,p) A_{h_2}(x,p_2) A_{h_3}(x,p_3) A_{h_4}(x,p_4) \nn
&\phantom{a}\hspace*{-2cm}
\times \left\{ N_{h_3}(x,p_3) N_{h_4}(x,p_4) \bar{f}_h(x,p)
   \bar{f}_{h_2}(x,p_2) - N_h(x,p) N_{h_2}(x,p_2)\bar{f}_{h_3}(x,p_3)
   \bar{f}_{h_4}(x,p_4) \right\}. \nno
}
Here $\partial^x_\mu \equiv (\partial_t, \vec\nabla_r)$,
$\partial^p_\mu \equiv (\partial_\varepsilon, \vec\nabla_p), (\mu=0,1,2,3)$
and $\bar{f}_h (x,p)=1 - N_h (x,p)$ for fermions.

The 'backflow' term in (\ref{Ehg24})
-- i.e. the Poisson bracket with the self-energy and retarded
Green function -- is given by
\eq{ \label{eq:backflow}
\phantom{a}\hspace*{-2cm}
 -\{ i{\Sigma}^{<}, Re{G}^{R} \}
& = \partial^\mu_p \left(\frac{ M_h(x,p)}{ M_h(x,p)^2
  + \Gamma_h(x,p)^2/4}\right) ~ \partial_\mu^x
         \left(N_h(x,p) ~ \Gamma_h(x,p) \right)  \\[0.2cm]
& - \partial_\mu^x \left(\frac{ M_h(x,p)}{ M_h(x,p)^2
  + \Gamma_h(x,p)^2/4}\right) ~ \partial^\mu_p
         \left( N_h(x,p) ~ \Gamma_h(x,p) \right). \nno
}
It stands for the off-shell evolution which vanishes in the on-shell
limit, when the spectral function $A_h(x,p)$ does not change its
shape during the propagation through the medium, i.e. ${\vec \nabla}_r
\Gamma(x,p)$=0 and ${\vec \nabla}_p \Gamma(x,p)$=0 with $\Gamma$ denoting
the local width of the spectral function.

In (\ref{Ehg24}) the trace over particles 2,3,4 reads explicitly for
fermions
\eq{ \label{trace} Tr_2 = \sum_{\sigma_2, \tau_2}
\frac{1}{(2 \pi)^4} \int d^3 p_2 \frac{d M^2_2}{2
\sqrt{\vec{p}^2_2+M^2_2}},
}
where $\sigma_2, \tau_2$
stand for the spin and isospin of particle 2. In case of bosons one has
\eq{ \label{trace2}
Tr_2 = \sum_{\sigma_2, \tau_2} \frac{1}{(2 \pi)^4} \int d^3 p_2 \frac{d p_{0,2}^2}{2},
}
since here the spectral function $A_B$ is normalized as
\eq{
\label{sb} \int \frac{d p_0^2}{4 \pi} A_B(x,p) = 1
}
whereas for fermions one obtains the normalization
\eq{ \label{sb1}
\int \frac{d p_0}{2 \pi} A_h(x,p) = 1.
}
For the bosons $B$ the mass function is Eq. (\ref{eq:backflow})
\eq{ \label{massfunction2}
M_B(x,p) = p_0^2 - \vec{p}^{~2} - m^2 - {\rm Re}{{\Sigma}(x,p)}
}
where the (local and nonlocal) self-energies are included in ${\rm
Re}{{\Sigma}(x,p)}$.

The Lorentz covariant mass-function for the fermions is
\eq{ \label{massfunction3}
{M}_h(x,p) ~=~  \Pi_0^2 - \vec{\Pi}^{~2} - m_h^{*2} ~ ,
}
with the effective mass and four-momentum given by
\eq{\label{Ehg26}
m_h^* (x,p)&~=~ m_h + U_h^{{S}}(x,p) \\[0.2cm]
\Pi^\mu (x,p)&~=~ p^\mu-U^\mu_h (x,p)~. \nno
}
In (\ref{Ehg26})  $U_h^S(x,p)$ and
$U_h^\mu(x,p)$ denote the real part of the scalar and vector
hadron self-energies, respectively, and $m_h$ stands for the bare
(vacuum) mass of baryon $h$.

In (\ref{Ehg24}) $[T^+ T]_{12\rightarrow 34}$ is the 'transition
rate' for the process $1+2\rightarrow 3+4$ (which is taken to be
on-shell in the default HSD approach). In the c.m.s. of the
colliding particles the transition rate is given by
\eq{
[T^\dagger T]_{12\rightarrow 34} ~=>~
 v_{12} \left. \frac{d\sigma}{d\Omega} \right|_{1+2\to 3+4},
}
where $d\sigma /d \Omega$ is the differential cross section of the
reaction and $v_{12}$ the relative velocity of particles 1 and 2.
Note, that a generalized collision term for $n \leftrightarrow m$
reactions is given in Ref. \cite{Ca02} and optionally included in
HSD, which is, however, not used in the present study.

The hadron quasi-particle properties in (\ref{Ehg24}) are defined via
the mass-functions (\ref{massfunction2}) or (\ref{massfunction3}) with
(\ref{Ehg26}) while the phase-space factors
\eq{ \label{Ehg26a}
\bar{f}_h (x,p)=1 \pm N_h (x,p)
}
are responsible for fermion Pauli-blocking or Bose enhancement,
respectively, depending on the type of hadron in the final/initial
channel.  The transport approach (\ref{Ehg24}) is fully specified
by $U_h^S(x,p)$ and $U_h^\mu(x,p)$, which determine the mean-field
propagation of the hadrons, and by the transition rates $T^\dagger
T$ in the collision term, that describe the scattering and hadron
production/absorption rates. The coupled set of hadronic transport
equations (\ref{Ehg24}) may be employed for a large variety of
hadronic systems at relativistic (and non-relativistic) energies.

\subsection{On-shell limit}

In our present study we will discard the scalar and vector
potentials $U_h^S$ and $U_h^\mu$ in (\ref{Ehg24}) and
(\ref{Ehg26}) which are of minor importance at SPS and RHIC
energies due to the dominance of string excitations and decay.
Furthermore, we will employ the on-shell limit for the spectral
functions
$$A_h(x,p) \sim  \delta(p^2-m_h^2),$$
where $m_h$ denotes the hadron on-shell mass. These (legitimate)
approximations - employed in the standard transport approximations
- will allow for a large reduction in the necessary CPU time and a
proper statistics for the fluctuation analysis. We mention that
each individual cascade simulation will correspond to a
micro-canonical non-equilibrium simulation of the dynamics with
exact conservation of energy and momentum, baryon number, charge
number, strangeness etc.

Also note that the transport equations -- in the limits discussed above --
yield the well-known statistical
distributions
\begin{eqnarray}
N_h(x,p) =\frac{1}{e^{(p_0-\mu_h)/T}-\eta}
\label{eq:Nh}\end{eqnarray}
with $\eta=1$ for mesons and $\eta=-1$ for fermions in the limit
$t\to\infty$, if the system is confined to a finite (and fixed)
volume. In (\ref{eq:Nh}) $\mu_h$ and $T$ denote the conventional
Lagrange parameters for the chemical potential of hadron $h$ and
equilibrium temperature. Accordingly, the transport approaches
allow to explore non-equilibrium dynamical effects for
fluctuations observables while they smoothly approach the grand
canonical ensemble (GCE) in the limit $t\to\infty$ for stationary
systems when averaging over many events with a different amount of
excitation energy and different number of produced particles.


\subsection{Basic concept of HSD and UrQMD}

We recall that in the HSD (v. 2.5) \cite{EhCa96,CaBr99,GeCaGr98}
approach nucleons, $\Delta$'s, N$^*$(1440), N$^*$(1535),
$\Lambda$, $\Sigma$ and $\Sigma^*$ hyperons, $\Xi$'s, $\Xi^*$'s
and $\Omega$'s as well as their antiparticles are included on the
baryonic side whereas the $0^-$ and $1^-$ octet states are
incorporated in the mesonic sector.  Inelastic baryon--baryon (and
meson-baryon) collisions with energies above $\sqrt s_{th}\simeq
2.6 GeV$ (and $\sqrt{s_{th}} \simeq 2.3 GeV$) are described by the
FRITIOF string model \cite{NiSt86} whereas low energy
hadron--hadron collisions are modelled in line with experimental
cross sections. Low energy cross sections such as threshold meson
production in proton-neutron ($pn$) collisions -- which are
scarcely available from experiments --  are fixed by proton-proton
($pp$) cross sections and isospin factors emerging from
pion-exchange diagrams.  Since we address ultra-relativistic
collisions at SPS and RHIC energies such 'low energy
uncertainties' are of minor relevance here. As pre-hadronic
degrees of freedom HSD includes 'effective' quarks (antiquarks)
and diquarks (antidiquarks) which interact with cross sections in
accordance with the constituent quark model (cf. Refs.
\cite{Fa04}).

The UrQMD (v.1.3) transport approach \cite{Ba98,Bl99} includes all
baryonic resonances up to  masses of 2 GeV as well as mesonic
resonances up to 1.9 GeV as tabulated by the Particle Data Group
\cite{Mo94}. For hadronic continuum excitations a string model is
used with hadron formation times in the order of 1-2~fm/c
depending on the momentum and energy of the created hadron.  The
transport approach -- by construction -- yields a good
reproduction of nucleon-nucleon, meson-nucleon and meson-meson
cross section data in a wide kinematical range \cite{Ba98,Bl99}.

The HSD and UrQMD transport models have been used for the
description of $pp$, $pA$ and $AA$ collisions from SIS to RHIC
energies and lead to a fair reproduction of 'bulk' properties such
as hadron abundances, rapidity distributions etc. We mention that
some rare probes (e.g. particles with high transverse momentum)
are not well described by transport models (cf.  Refs.
\cite{WeBrCaSt03,BrSoStLeCa04,Br04}), however, this issue is of
minor impact on the fluctuation analysis to be presented below.

\section{Multiplicity Fluctuations in Nucleus-Nucleus Collisions at the SPS}
\label{ch:mult}

This Section contains the results for the particle number
fluctuations in Pb+Pb collisions at 158~AGeV within HSD and
UrQMD~\cite{Ko06}.
The results of these transport models are compared with the model
of independent sources and with the experimental data of the NA49
Collaboration.

\subsection{Fluctuations in the Number of Participants}

\begin{figure}[htp]
\centerline{\epsfig{file=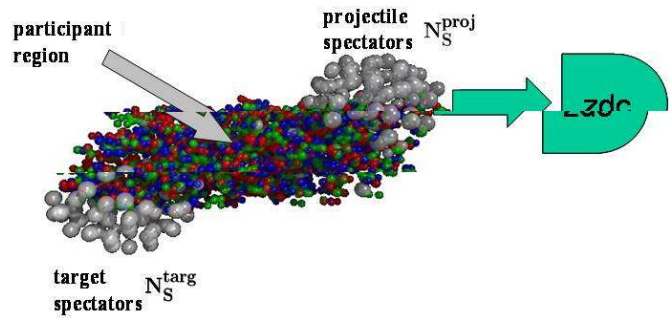,width=0.6\textwidth}}
\caption{Non-central heavy ion collision: Spectators in the
projectile and target are denoted as $N_P^{proj}$ and
$N_P^{targ}$. 'Participant region' indicates the region where the
particle production occurs due to the interaction of participants
from target and projectile. Note that in a fixed target experiment
it is possible to fix the number of projectile participants by
detecting projectile spectators via a Zero Degree Calorimeter
(ZDC).} \label{fig:participants}
\end{figure}

Fig.~\ref{fig:participants} demonstrates a typical non-central A+A
collision. In each collision only a fraction of all 2$A$ nucleons
interact. These are called participant nucleons and are denoted as
$N_P^{proj}$ and $N_P^{targ}$ for the projectile and target
nuclei, respectively. The corresponding nucleons, which do not
interact, are called the projectile and target spectators,
$N_S^{proj} = A - N_P^{proj}$ and $N_S^{targ} = A - N_P^{targ}$.
In the transport models the nucleon spectators are defined
according to the criteria, $|y-y_{beam(target)}|\le 0.32$, which
restricts their rapidities.

The fluctuations in high energy A+A collisions are dominated by a
geometrical  variation of the impact parameter. However, even for
the fixed impact parameter the number of participants, $N_P\equiv
N_P^{proj}+N_P^{targ}$, fluctuates from event to event. This is
due to the fluctuations of the initial states of the colliding
nuclei and the probabilistic character of the interaction process.
The fluctuations of $N_P$ form usually a large and uninteresting
background. In order to minimize its contribution the NA49
Collaboration has selected samples of collisions with a fixed
numbers of projectile participants. This selection is possible due
to a measurement of $N_S^{proj}$ in each individual A+A collision by a
calorimeter which covers the projectile fragmentation domain.
However, even in the samples with $N_P^{proj} = const$ the number
of target participants fluctuates considerably. Hence, an
asymmetry between projectile and target participants is
introduced, i.e. $N_P^{proj}$ is constant by constraint, whereas
$N_P^{targ}$ fluctuates.

\begin{figure}[ht!]
\centerline{\epsfig{file=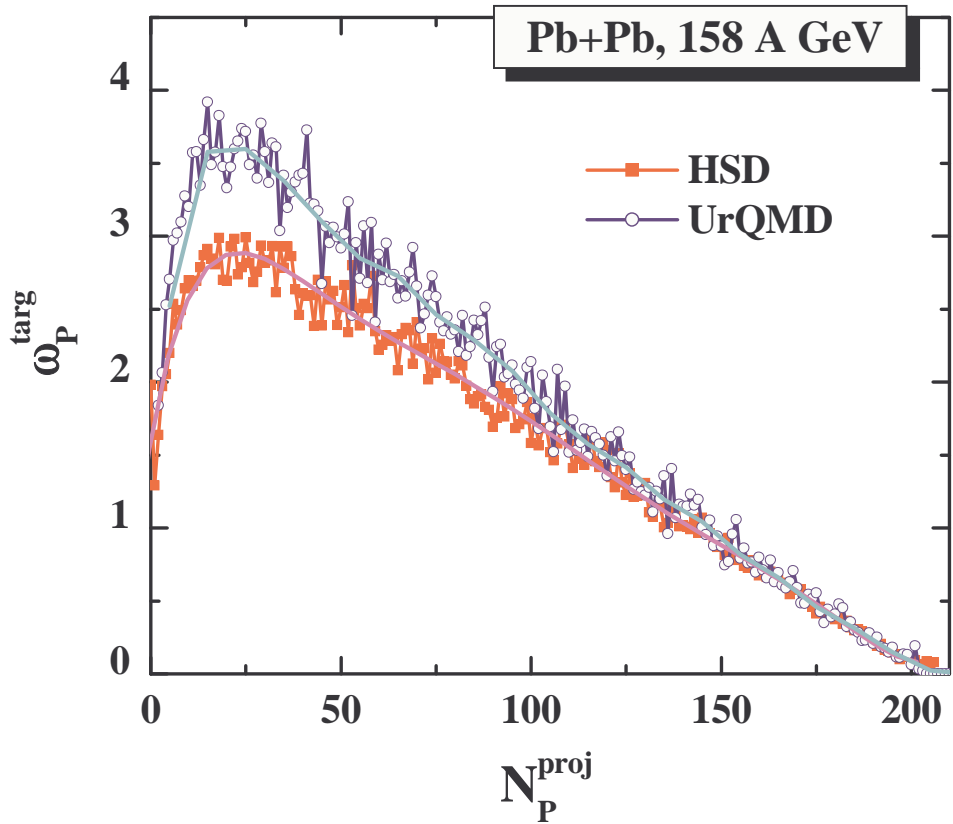,width=0.8\textwidth}}
\caption{The HSD and UrQMD results for the scaled variance
$\omega_P^{targ}$ as the function of $N_P^{proj}$.}
\label{fig:wNtarg}
\end{figure}

From an output of the HSD minimum bias simulations of Pb+Pb
collisions at 158~AGeV the samples of events with fixed values of
$N_P^{proj}$ have been formed. For the average number of target
participants in these samples one finds, $\langle
N_P^{targ}\rangle\cong  N_P^{proj}$. The deviations are only seen
at very small ($N_P^{proj}\approx 1$) and very large
($N_P^{proj}\approx A$) numbers of projectile participants. In
each sample with fixed value of  $N_P^{proj}$ the $N_P^{targ}$
number fluctuates around its mean value.
The scaled variance of these fluctuations is denoted as $\omega_P^{targ}$.
Fig.~\ref{fig:wNtarg} shows both the HSD and UrQMD results
for the scaled variance $\omega_P^{targ}$ as the function of $N_P^{proj}$.
The fluctuations of $N_ P^{targ}$ are quite strong;
the largest value of $\omega_P^{targ} = 3\div 3.5$ occurs at
$N_{P}^{proj}=20\div 30$.

\subsection{HSD and UrQMD Results for Pb+Pb at 158 AGeV }

\begin{figure}[t!]
\centerline{\epsfig{file=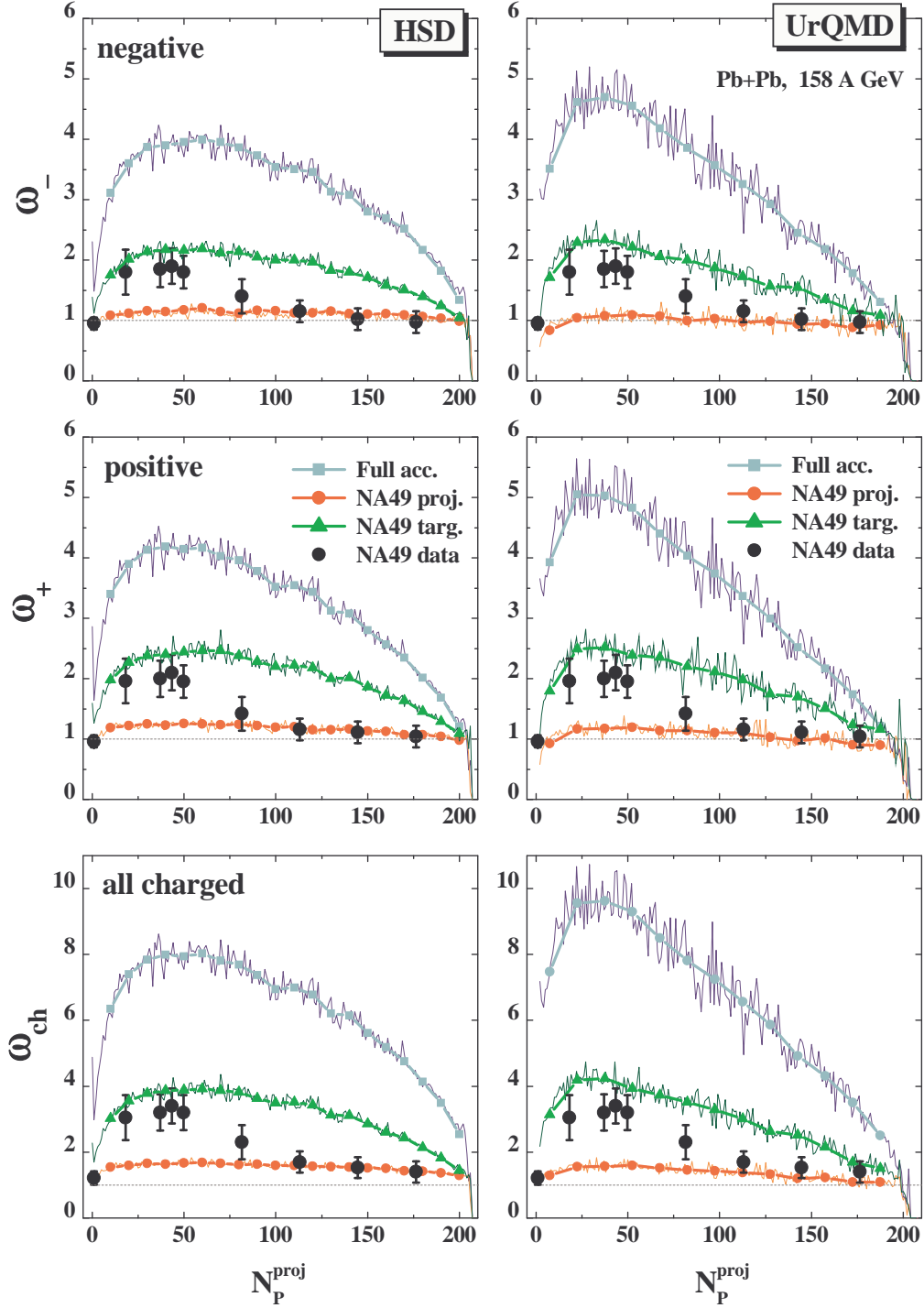,width=0.85\textwidth}} \caption{The
results of the HSD () and UrQMD (right) simulations are shown for
$\omega_-$, $\omega_+$, and $\omega_{ch}$ in Pb+Pb collisions at
158~AGeV as functions of $N_P^{proj}$. The black points are the
NA49 data. The different lines correspond to the model simulations
with the original NA49 acceptance, $1.1<y<2.6$, in the projectile
hemisphere  (lower lines), the NA49-like  acceptance in the mirror
rapidity interval, $-2.6<y<-1.1$, in the target hemisphere (middle
lines), and in full $4\pi$ acceptance (upper lines).}
\label{fig:w_Np}
\end{figure}

The NA49 Collaboration has minimized the event by event fluctuations of 
the number of nucleon participants in measuring the multiplicity 
fluctuations by selecting the samples of collisions with a fixed number 
of projectile spectators, $N_S^{proj} = const$, and thus a fixed number 
of projectile participants, $N_P^{proj} = A- N_S^{proj}$.  From an 
output of the HSD and UrQMD minimum bias simulations the samples of 
Pb+Pb events with fixed values of $N_P^{proj}$ have been formed. 
Fig.~\ref{fig:w_Np} presents the HSD and UrQMD results for the scaled 
variances of negatively, positively, and all charged particles, 
$\omega_i$ (with $i=+,-,ch$), in Pb+Pb collisions at 158 AGeV 
calculated at fixed $N_P^{proj}$.

The final particles in the HSD and UrQMD simulations are accepted
at rapidities $1.1< y <2.6$ ($y$ corresponds to the particle
rapidity in the Pb+Pb c.m.s. frame) in accordance to the NA49
transverse momentum filter \cite{Ry05}. This is done to compare
the HSD and UrQMD results with the NA49 data. The HSD and UrQMD
simulations both show flat $\omega_i$ values, $\omega_-\approx
\omega_+\approx 1.2$, $\omega_{ch}\approx 1.5$, and exhibit almost
no dependence on $N_P^{proj}$. The NA49 data, in contrast, exhibit
an enhancement in $\omega_i$ for $N_P^{proj}\approx 50$. The data
show maximum values, $\omega_-\approx \omega_+\approx 2$ and
$\omega_{ch}\approx 3$, and a rather strong dependence on
$N_P^{proj}$.

Fig.~\ref{fig:w_Np} also shows results of the HSD and UrQMD
simulations for the full 4$\pi$ acceptance for final particles,
and shows the NA49-like acceptance in the mirror rapidity
interval, $-2.6<y<-1.1$ of the target hemisphere. HSD and UrQMD
both result in large values of $\omega_i$, i.e. large fluctuations
in the backward hemisphere: in the backward rapidity interval
$-2.6<y<-1.1$ (target hemisphere) the fluctuations are much larger
than those calculated in the forward rapidity interval $1.1<y<2.6$
(projectile hemisphere, where the NA49 measurements have been
done). Even much larger fluctuations seen in Fig.~\ref{fig:w_Np}
follow from the HSD and UrQMD simulations for the full acceptance
of final particles.

At fixed values of the numbers of participants $N_P^{proj}$ and
$N_P^{targ}$ one can introduce
the probability $W_i(N_i; N_ P^{targ},N_P^{proj})$ for
producing  $N_i$ ($i=-,+,ch$) final hadrons.
At fixed $N_P^{proj}$
the averaging procedure is defined as,
\eq{\label{dynav}
\langle \cdots \rangle~\equiv~
\sum_{N_P^{targ}\ge 1}^{A}\sum_{N_i\ge 0} \cdots ~W_P(N_P^{targ}; N_P^{proj})~
W_i(N_i; N_ P^{targ},N_P^{proj})~,}
where $W_P(N_P^{targ}; N_P^{proj})$  is the probability for a
given value of $N_ P^{targ}$ in a sample of events with fixed
number of the projectile participants $N_P^{proj}$.
Under certain simplifications, the variance can be presented as,
\eq{
Var(N_i)~
\equiv~\langle N_i^2\rangle~-~\langle N_i\rangle^2~=
~\omega_i^*~\langle N_i\rangle~+~\omega_P~n_i~\langle N_i\rangle~,
\label{VarNi}
}
where the scaled variance $\omega_i^*$ corresponds to the fluctuations
of $N_i$ at fixed values of $N_P^{proj}$ and $N_P^{targ}$, $n_i
\equiv \langle N_i\rangle /\langle N_P \rangle$~, and
$N_P=N_P^{targ}+N_P^{proj}$ is the total number of participants.
In obtaining of Eq.~(\ref{VarNi}) two assumptions have been made.
First, it is assumed that $\omega_i^*$ does not depend on $N_P$.
The second assumption is that the average  multiplicities $\langle
N_i \rangle $ are proportional to the number of participating
nucleons.
Finally, the scaled variances, $\omega_i$, can be presented as:
\eq{ \label{omegai}
\omega_i~\equiv~\frac{Var(N_i)}{\langle N_i\rangle}~=~\omega_i^*~+~\omega_P~n_i~.
}
The average values are $\langle N_P^{targ}\rangle\cong
N_P^{proj}$, thus, $\langle N_P^{targ}\rangle\cong
\langle N_P \rangle /2$.
Therefore, the scaled variance $\omega_P$
for the total number of participants in Eq.~(\ref{omegai})
equals to $\omega_P=\omega_P^{targ}/2$ as only a half of the total
number of participants does fluctuate.
The value of $\omega_P^{targ}$ depends on $N_P^{proj}$, as shown
by the HSD and UrQMD results in Fig.~\ref{fig:wNtarg}. The average
particle number $n_i$ ($i=+,-,ch$) per participant calculated within the HSD and UrQMD
models for full $4\pi$-acceptance
show a good agreement with each other as well as with the
extrapolated to $4\pi$ NA49 data.

Eq.~(\ref{omegai}) corresponds to the so called
model of independent sources (MIS), see, e.g., Ref~\cite{He01}.
One assumes that final particles are produced by
the sources. The numbers of the sources are taken to be
proportional to the number of projectile and target participant
nucleons,
and the sources are assumed to be independent of each other.
The physical meaning of the sources depends on the model under
consideration (e.g.,  wounded nucleons \cite{BiBlCz76}, strings
and  resonances \cite{EhCa96,CaBr99,Ba98,Bl99}, or the fluid cells
at chemical freeze-out in the hydrodynamical models). The
Eq.~(\ref{omegai}) presents the final multiplicity fluctuations as
a sum of two terms: the fluctuations from one source,
$\omega_i^*$, and the contribution due to the fluctuations of the
number of sources, $\omega_P n_i$.

In peripheral A+A collisions there are only few nucleon-nucleon
(N+N) collisions, and
rescatterings are rare. The picture of independent N+N
collisions looks thus reasonable.  In this case, a hadron production source
can be associated with a N+N collision. The multiplicities
and scaled variances of N+N is given in terms of proton-proton (p+p),
proton-neutron (p+n), and neutron-neutron
(n+n) collisions:
\eq{ \label{mult-NN}\fl\hskip3em
\langle N_i^{NN}\rangle~=~\alpha_{pp}~\langle
N_i^{pp}\rangle~+~\alpha_{pn}~\langle
N_i^{pn}\rangle~+~\alpha_{nn}~\langle N_i^{nn}\rangle~, }
\eq{
 \label{omega-NN}\fl\hskip3em
\omega_i^{NN}~=~\frac{1}{\langle
N_i^{NN}\rangle}~\left[\alpha_{pp}~\omega_i^{pp}~\langle
N_i^{pp}\rangle~+~ \alpha_{pn}~\omega_i^{pn}~\langle
N_i^{pn}\rangle~ +~\alpha_{nn}~\omega_i^{nn}~\langle
N_i^{nn}\rangle\right]~,
}
where
$\alpha_{pp}=Z^2/A^2=0.155,~~\alpha_{pn}=2Z(A-Z)/A^2=0.478,~~
\alpha_{nn}=(A-Z)^2/A^2=0.367~$
are the probabilities of p+p, p+n, and
n+n collisions in Pb+Pb reactions (A=208, Z=82).  The
average multiplicities and scaled variances for elementary collisions
calculated within the HSD simulations at 158 GeV are equal to:
\eq{ \langle N_{ch}^{pp}\rangle ~&=~6.2~,~~~~ \langle
N_{ch}^{pn}\rangle ~=~5.8~,~~~~
\langle N_{ch}^{nn}\rangle ~=~5.4~,\label{NchNN}\\
      \omega_{ch}^{pp}~&=~2.1~,~~~~
      \omega_{ch}^{pn} ~=~2.4~,~~~~
      \omega_{ch}^{nn} ~=~2.9~.      \label{omegachNN}
}
For negatively and positively charged hadrons, the average
multiplicities and scaled variances  in elementary reactions can
be presented in terms of corresponding quantities for all charged
particles:
\eq{\label{ppplusmin}
\langle N_{\pm}\rangle ~=~\frac{1}{2}~\langle N_{ch}\rangle ~\pm~
\gamma)~,~~~~ \omega_{\pm}~=~\frac{1}{2}~\omega_{ch}
~\frac{\langle N_{ch}\rangle }{\langle N_{ch}\rangle ~\pm~
\gamma}~,
}
with $\gamma = 2, 1,0$ for pp, pn and nn reactions, respectively.
Thus, using Eqs.~(\ref{mult-NN}-\ref{ppplusmin}) one finds the HSD
results for $\omega_i^*$ per N+N  collision at 158 GeV:
\eq{
\omega_{ch}^*~=~2.5~,~~~~\omega_-^*~=~1.5~,~~~~\omega_+^*~=~1.1~.~~~~
\label{omegaiNN}
}

The above arguments  are not applicable for central A+A
collisions, where a large degree of thermalization is expected. In
the limit of $N_P^{proj}=A$ one can take the values of
$\omega_i^*$ from the Pb+Pb data or model simulations. In this
limit, $\omega_P=\omega_P^{targ}/2\approx 0$ (see
Fig.~\ref{fig:wNtarg}), and thus $\omega_i\approx\omega_i^*$. It
has been found that (\ref{omegaiNN}) gives a reasonable
description of $\omega_i$ in the HSD simulations for central Pb+Pb
collisions at 158 AGeV. Therefore, one can extrapolate
(\ref{omegai}) and (\ref{omegaiNN}) to all values of $N_P^{proj}$.
This leads to a good agreement for $\omega_i$ (\ref{omegai})
in the full 4$\pi$ acceptance with the transport
model simulations \cite{Ko06}.

The fluctuations of the total hadron multiplicities  generated by
the transport model dynamics  are large, i.e. the $\omega_i$ for
$4\pi$-acceptance are essentially larger than 1. The main
contributions to $\omega_i$ come from the second terms in
(\ref{omegai}), which are due to the fluctuations of $N_P^{targ}$.
These fluctuations of the target nucleon participants presented in
Fig.~\ref{fig:wNtarg} explain both, the large values of $\omega_i$
and their strong dependence on $N_P^{proj}$.

\begin{figure}[ht!]
\begin{center}
\epsfig{file=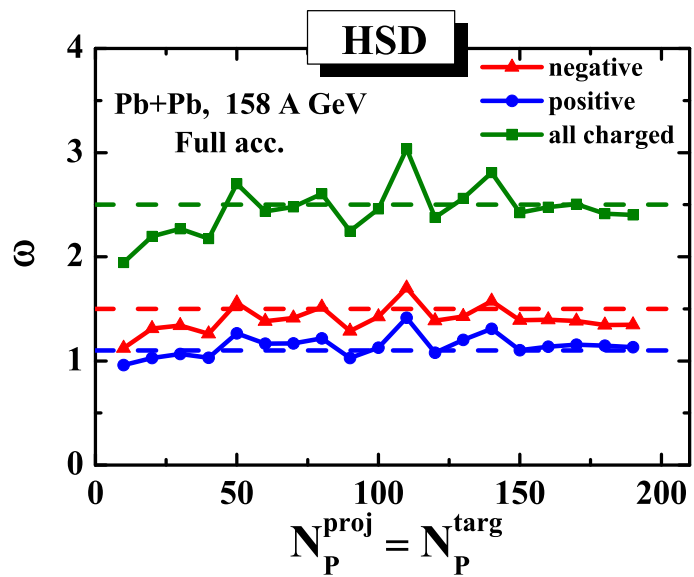,width=0.6\textwidth}
\end{center}
\caption{The circles, triangles, and boxes are the results of the
HSD simulations for $\omega_i$ in full $4\pi$ acceptance with
$N_P^{targ}=N_P^{proj}$. This condition yields
$\omega_P^{targ}=0$, and Eq.~(\ref{omegai}) is reduced to
$\omega_i=\omega_i^*$. The dashed lines correspond to $\omega_i^*$
taken from Eq.~(\ref{omegaiNN}).} \label{fig:w_NpNt}
\end{figure}

Fig.~\ref{fig:w_NpNt} supports the previous findings.  The HSD
events with fixed target participant number,
$N_P^{targ}=N_P^{proj}$, exhibit much smaller multiplicity
fluctuations. This is due to the fact that terms proportional to
$\omega_P^{targ}$ in (\ref{omegai}) do not contribute,  $\omega_i$
become approximately independent of the number of participants and
equal to $\omega_i^*$ (\ref{omegaiNN}).

\subsection{Transparency and Mixing in Nucleus-Nucleus Collisions
\label{sub:TMR}}

\begin{figure}[ht!]
\begin{center}
\epsfig{file=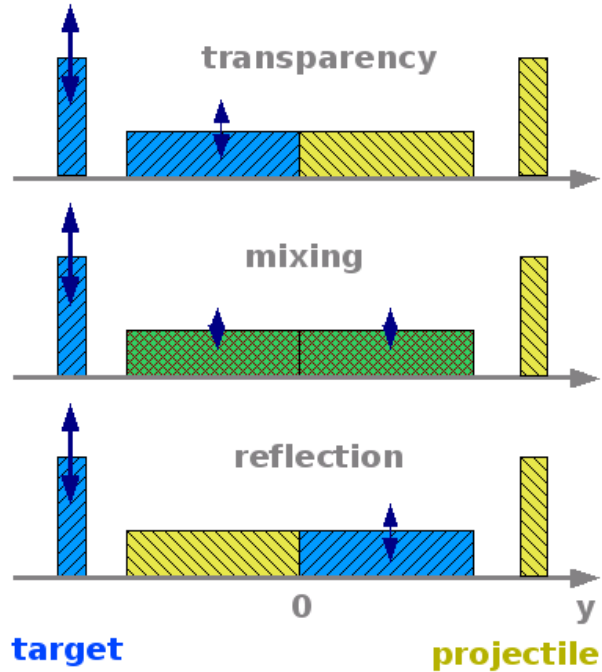,width=0.5\textwidth}
\end{center}
\caption{The sketch of the rapidity distributions of the baryon
number or the particle production sources (horizontal rectangles)
in nucleus-nucleus collisions resulting from the transparency,
mixing and reflection models. The spectator nucleons are indicated
by the vertical rectangles. In the collisions with a fixed number
of projectile spectators only matter related to the target shows
significant fluctuations (vertical arrows). See Ref.~\cite{GaGo06}
for more details.} \label{fig:sketch}
\end{figure}

Different models of hadron production in relativistic A+A
collisions can be divided into three limiting groups:
transparency, mixing, and reflection models (see
Ref.~\cite{GaGo06}). The first group assumes that the final
longitudinal flows of the hadron production sources related to
projectile and target participants follow in the directions of the
projectile and target, respectively. One calls this group of
models as transparency (T-)models. If the projectile and target
flows of hadron production sources are mixed, these models are
called as mixing (M-)models. Finally, one may assume that the
initial flows are reflected in the collision process. The
projectile related matter  then flows in the direction of the
target and the target related matter flows in the direction of the
projectile. This class of models corresponds to the reflection
(R-)models. The rapidity distributions resulting from the T-, M-,
and R-models are sketched in Fig.~\ref{fig:sketch} taken from
Ref.~\cite{GaGo06}.

\begin{figure}[t!]
\begin{center}
\epsfig{file=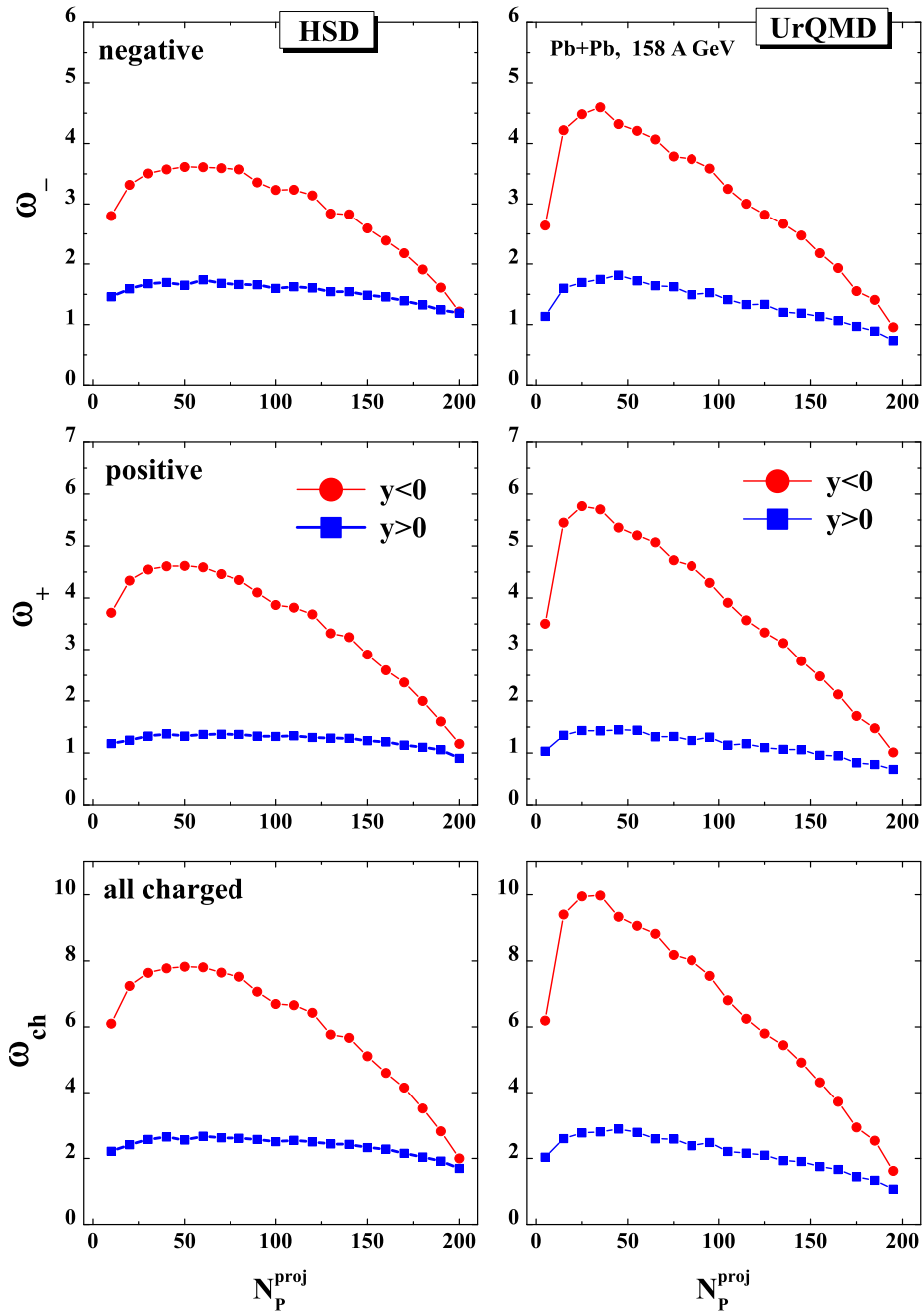,width=0.8\textwidth}
\end{center}
\caption{The scaled variances $\omega_i$ for the projectile
(boxes)
 and target (circles) hemispheres in the HSD ({\it left}) and UrQMD ({\it right}) simulations.}
\label{fig:proj-targ}
\end{figure}

An asymmetry between the projectile and target participants
introduced by the experimental selection procedure in a fix target
experiment can be used to distinguish between projectile related
and target related final state flows of hadron production sources.
The multiplicity fluctuations measured in the target momentum
hemisphere clearly are larger than those measured in the
projectile hemisphere in T-models. The opposite relation is
predicted for R-models, whereas for M-models the fluctuations in
the projectile and target hemispheres are expected to be the same.
Note that there are models which assume
the mixing of hadron production sources, however,
the transparency of baryon flows, e.g. three-fluid
hydrodynamical model \cite{Ka93,IvRuTo05}.

\begin{figure}[t!]
\begin{center}
\epsfig{file=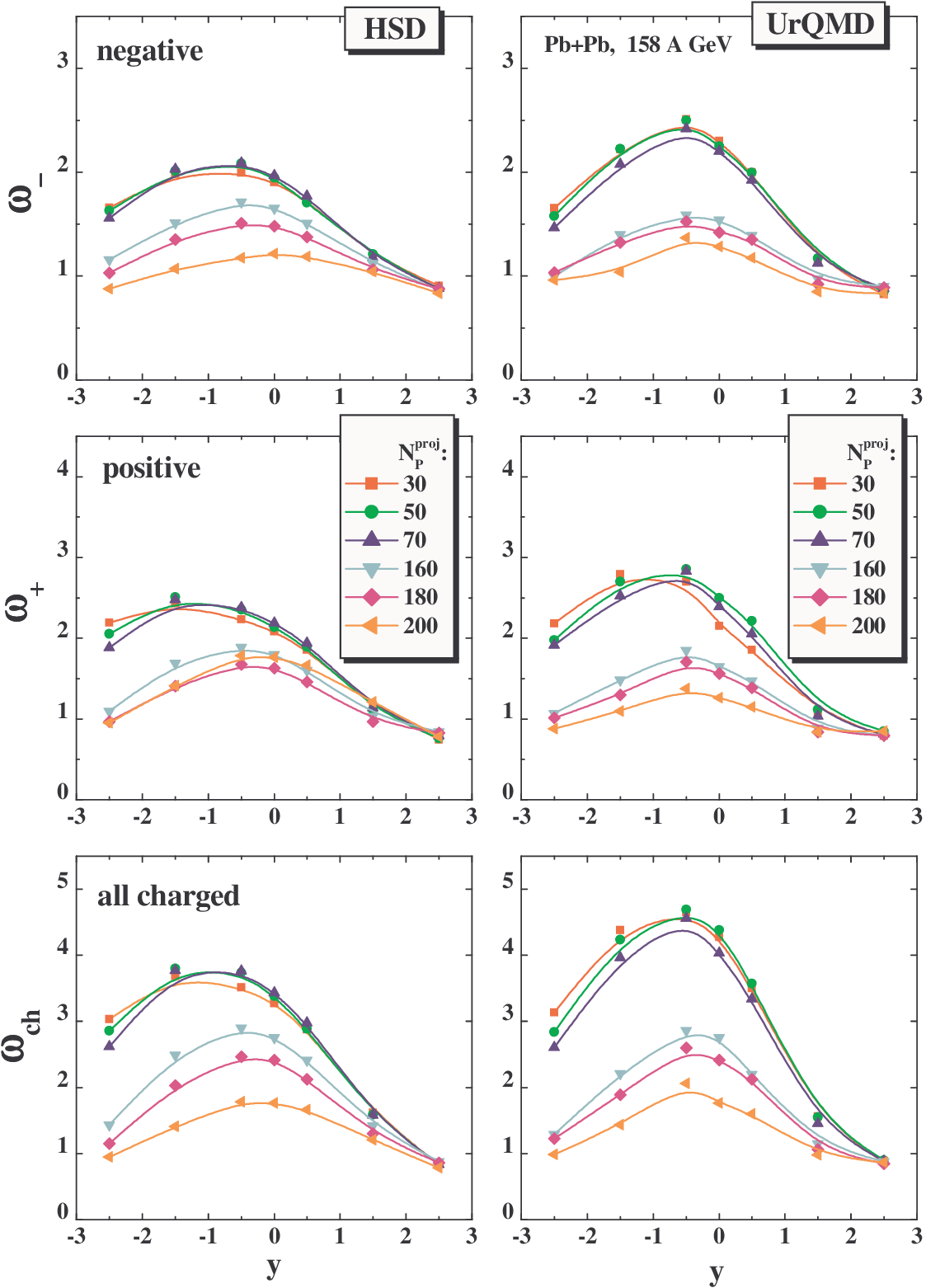,width=0.8\textwidth}
\end{center}
\caption{Scaled variances $\omega_-, \omega_+$, and $\omega_{ch}$
from the HSD ({\it left}) and UrQMD (right) in the rapidity
intervals $\Delta y=1$ around $y=0,\pm 0.5,\pm 1.5,\pm 2.5$
for different number of projectile participants $N_p^{part}$.}
\label{fig:w-yNproj}
\end{figure}

Now the fluctuations of the particle multiplicities in the
projectile ($y>0$) and target ($y<0$) hemispheres are considered.
As one can see from Fig.~\ref{fig:wNtarg}, in samples with
$N_P^{proj} = const$ the number of target participants,
$N_P^{targ}$, fluctuates considerably.  Of course, this event
selection procedure introduces an asymmetry between projectile and
target participants: $N_P^{proj}$ is constant, whereas
$N_P^{targ}$ fluctuates. Under this restriction the transport
models give very different results for the particle number
fluctuations in the projectile and target hemispheres. As clearly
seen from Fig.~\ref{fig:proj-targ} the particle number
fluctuations in the target hemispheres are much stronger  than
those in the projectile hemispheres. There is also a strong
$N_P^{proj }$-dependence of  $\omega_i$ in the target hemisphere,
which is almost absent for the $\omega_i$ in the projectile
hemisphere.
Thus, the fluctuations of $N_P^{targ}$ have a small influence on
the final multiplicity fluctuations in the projectile hemisphere,
but they contribute very strongly to those in the target
hemisphere.

A selection of collisions with a fixed number of $N_P^{proj}$ and
fluctuating number of $N_P^{targ}$ means that the projectile and
target initial flows are marked in fluctuations \cite{GaGo06}
in the number of colliding nucleons. The projectile and target
related matters in the final state of collisions can be then
distinguished by an analysis of fluctuations of extensive
quantities.
The analysis of the fluctuations applied to collisions of
identical nuclei gives a unique possibility to investigate the
flows of particle production sources.

Fig.~\ref{fig:w-yNproj} shows the scaled variances $\omega_-, \omega_+$
and $\omega_{ch}$ in the HSD and UrQMD simulations as  functions of
rapidity for different $N_p^{proj}$ values. It is clearly seen that the
bias on a fixed number of projectile participants reduces strongly the
particle fluctuations in the forward hemisphere, in particular within
the NA49 acceptance ($1.1 <y <2.6$).  The fluctuations of the target
participant numbers influence strongly the hadron production sources in
the target hemispheres. They also contribute to the projectile
hemisphere, but this contribution is only important in the rapidity
interval $0<y<1$, i.e. close to midrapidity.  It turns out that this
''correlation length'' in rapidity,  $\Delta y\approx 1$, as seen in
Fig.~\ref{fig:w-yNproj}, is not large enough to reproduce the data.
The large values of $\omega_i$  and their strong
$N_P^{proj}$-dependence in the NA49 data (cf. Fig.~\ref{fig:w_Np})  in
the projectile rapidity interval, $1.1<y<2.6$,  thus demonstrate a
significantly larger amount of mixing in peripheral reactions than
generated in simple hadron/string transport approaches.

\section{Multiplicity Fluctuations in Central Nucleus-Nucleus Collisions}
\label{ch:ppAA}

This Section presents the HSD results
on the excitation function of the multiplicity fluctuations
in central A+A collisions.
They will be compared with corresponding results in p+p collisions.
Note, that some observables in p+p and A+A collisions are rather close to each other.
For example, the charged hadron multiplicity per participating nucleon,
$n_{ch}\equiv \langle N_{ch}\rangle/\langle N_P\rangle$,
at SPS energies of $20\div 158$~AGeV are not  much
different in central Pb+Pb and inelastic p+p collisions
\cite{Af02,An04a}, $R_{ch}\equiv \left(n_{ch}\right)_{AA}/
\left(n_{ch}\right)_{pp}=1\div 1.5$~\footnote{Note that $R_{ch}<1$
at low collision energies. The change in behavior of $R_{ch}$  is
discussed in Ref.~\cite{GaGo99}.}. This explains a vitality of the
wounded nucleon model (WNM) \cite{BiBlCz76} which treats the final
state in A+A collision as the result of independent
N+N collisions (see, e.g., Ref.~\cite{BiCz05}
which discusses the recent data \cite{Ba05} on d+Au collisions at
the RHIC energies of $\sqrt{s_{NN}}=200$~GeV). However, the basic concept
of the WNM is in a severe conflict with many other data, e.g.,
with multi-strange baryon production,
$R_{\Omega}\equiv\left(n_{\Omega}\right)_{AA}/
\left(n_{\Omega}\right)_{pp}\cong 12.5$ \cite{Al05} in Pb+Pb at
158~AGeV. The $R_{\Omega}$ enhancement is expected to be even
stronger at smaller collision energies. The search for
quark-gluon plasma signatures in A+A collisions is usually based
on the expectation of a very different behavior of special
physical observables  in A+A and p+p collisions. Famous examples
of QGP signatures are the `strangeness enhancement', `$J/\psi$
suppression', and `jet quenching'. In all these cases one compares
a suitably normalized physical quantity in A+A and in p+p
reactions at the same collision energy per nucleon.

In general, one can define two groups of hadron observables. The first
group includes observables which are rather similar in  A+A and p+p
collisions, thus, they can be reasonably described within the WNM.  The
second group consists of A+A observables which are very different from
those in p+p collisions.  The question arises: are the multiplicity
fluctuations in A+A collisions close to those in p+p reactions, or are
they very different? The aim of present Section is to study this
question \cite{KoGoBr07}.

\subsection{Inelastic N+N and Central A+A Collisions}

\begin{figure}[ht!]
\centerline{\epsfig{file=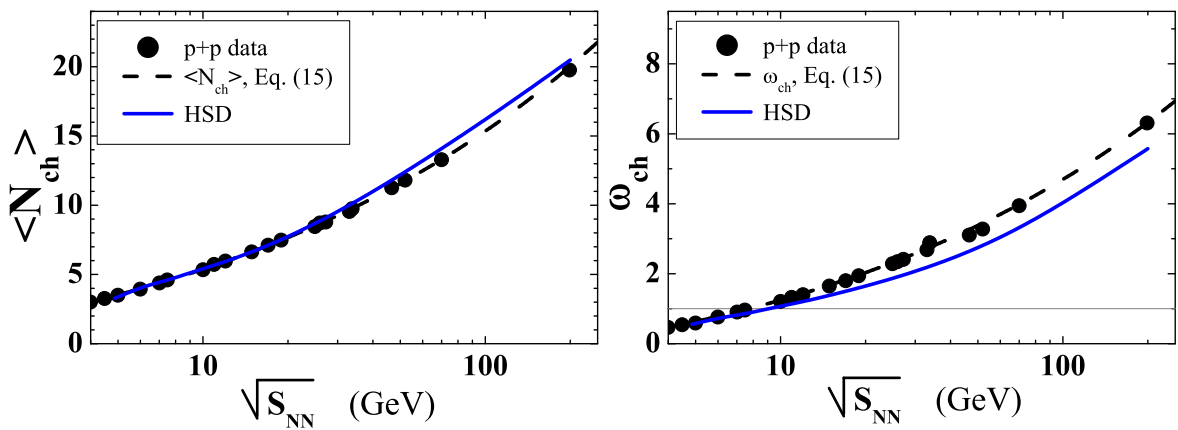,width=\textwidth}}
\caption{ The multiplicity {\it (left)} and scaled variance {\it
(right)} of all charged hadrons in p+p inelastic collisions as
functions of collision energy. The dashed lines correspond to the
parametrization (\ref{Nchpp}) from Ref.~\cite{He01}. The solid
lines are the HSD results.}
\label{fig:flucP1}
\end{figure}

The compilation of p+p data for $\langle N_{ch}\rangle$ and
$\omega_{ch}$ are taken from Ref.~\cite{He01} and presented in
Fig.~\ref{fig:flucP1}. The energy dependence can be parameterized by
the functions \cite{He01}:
\eq{\label{Nchpp}\fl\hskip3em
\langle N_{ch}\rangle~\cong~-4.2~+~4.69~\left(\frac{\sqrt{s_{NN}}}{\mbox{GeV}}\right)^{0.31}~,~~~~
\omega_{ch}~\cong~ 0.35~\frac{(\langle N_{ch}\rangle~-~1)^2}{\langle N_{ch}\rangle}~.
}
At high collision energies the KNO scaling \cite{KoNiOl72} holds which implies
that the multiplicity distribution $P(N_{ch})$ behaves as
$$\langle N_{ch}\rangle P(N_{ch})=\psi(N_{ch}/\langle N_{ch}\rangle)$$
(see also Ref.~\cite{Go78,GaSzWrWr91}).
For $\langle N_{ch}\rangle\gg 1$ it follows,
$$\langle N_{ch}^k\rangle =C_k\langle N_{ch}\rangle^k.$$
In particular, $\omega_{ch} \propto \langle N_{ch}\rangle$ \cite{Wr73}
as also seen from the parametrization (\ref{Nchpp}).

The HSD model description of the p+p data (for p+p reaction this
is almost equivalent to the Lund-String model \cite{AnGuPi93}) is shown in
Fig.~\ref{fig:flucP1} by the solid lines. It gives a good reproduction of
the p+p data for $\langle N_{ch}\rangle$, but slightly
underestimates $\omega_{ch}$ at high collision energies. For
negatively and positively charged hadrons the average
multiplicities and scaled variances  in p+p collisions can be
presented in terms of the corresponding quantities for all charged
particles, see Eq.~(\ref{ppplusmin}).

We compare now the central collisions of heavy nuclei and N+N inelastic
collisions (\ref{mult-NN},\ref{omega-NN}) within the HSD model.  A
small difference between p+p and N+N collisions is only present at the
SPS energies and gradually disappears at  RHIC energies.

\begin{figure}[ht!]
\centerline{\epsfig{file=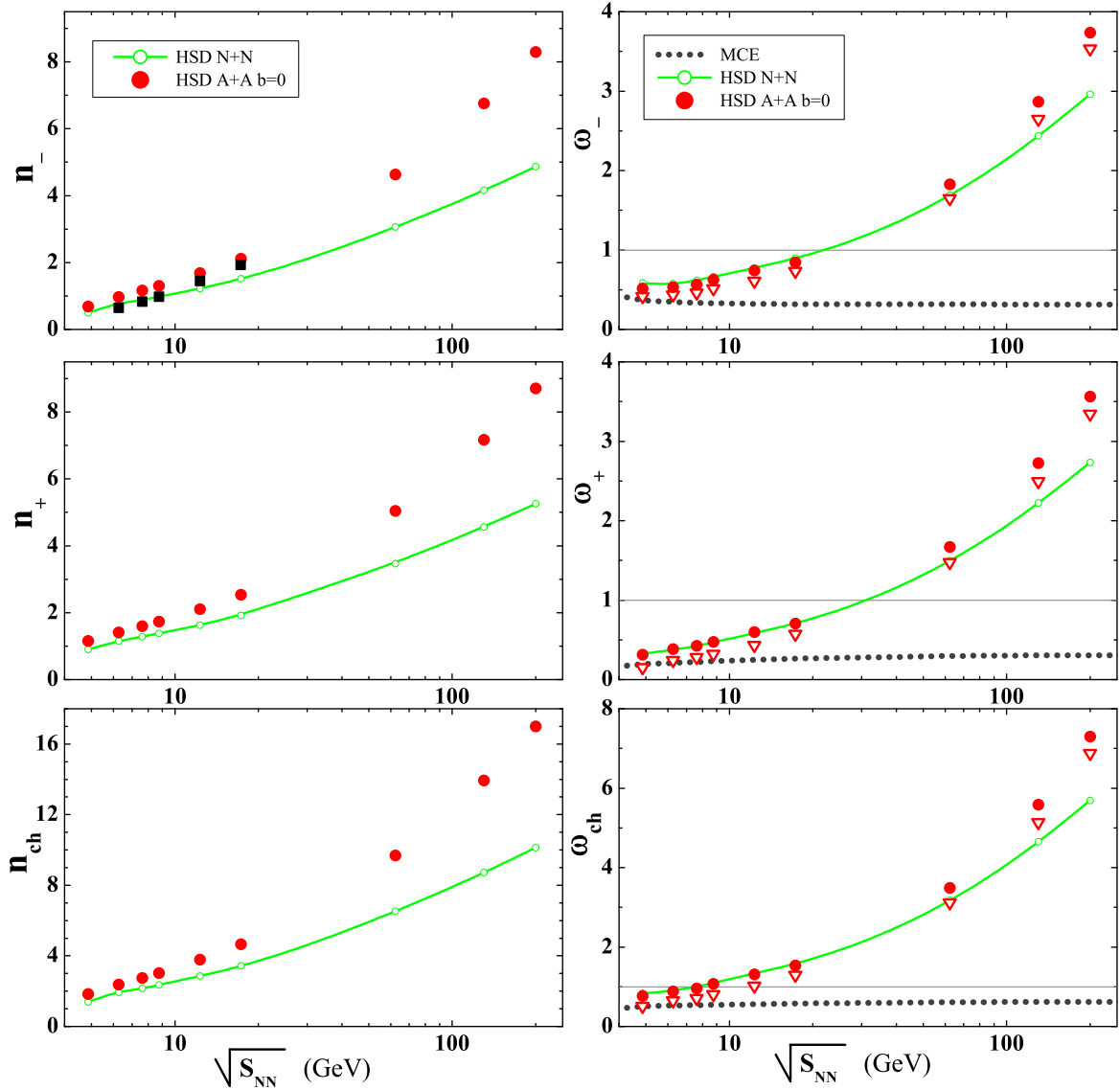,width=\textwidth}} \caption{The
multiplicities per participant, $n_i$ ({\it left}), and scaled
variances, $\omega_i$ ({\it right}). The solid lines are the HSD
results for N+N collisions according to (\ref{mult-NN}). The full
circles are the HSD results for central A+A collisions for zero
impact parameter, $b=0$.  The full squares for $n_-$ are the NA49
data \cite{Af02,An04a} for $(\langle \pi^-\rangle +\langle
K^-\rangle)/\langle N_P\rangle$ in the samples of 7\% most central
Pb+Pb collisions. The HSD results for $\omega_i$ after the
subtraction of the contributions $\Delta \omega_i=n_i\omega_P$.
are shown by open triangles. The dotted lines are the MCE HG model
results for $\omega_i$~\cite{Be07}. The HG parameters correspond
to the chemical freeze-out conditions  found from fitting the
hadron yields.} \label{fig:fluc-A}
\end{figure}

In Fig.~\ref{fig:fluc-A} the HSD model results are shown for the
multiplicities per participating nucleons, $n_i=\langle
N_i\rangle/\langle N_P\rangle$, and for the scaled variances,
$\omega_i$, in central collisions (zero impact parameter, $b=0$)
of Pb+Pb at $E_{lab}$=10, 20, 30, 40, 80, 158~AGeV and  Au+Au at
$\sqrt{s_{NN}}$=62, 130, 200~GeV. From Fig.~\ref{fig:fluc-A} one
concludes that the HSD results for the scaled variances in central
A+A collisions are close to those in inelastic N+N collisions.
The participant number fluctuations are found to be rather small
for central A+A collisions with zero impact parameter $b=0$.
For example, in Pb+Pb collisions with
$b=0$ at 158~AGeV the mean number of participants is
$\langle N_P\rangle \cong 392$, and the scaled variance is
$\omega_P\cong 0.055$ . The additional fluctuations, $\Delta
\omega_i$, of $i$th hadrons due to participant number fluctuations
can be estimated as,
$\Delta \omega_i ~=~ n_i~\omega_P~$.
The HSD results for $\omega_i$ after subtraction of the
contributions $\Delta \omega_i$
are shown in
Fig.~\ref{fig:fluc-A} by open triangles. These contributions to
$\omega_i$ due to participant number fluctuations
appear to be rather small at $b=0$, and they do not explain the
(positive) difference, $\omega_i($AA$)-\omega_i($NN$)$ seen in
Fig.~\ref{fig:fluc-A} at $\sqrt{s_{NN}}=200$~GeV.

On the other hand in the statistical model the scaled variances
$\omega_i = 1$ for the ideal Boltzmann gas in the grand canonical ensemble
(GCE). The deviations of $\omega_i$ from unity in the
hadron-resonance gas (HG) model stem from Bose and Fermi
statistics, resonance decays, and exactly enforced conservations
laws within the canonical ensemble (CE) or micro-canonical
ensemble (MCE) \cite{Be07,BeGaGoZo04,BeGoHaKoZo06}. Note that the statistical model
gives no predictions for the energy dependence of hadron
multiplicities. All yields are proportional to the system volume
$V$ which is a free model parameter fitted to the multiplicity
data at each collision energy. However, the statistical model does
predict the scaled variances $\omega_i$ as they become to be independent of
the system volume for large systems. In Fig.~\ref{fig:fluc-A} the
scaled variances $\omega_i$ calculated within the MCE HG model
along the chemical freeze-out line (see Ref.~\cite{Be07} for
details) are presented by the dotted lines:  $\omega_i$ reach
their asymptotic values at RHIC energies,
$\omega_{\pm}$(MCE)$\cong 0.3$ and $\omega_{ch}$(MCE)$\cong 0.6$.
The corresponding results in the GCE and CE are the following:
$\omega_{\pm}$(GCE)$\cong 1.2$ and $\omega_{ch}$(GCE)$\cong 1.6$,
$\omega_{\pm}$(CE)$\cong 0.8$ and $\omega_{ch}$(CE)$\cong 1.6$.
The HSD results for $\omega_i$ in central A+A collisions are very
different. They remain close to the corresponding values in p+p
collisions and, thus, increase with collision energy as
$\omega_i\propto n_i$. One observes no indication for
`thermalization' of fluctuations in the HSD results. This is
especially seen for RHIC energies:
$\omega_{i}$(HSD)$/\omega_i$(MCE)$\ge 10$ at
$\sqrt{s_{NN}}=200$~GeV.

\subsection{Comparison with the NA49 Data}

The fluctuations of the number of nucleon participants correspond to volume fluctuations,
hence, they translate directly to the final multiplicity fluctuations.
To avoid these `trivial' fluctuations, one has to select a sample of very central,
$\leq 1 \%$, collisions.
Such a rigid centrality selection has been recently done
for the NA49 data \cite{Lu06} by fixing the number
of projectile participants close to its maximal value $N_P^{proj}= A$.

\begin{figure}[ht!]
\centerline{\epsfig{file=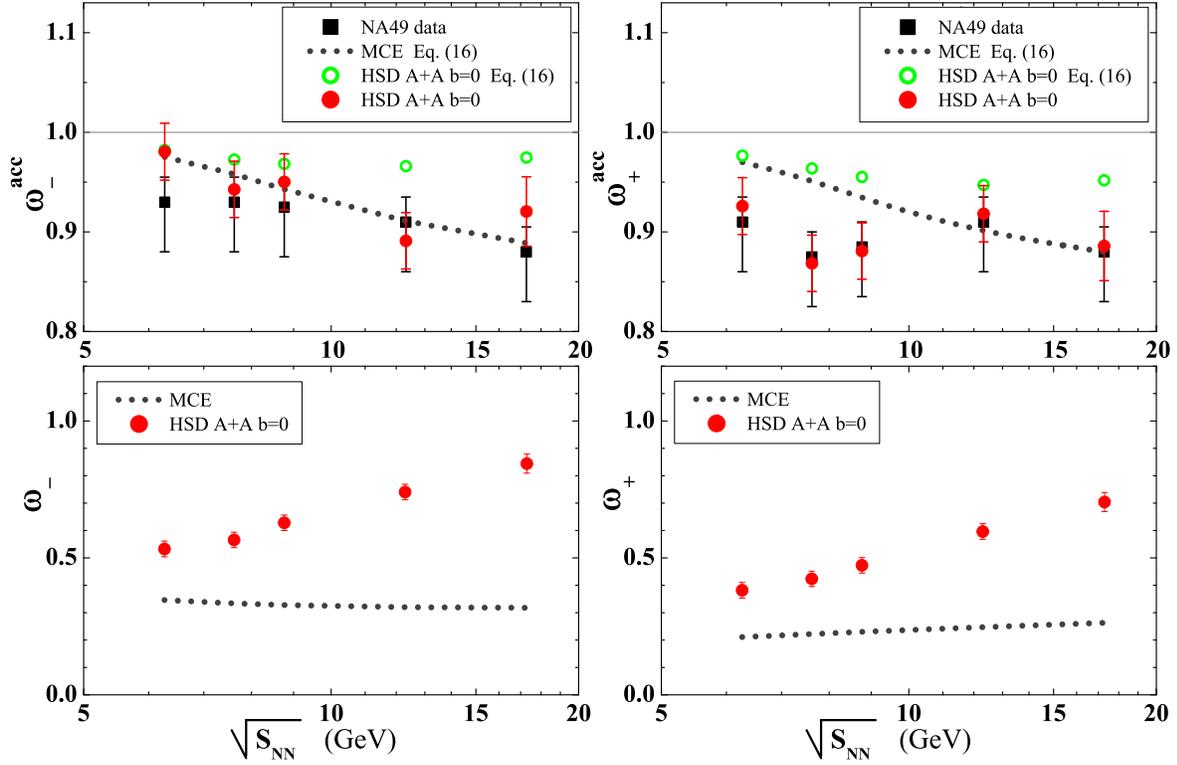,width=\textwidth}} \caption{{\it
Upper panel.} The scaled variances $\omega^{acc}_{\pm}$ for
central Pb+Pb collisions. The squares with error bars are the NA49
data for  1\% most central collisions \cite{Lu06}. The dotted
lines show the MCE HG model results calculated from full 4$\pi$
scaled variances using (\ref{acc}). The full circles present the
HSD results in Pb+Pb collisions for $b=0$ with the NA49
experimental acceptance conditions, while the open circles are
obtained from the 4$\pi$ HSD scaled variances using (\ref{acc}).
{\it Lower panel.} The MCE HG (dotted line) and HSD (full circles,
the same as in Fig.~\ref{fig:fluc-A}) for the 4$\pi$ scaled
variances $\omega_{\pm}$ at the SPS energies.} \label{fig:NA49}
\end{figure}

The HG model was compared with the NA49 data~\cite{Lu06} for
the sample of 1\% most central collisions at the SPS energies,
$20\div 158$~AGeV in Ref.~\cite{Be07}. It was found  that the
MCE results for $\omega_{\pm}$ are very close to the data, they
are shown by the dashed lines in Fig.~\ref{fig:NA49}. The
NA49 acceptance probabilities for positively and negatively
charged hadrons are approximately equal, and their numerical
values are: $q=0.038$, 0.063, 0.085, 0.131, 0.163, at the SPS
energies of 20, 30, 40, 80, 158~AGeV, respectively. In the
statistical model the scaled variances $\omega_{\pm}^{acc}$ for
the accepted particles are calculated from $\omega_{\pm}$ in the
full space according to the acceptance scaling formulae (see
Ref.~\cite{Be07} for details):
\eq{\label{acc}
\omega_{\pm}^{acc}~=~1~-~q~+~q~\omega_{\pm}~.
}
Note that the energy dependence of $\omega^{acc}_{\pm}$ seen in
Fig.~\ref{fig:NA49} is strongly influenced by an increase with energy
of the acceptance parameter $q$: only about 4\% of the hadrons are
detected at 20~AGeV and 16\% at 158~AGeV.

The comparison of the HSD results for central Pb+Pb collisions
(zero impact parameter, $b=0$) with the NA49 data of 1\% most
central collisions, selected by the number of projectile
spectators, is presented in Fig.~\ref{fig:NA49}. It demonstrates a
good agreement of the HSD results with the NA49 data. There are
also no essential differences between the MCE HG model and the HSD
transport model results. Several comments are needed at this
point. The HSD results within the NA49 acceptance demonstrate that
the acceptance scaling formulae (\ref{acc}) is violated. The
straightforward calculations (full circles in Fig.~\ref{fig:NA49})
lead to smaller values of $\omega_{\pm}^{acc}$ than those obtained
with the acceptance scaling formulae (\ref{acc}) (open circles in
Fig.~\ref{fig:NA49}). This difference may lead to a 10\% effect in
$\omega_{\pm}^{acc}$ for the NA49 acceptance conditions. Thus, the
MCE results for $\omega_{\pm}^{acc}$ may also be about 10\%
smaller than those obtained from (\ref{acc}) and shown in the
upper panel of Fig.~\ref{fig:NA49}. The lower panel of
Fig.~\ref{fig:NA49} demonstrates that the MCE and HSD results for
$\omega_{\pm}$ at the lowest  SPS energy 20~AGeV are
`occasionally' rather close to each other. They both are also
close to $\omega_{\pm}$ in p+p collisions.
The HSD scaled variances $\omega_{i}$ increase with collision
energy. In contrast, the MCE $\omega_i$ values remain
approximately constant. The ratio of the HSD to MCE values of
$\omega_{\pm}$ reaches about the factor of 2 at the highest SPS
energy 158~AGeV. It becomes a factor of 10 at the top RHIC energy
$\sqrt{s_{NN}}=200$~GeV. However, the rigid centrality selection
is absent for the available RHIC fluctuation data. Due to this
reason the participant number fluctuations give a dominant
contribution to $\omega_i$. On the other hand, for the SPS data
the small values of the acceptance, $q=0.04\div 0.16$, and 10\%
possible ambiguities coming from (\ref{acc}) almost mask the
difference between the HSD and MCE results (Fig.~\ref{fig:NA49},
upper panel).

\begin{figure}[ht!]
\includegraphics[height=5cm]{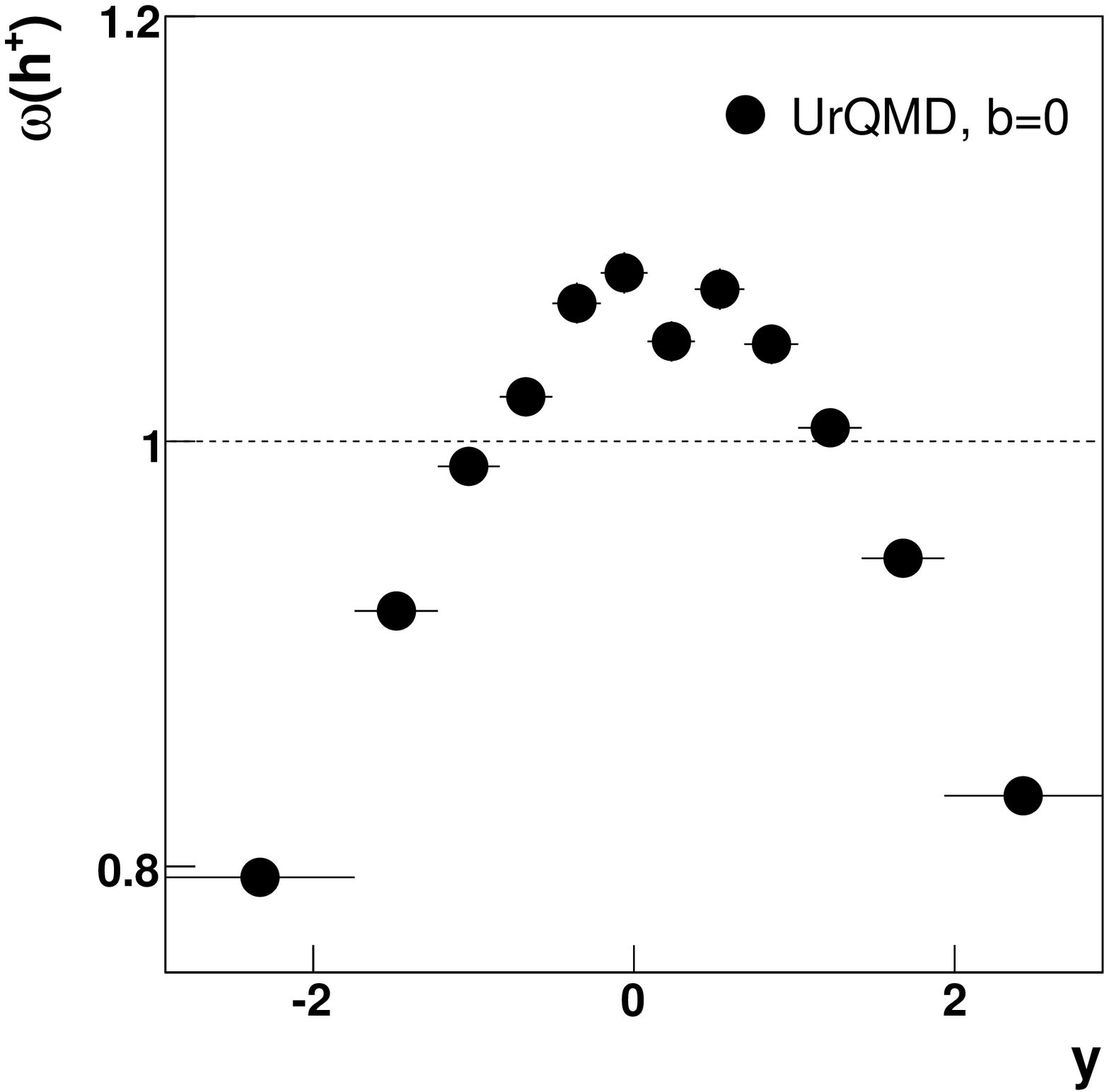}
\includegraphics[height=5cm]{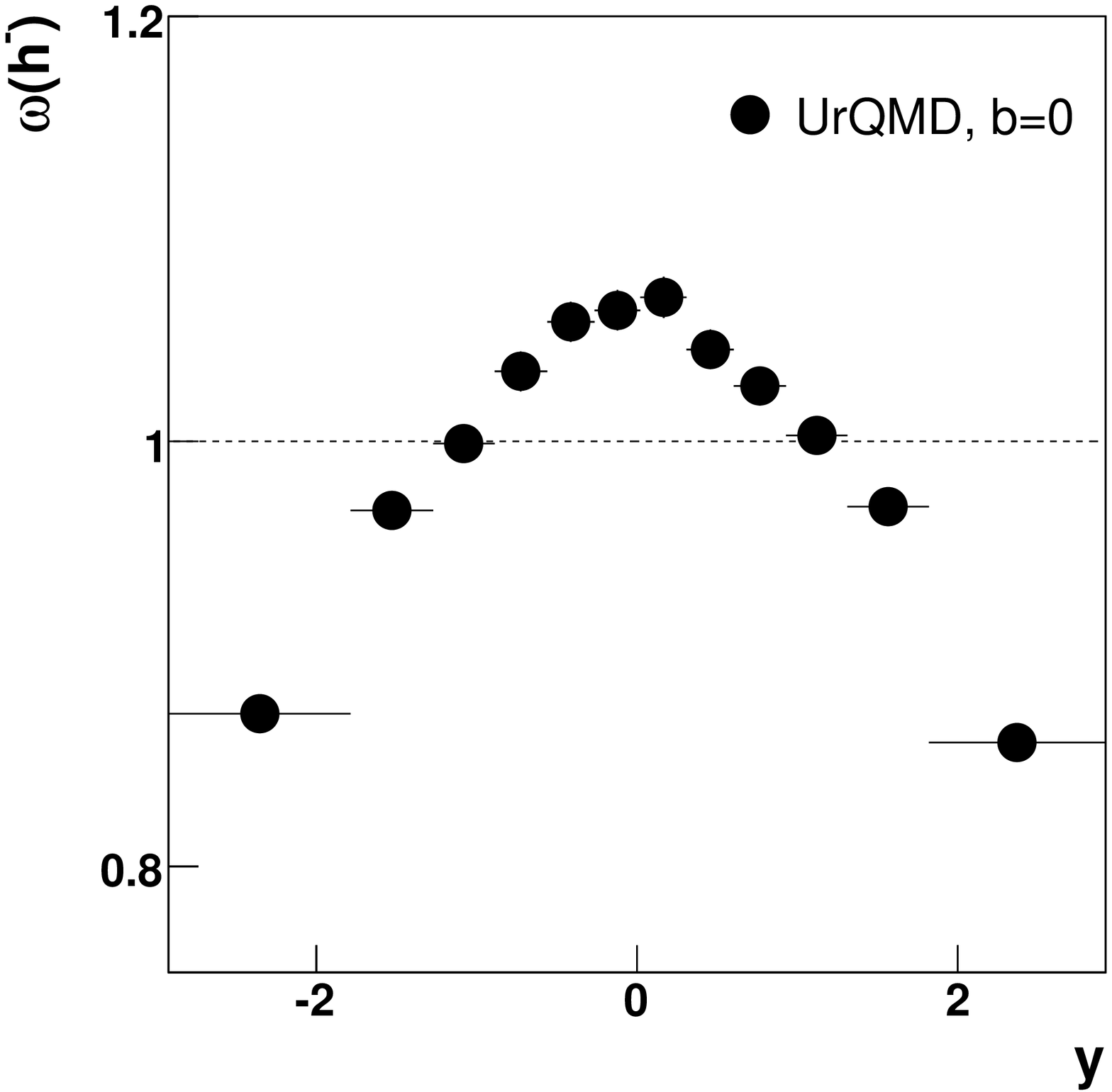}
\includegraphics[height=5cm]{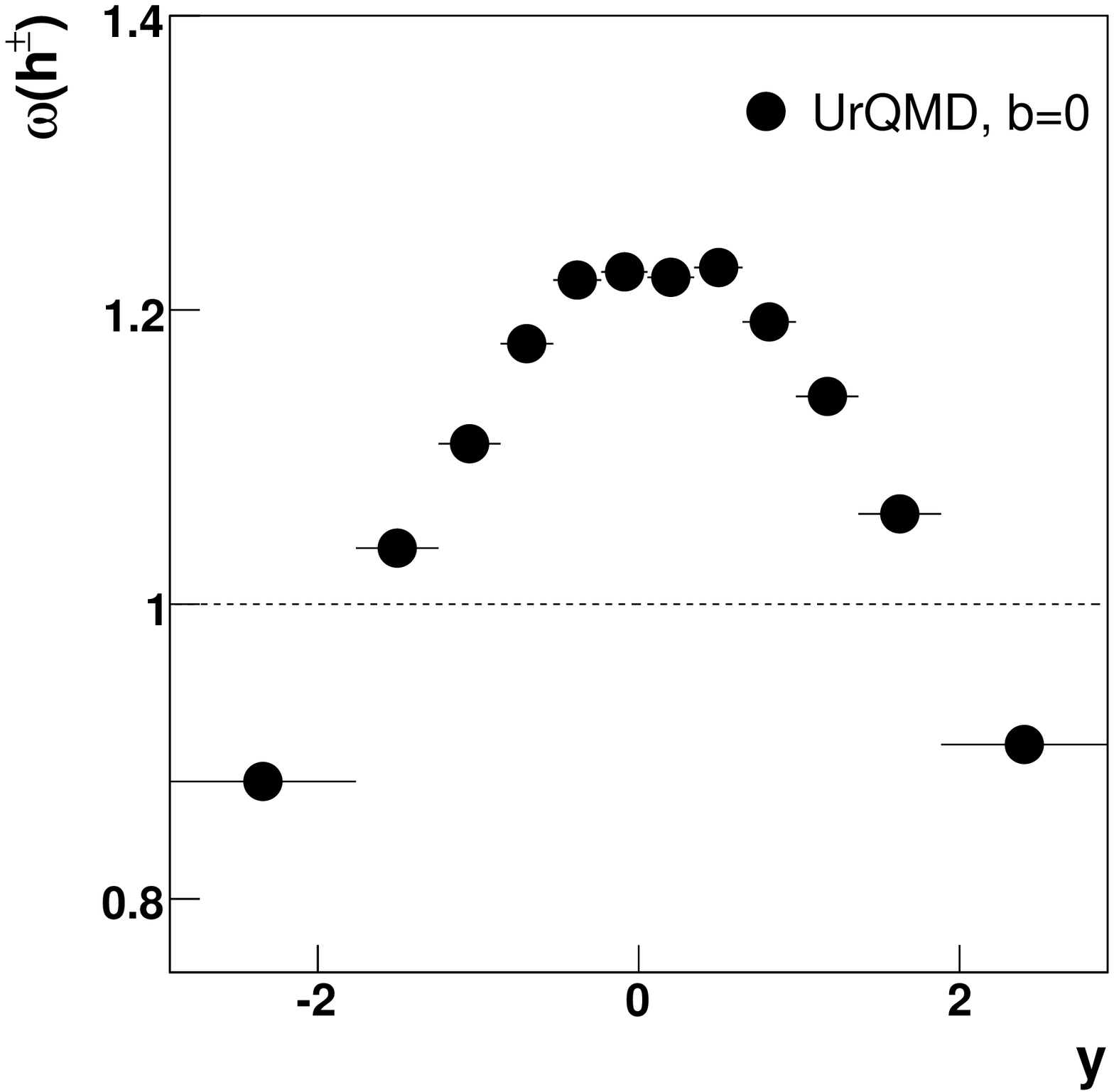}
\caption{\label{ydep_4pi}Rapidity dependence of scaled variance in
UrQMD simulation performed in full acceptance of positive (top),
negative (middle) and all charged (bottom) hadrons in central
Pb+Pb collisions at 158 AGeV. The rapidity bins are constructed in
such a way that the mean multiplicity in each bin is the same.
The figure is taken from Ref.~\cite{LuBl07}.
}
\end{figure}

Note that the scaled variance is a non-trivial function of the
selected phase-space. This issue has been addressed in
Ref.~\cite{LuBl07}. In order to study the dependence of scaled
variance on rapidity, 12 different rapidity intervals have been
constructed in such a way that the mean multiplicity in each
interval is the same. If the scaled variance would follow the
acceptance scaling formula~(\ref{acc}), the scaled variance would
be the same in each interval. However, as it has been shown in
Ref.~\cite{LuBl07}, the scaled variance is much higher near
midrapidity than in forward and backward rapidities, which is illustrated
in Fig.~\ref{ydep_4pi} taken from Ref.~\cite{LuBl07}.

\begin{figure}[ht!]
\includegraphics[height=5cm]{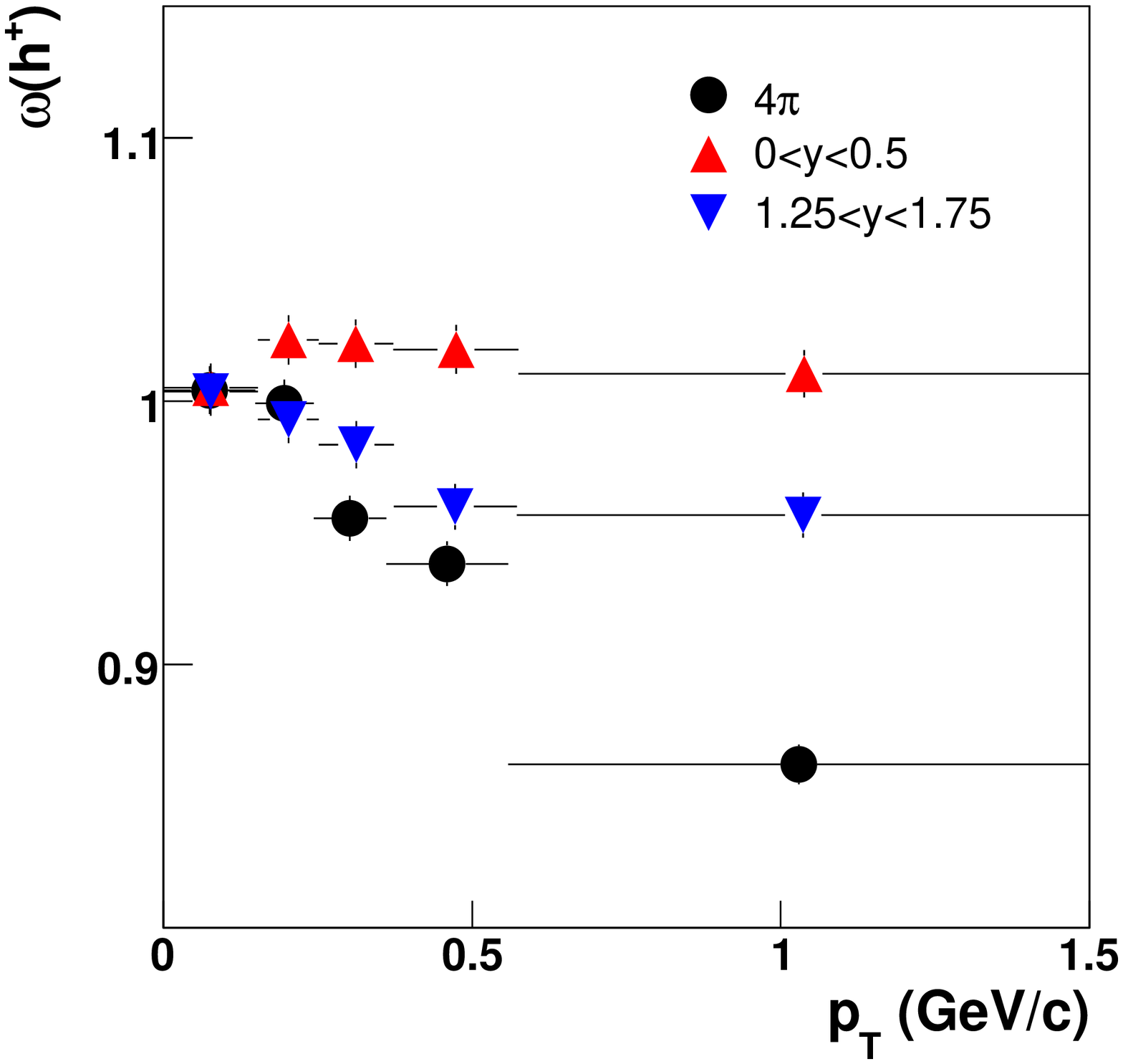}
\includegraphics[height=5cm]{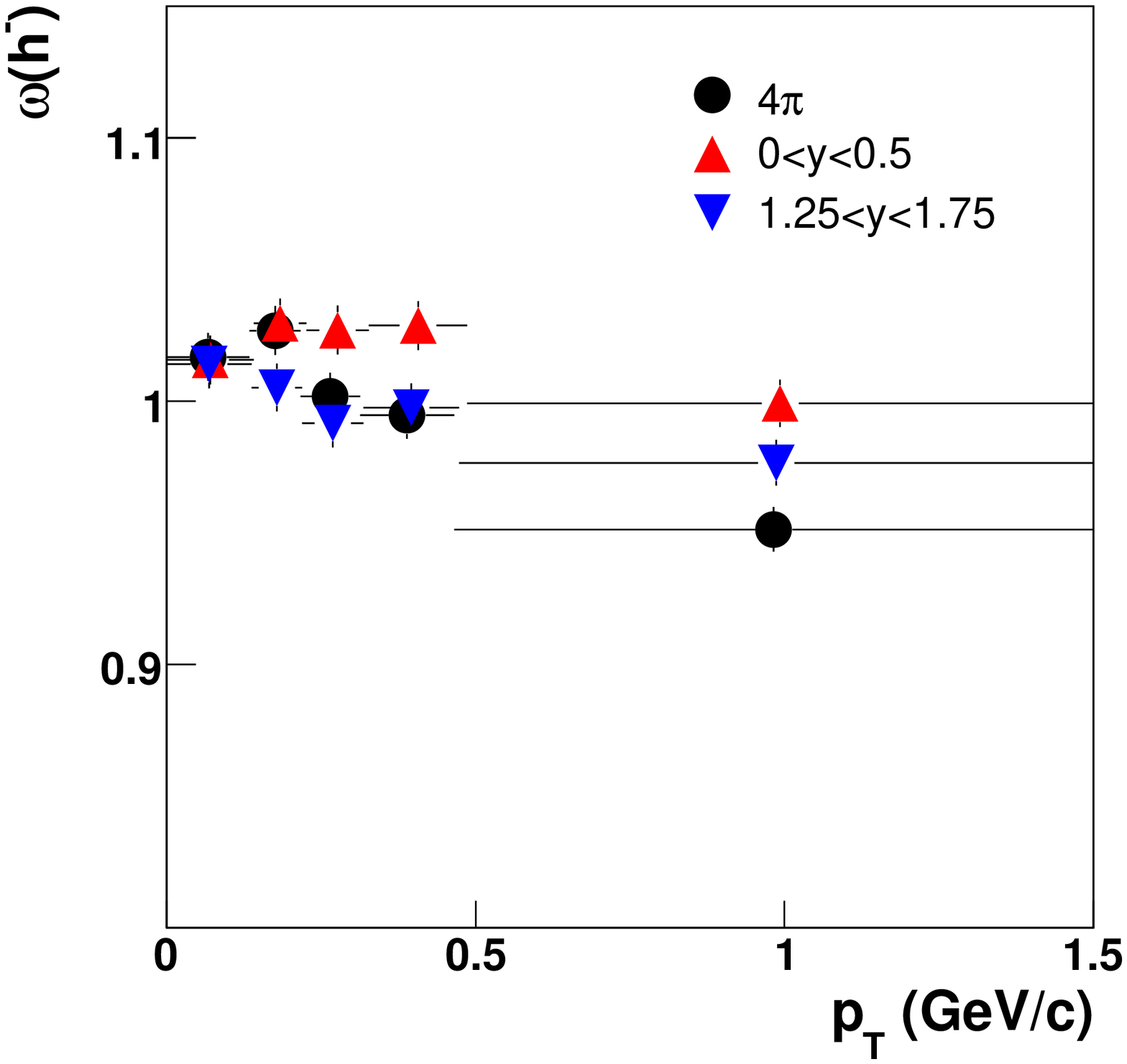}
\includegraphics[height=5cm]{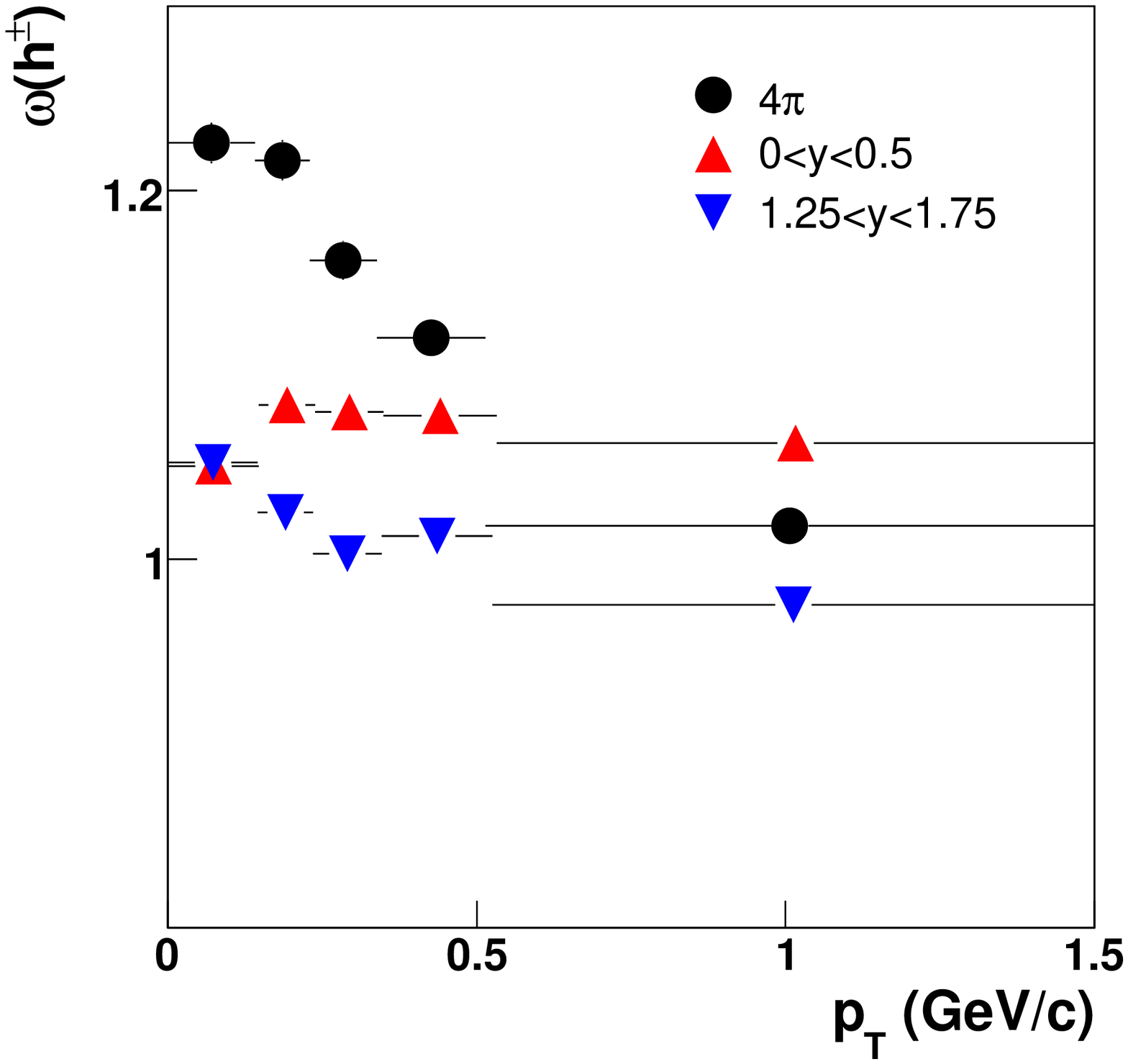}
\caption{\label{ptdep}(Color online) Transverse momentum
dependence of multiplicity fluctuations of positively (top),
negatively (middle) and all charged hadrons (bottom) for all
rapidities), $0<y<0.5$ and $1.25<y<1.75$. The transverse momentum
bins are constructed in such a way that the mean multiplicity in
each bin is the same.
The figure is taken from Ref.~\cite{LuBl07}.
}
\end{figure}

The transverse momentum dependence of scaled variance is shown in
Fig.~\ref{ptdep} for the full longitudinal phase-space and for a
midrapidity and a forward rapidity interval. The scaled variance
decreases with increasing transverse momentum for the full
acceptance and at forward rapidity. At midrapidity it stays
approximately constant. The decrease of scaled variance is
stronger for positively charged hadrons than for negatively
charged ones because the protons, which have smaller relative
fluctuations due to the large number of protons which enter the
collision, have a larger mean transverse momentum.

A similar effect of decreasing fluctuations for larger rapidities
and transverse momenta is observed as a result of energy- and
momentum conservation in a hadron gas model using the
micro-canonical ensemble~\cite{Ha08a}. It costs more energy to
create a particle with high momentum, therefore their number is
expected to fluctuate less.

\section{ Dependence on Energy and Atomic Number}
\label{ch:NA61}

\begin{figure}[ht!]
\begin{center}
\epsfig{file=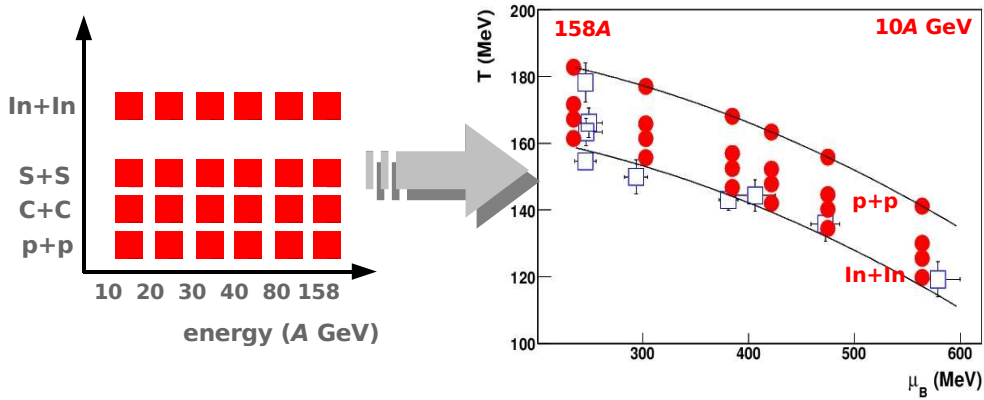, width=0.85\textwidth }
\end{center}
\caption{Left: The data sets on central A+A collisions planned to
be registered by NA61 in a search for the critical point of
strongly interacting matter and a study of the properties of the
onset of deconfinement. Right: Hypothetical positions of the
chemical freeze-out points of the reactions (In+In, S+S, C+C and
p+p from bottom to top at 158$A$, 80$A$, 40$A$, 30$A$, 20$A$ and
10$A$ GeV from left to right) to be studied by NA61 in the
(temperature)-(baryon-chemical potential) plane are shown by full
dots. The open squares show the existing Pb+Pb NA49 data.}
\label{fig:beam_phase}
\end{figure}

An ambitious experimental program for a search of the QCD critical
point has been started by the NA61 Collaboration  at the SPS
\cite{Ga06,La07}. The program includes a variation in the atomic
mass number $A$ of the colliding nuclei as well as an energy scan.
This allows to scan the phase diagram in the plane of temperature
$T$ and baryon chemical potential $\mu_B$ near the critical point
as argued in Ref.~\cite{Ga06,La07} and shown in
Fig.~\ref{fig:beam_phase} which is taken from Ref.~\cite{Ga08}.
One expects to `locate' the position of the critical point by
studying its `fluctuation signals'. High statistics multiplicity
fluctuation data will be taken for p+p, C+C, S+S, In+In, and Pb+Pb
collisions at bombarding energies of $E_{lab}$=10, 20, 30, 40, 80,
and  158~AGeV.

The aim of the Section is to present the results of studies on
the energy and system size dependence of  event-by-event
multiplicity fluctuations within the HSD and UrQMD microscopic
transport approaches.
Our study thus is in full correspondence to the experimental
program of the NA61 Collaboration \cite{Ga06,La07}.

The QCD critical point is expected to be experimentally seen as a
non-monotonic dependence of the multiplicity fluctuations, i.e. a
specific combination of atomic mass  number $A$ and bombarding
energy $E_{lab}$ could move the chemical freeze-out of the system
close to the critical point and show a `spike' in the multiplicity
fluctuations. Since  HSD and UrQMD do not include explicitly a
phase transition from a hadronic to a partonic phase, a clear
suggestion for the  location of the critical point can not been
made -- it is beyond the scope of such hadron-string models.
However, this study might be helpful in the interpretation of the
upcoming experimental data since it will allow to subtract simple
dynamical and geometrical effects from the expected QGP signal.
The deviations of the future experimental data from the HSD and
UrQMD predictions may be considered as an indication for the
critical point signals.

Theoretical estimates give about 10\%  increase of the
multiplicity fluctuations due to the critical point
\cite{StRaSh98,StRaSh99,St04}. It is large enough to be observed
experimentally within the statistics of NA61 \cite{Ga06,La07}. To
achieve this goal, it is necessary to have a control on other
possible sources of fluctuations. One of such sources is the
fluctuation of the number of nucleon participants. It has been
shown in Section~\ref{ch:mult} that these fluctuations give a
dominant contribution to hadron multiplicity fluctuations in A+A
collisions.
It was demonstrated that one can suppress the participant number
fluctuations by selecting most central A+A collisions. That's why
the NA61 Collaboration plans to measure central collisions of
light and intermediate ions instead of peripheral Pb+Pb
collisions.  It is important to stress, that the conditions for
the centrality selection in the measurement of fluctuations are
much more stringent than those for mean multiplicity measurements.

\subsection{Why Does One Need Central Collisions of Light Ions? 
\label{Np-fluc}}

To minimize the event-by-event fluctuations of the number of
nucleon parti\-cipants in measuring the multiplicity fluctuations
the NA49 Collaboration has been trying to fix $N_P^{proj}$ in Pb+Pb
collisions. Samples of collisions with a fixed number of
projectile spectators, $N_S^{proj} = const$ (and thus a fixed
number of projectile participants, $N_P^{proj}$), have been selected.
A similar centrality selection is expected to be implemented in the
NA61 experiment.
\begin{figure}[ht!]
\begin{center}
\epsfig{file=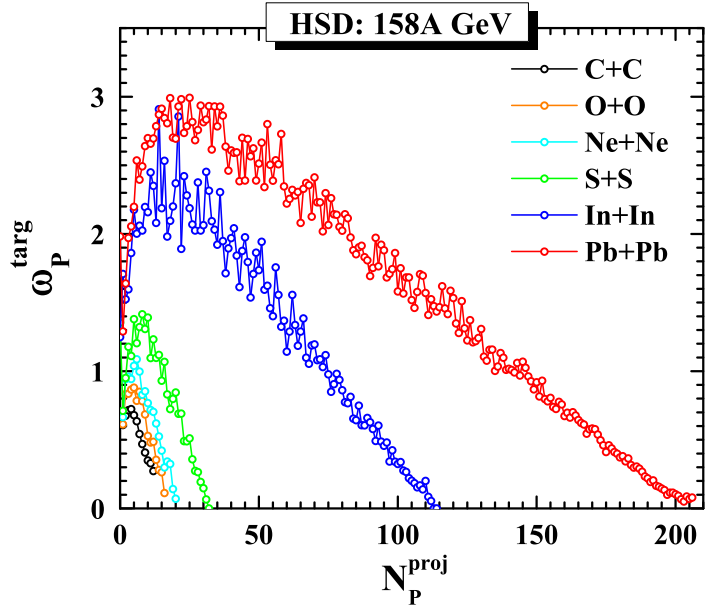,width=0.6\textwidth}
\end{center}
\caption{The scaled variance $\omega_P^{targ}$  for the
fluctuations of the number of target participants, $N_P^{targ}$.
The HSD simulations of $\omega_P^{targ}$ as a function of
$N_P^{proj}$ are shown for different colliding nuclei, In+In, S+S,
Ne+Ne, O+O and C+C at $E_{lab}$=158~AGeV.} \label{fig:w_targ}
\end{figure}

Fig.~\ref{fig:w_targ} presents the HSD scaled variances
$\omega_P^{targ}$ for C+C, O+O, Ne+Ne, S+S, In+In, and Pb+Pb
collisions at 158~AGeV as a function of $N_P^{proj}$. The
fluctuations of $N_P^{targ}$ are quite strong for mid-peripheral
reactions with $N_P^{proj}=20\div 30$ and negligible for the most
central collisions.
A vanishing of $\omega_P^{targ}\cong 0$ at $N_P^{proj}\cong $A
does not, however, show up  in collisions of light nuclei (from C
to S). Even for the  maximal values of $N_P^{proj}=A$ the
fluctuations $\omega_P^{targ}$ do not vanish and increase with
decreasing atomic mass  number $A$. For example in C+C collisions
for $N_P^{proj}=A=12$  the number of participants from the target
still fluctuates and the scaled variance amounts to
$\omega_P^{targ}\cong 0.25$.
Some combination of the system size $N_P$ and collision energy
$E_{lab}$ might move the chemical freeze-out point close the QCD
critical point. One could then expect an increase of multiplicity
fluctuations in comparison to their `background values'.

Why does one need central collisions of light and intermediate
ions instead of studying peripheral Pb+Pb collisions for a search
of the critical point?  Fig.~\ref{fig:w_targ} explains this issue. At
fixed $N_P^{proj}$ the average total number of participants,
$N_P\equiv N_P^{proj}+N_P^{targ}$, is equal to $\langle N_P\rangle
\cong 2 N_P^{proj}$, and, thus, it fluctuates as
$\omega_P=0.5\omega^{targ}_P$. Then, for example, the value of
$N_P^{proj}\cong 30$ corresponds to almost zero participant number
fluctuations, $\omega_P\cong 0$, in S+S collisions while
$\omega_P$ becomes large and is close to 1 and 1.5 for In+In and
Pb+Pb, respectively. Even if $N_P^{proj}$ is fixed exactly, the
sample of the peripheral collision events in the heavy-ion case
contains large fluctuations of the participant number:  this would
`mask' the critical point signals. As also seen in
Fig.~\ref{fig:w_targ} ({\it right}), the picture becomes actually more
complicated if the atomic mass  number $A$ is too small. In this case,
the number of participants from a target starts to fluctuate
significantly even for the largest and fixed value of
$N_P^{proj}=$A.

\subsection{Multiplicity Fluctuations at Zero Impact Parameter}

The importance of a selection of the most central collisions for
studies of hadron multiplicity fluctuations has been stressed in
the previous sections. Due to its convenience in theoretical
studies (e.g., in hydrodynamical models) one commonly uses the
condition on impact parameter $b$ for the selection of the  `most
central' collisions in model calculations. However, the number of
participant even at $b=0$ is not strictly fixed and fluctuates
according to some distributions.
It should be also stressed that the conditions $b<b_{max}$ can not
be fixed experimentally since the impact parameter itself can not
be measured in a straightforward way. Actually, in experiments one
accounts for the 1\%, 2\% etc. most central events selected by the
measurement of spectators in the Veto calorimeter, which
corresponds to the event class with the largest $N_P^{proj}$.  As
it will be demonstrated below  the multiplicity fluctuations are
very sensitive to the centrality selection criteria. In
particular, the transport model results for $b=0$ and for 1\%
events with the largest $N_P^{proj}$ are rather different (see
below).

\begin{figure}[ht!]
\begin{center}
\epsfig{file=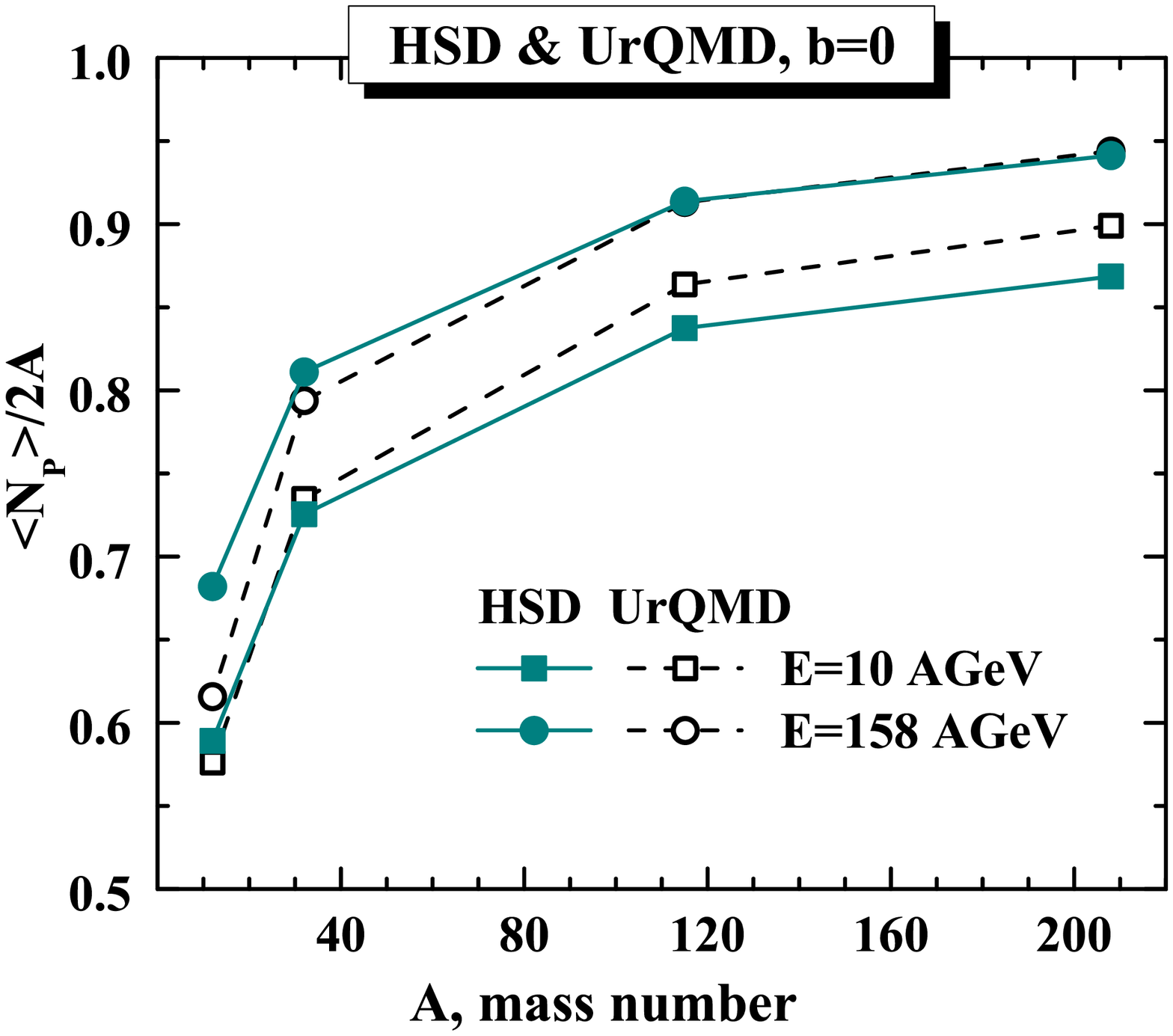,width=0.45\textwidth}
\epsfig{file=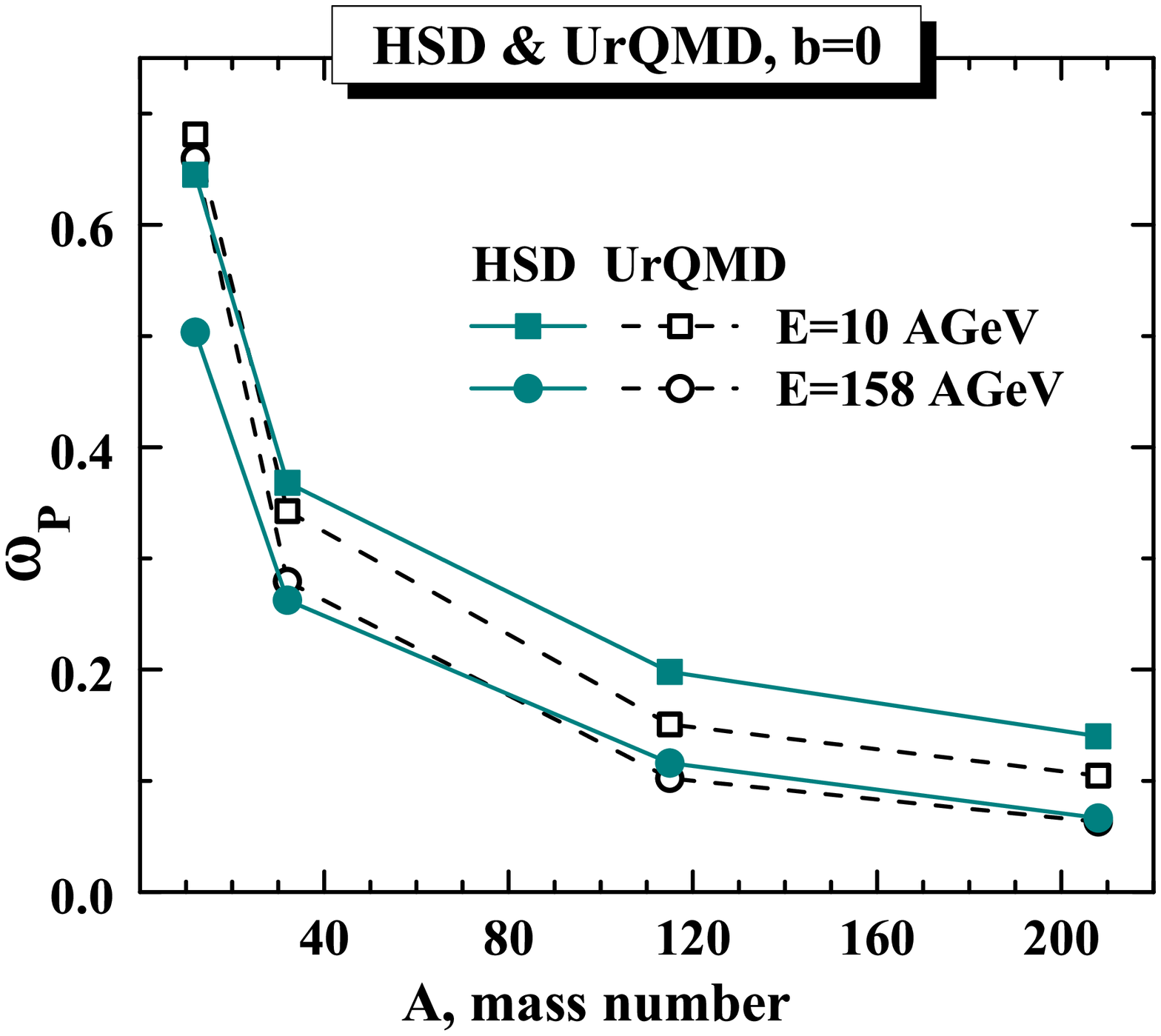,width=0.45\textwidth}
\end{center}
\caption{{\it Left:} Mean $\langle N_P\rangle$, divided by the
maximum number of  participants $2A$ in events with $b=0$ for
different nuclei at collision energies $E_{lab}$=10 and 158~AGeV.
{\it Right:} The scaled variance $\omega_P$ in events with $b=0$
for different nuclei at collision energies $E_{lab}$=10 and
158~AGeV.} \label{fig:Np_distr}
\end{figure}

Let's start with the $b=0$ centrality selection criterium.
First, we estimate the average number of participants, $\langle
N_P\rangle$, and the scaled variances of its fluctuations,
$\omega_P$, in $A+A$ collision events  which satisfy the $b=0$
condition. The left panel in Fig.~\ref{fig:Np_distr} shows  the
ratio, $\langle N_P\rangle/2$A, in $A+A$ collisions with $b=0$ for
different nuclei at collision energies $E_{lab}=10$ and 158~AGeV.
Both transport models (HSD and UrQMD) show a monotonous increase
of $\langle N_P\rangle/2$A with collision energy for all nuclei in
the energy range $10\div 158$~AGeV (Fig.~\ref{fig:Np_distr}, {\it
left}).  Correspondingly, the fluctuations of the number of
participants $\omega_P$ for all nuclei become smaller with
increasing collision energy (Fig.~\ref{fig:Np_distr}, {\it
right}.).  As seen from Fig.~\ref{fig:Np_distr} ({\it left}) about
90\% of all nucleons are participants for Pb+Pb collisions with
$b=0$. This number becomes essentially smaller, about 60-70\%, for
C+C collisions. One can therefore expect that participant number
fluctuations at $b=0$ are small for heavy nuclei but strongly
increase for light systems. This is demonstrated in
Fig.~\ref{fig:Np_distr} ({\it right}): $\omega_P$ is about
0.1$\div 0.2$ in Pb+Pb and In+In but becomes much larger,
$0.5\div$0.7, in C+C collisions.

One can conclude that the condition $b=0$ corresponds to `most
central' $A+A$ collisions only for nuclei with large atomic mass
number (In and Pb). In this case the average number of
participants is close to its maximum value and its fluctuations
are rather small. However, in the studies of event-by-event
multiplicity fluctuations in the collisions of light nuclei (C and
S) the criterium $b=0$ is far from selecting the `most central'
$A+A$ collisions.

\begin{figure}[ht!]
\begin{center}
\epsfig{file=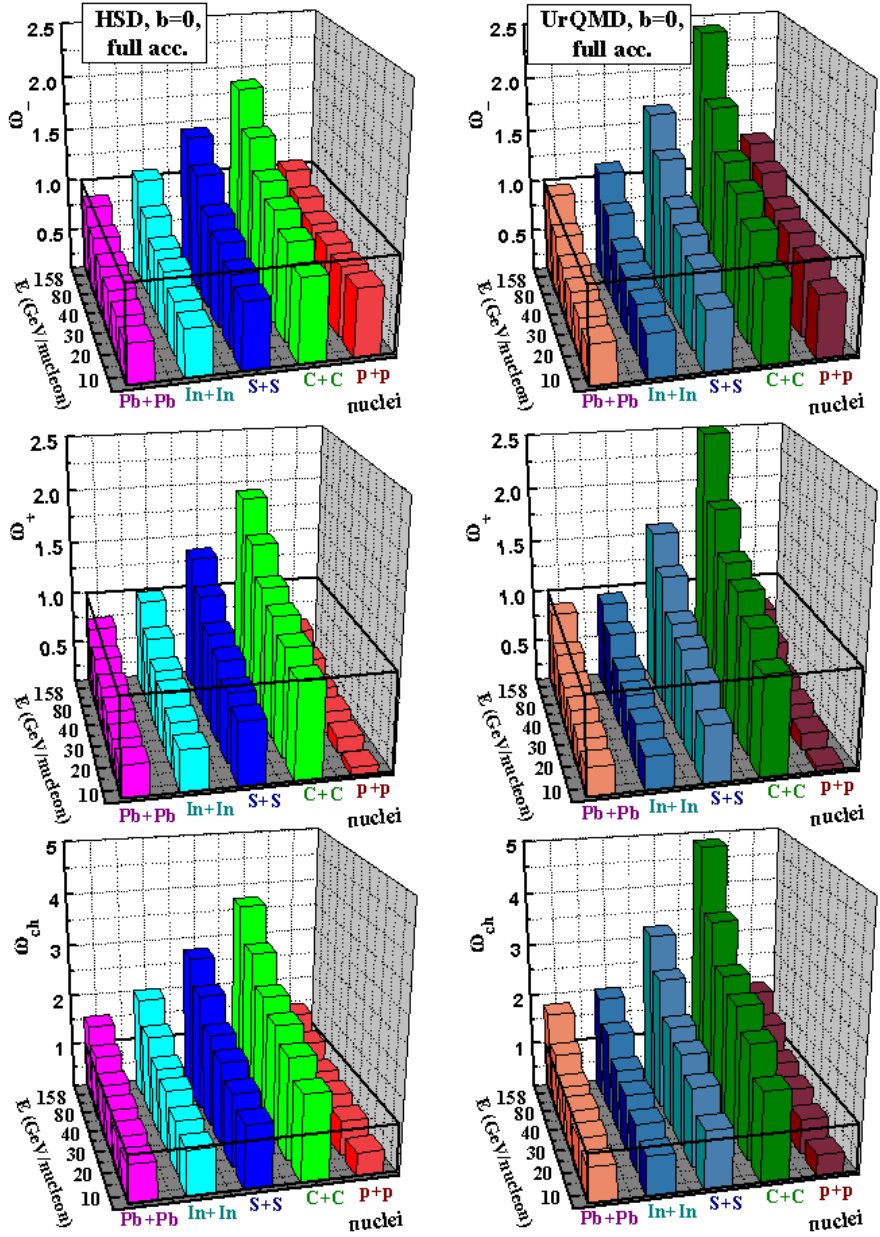,width=0.8\textwidth}
\end{center}
\caption{The results of HSD ({\it left})  and UrQMD ({\it right})
simulations for $\omega_-$ (top panel), $\omega_+$ (middle panel),
and $\omega_{ch}$ (lower panel) in p+p and central C+C, S+S,
In+In, Pb+Pb collisions at $E_{lab}=$~10, 20, 30, 40, 80,
158~AGeV. The condition $b=0$ is used here as a criterium for
centrality selection.  There are no cuts in acceptance.}
\label{fig:wi_b0_f}
\end{figure}

Results of HSD and UrQMD transport model calculations for the
scaled variance of negative, $\omega_-$, positive, $\omega_+$, and
all charged, $\omega_{ch}$, hadrons are shown in
Fig.~\ref{fig:wi_b0_f}
at different collision energies, $E_{lab}=$~10, 20, 30, 40, 80,
158~AGeV, and for different colliding nuclei,  C+C, S+S, In+In,
Pb+Pb. The transport model results correspond to collision events
for zero impact parameter, $b=0$. To make the picture more
complete, the transport model results for inelastic p+p collisions
are shown too, for reference. Note that the proton spectators are
not accounted for in the calculation of $N_+$ and $N_{ch}$. Thus,
proton spectators do not contribute to $\omega_+$ and
$\omega_{ch}$.

One sees a monotonic dependence of the multiplicity fluctuations
on both $E_{lab}$ and $A$: the scaled variances $\omega_-$,
$\omega_+$, and $\omega_{ch}$ increase with $E_{lab}$ and decrease
with $A$. The results for p+p collisions are different from those
for light ions. Note that within HSD
a detailed comparison of the multiplicity fluctuations in N+N
inelastic collisions and $b=0$ heavy-ion collisions (Pb+Pb and
Au+Au), including the energy dependence up to
$\sqrt{s_{NN}}=200$~GeV, has been presented in Sec.~\ref{ch:ppAA}
and in Refs.~\cite{KoGoBr07,LuBl07}.
Fig.~\ref{fig:wi_b0_f} corresponds to the full $4\pi$ acceptance.

The combination of the multiplicity fluctuations in elementary N+N 
collisions and fluctuations of the number of nucleon participants 
explains the main features of hadron multiplicity fluctuations in $A+A$ 
collisions. In particular, within the MIS (\ref{omegai}) one obtains 
the dependence on collision energy and atomic mass  number shown in 
Fig.~\ref{fig:wi_b0_f}.  The value of $n_i$ in Eq.~(\ref{omegai}) is  
the average number of $i$'th particles per participant, $n_i=\langle 
N_i\rangle /\langle N_P\rangle$, and $\omega_P$ equals the scaled 
variance for the number of nucleon participants. The N+N collisions 
define the fluctuations $\omega_i^*=\omega_i^{NN}$ from a single 
source.

\begin{figure}[ht!]
\begin{center}
\epsfig{file=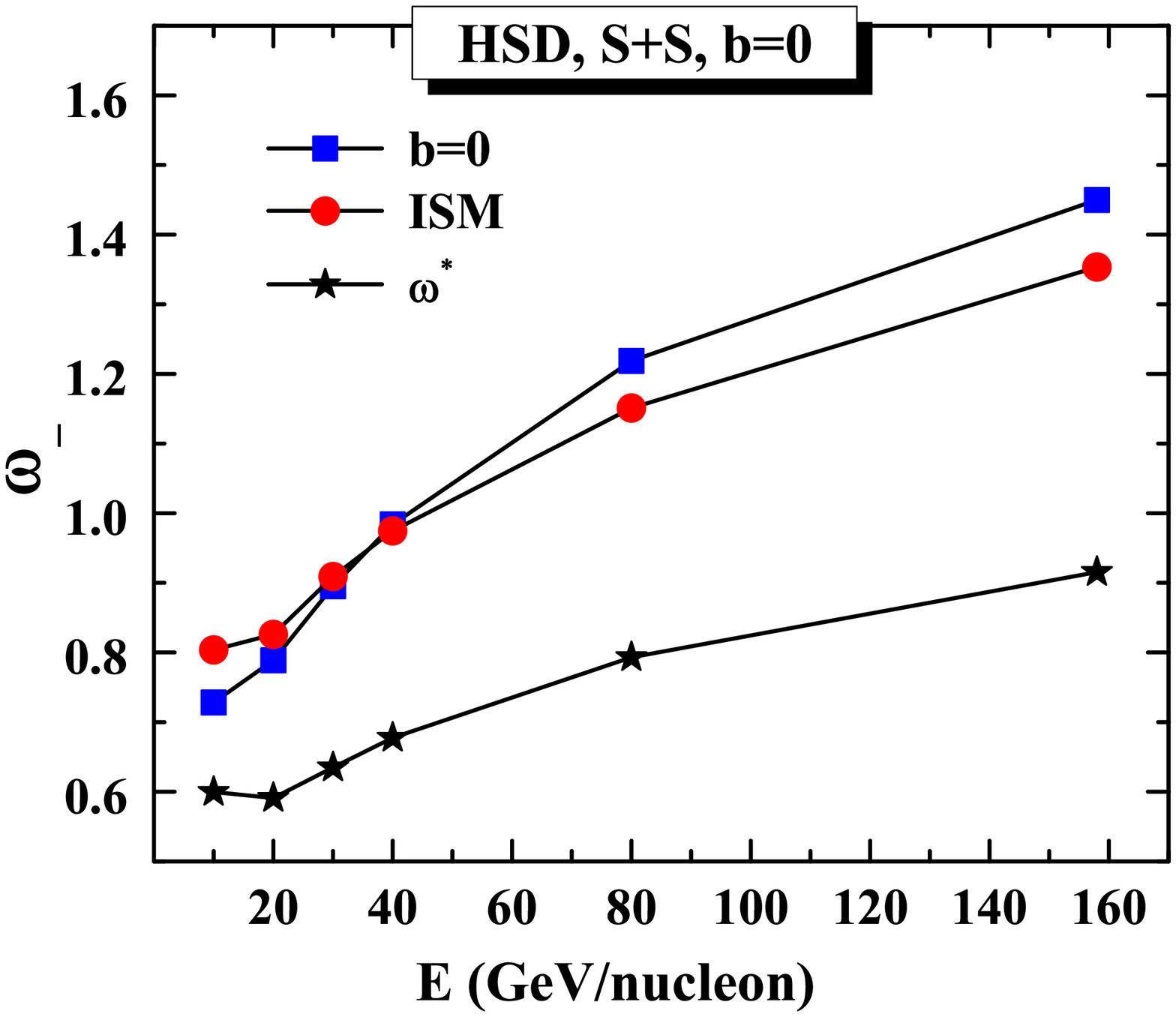,width=0.45\textwidth}
\epsfig{file=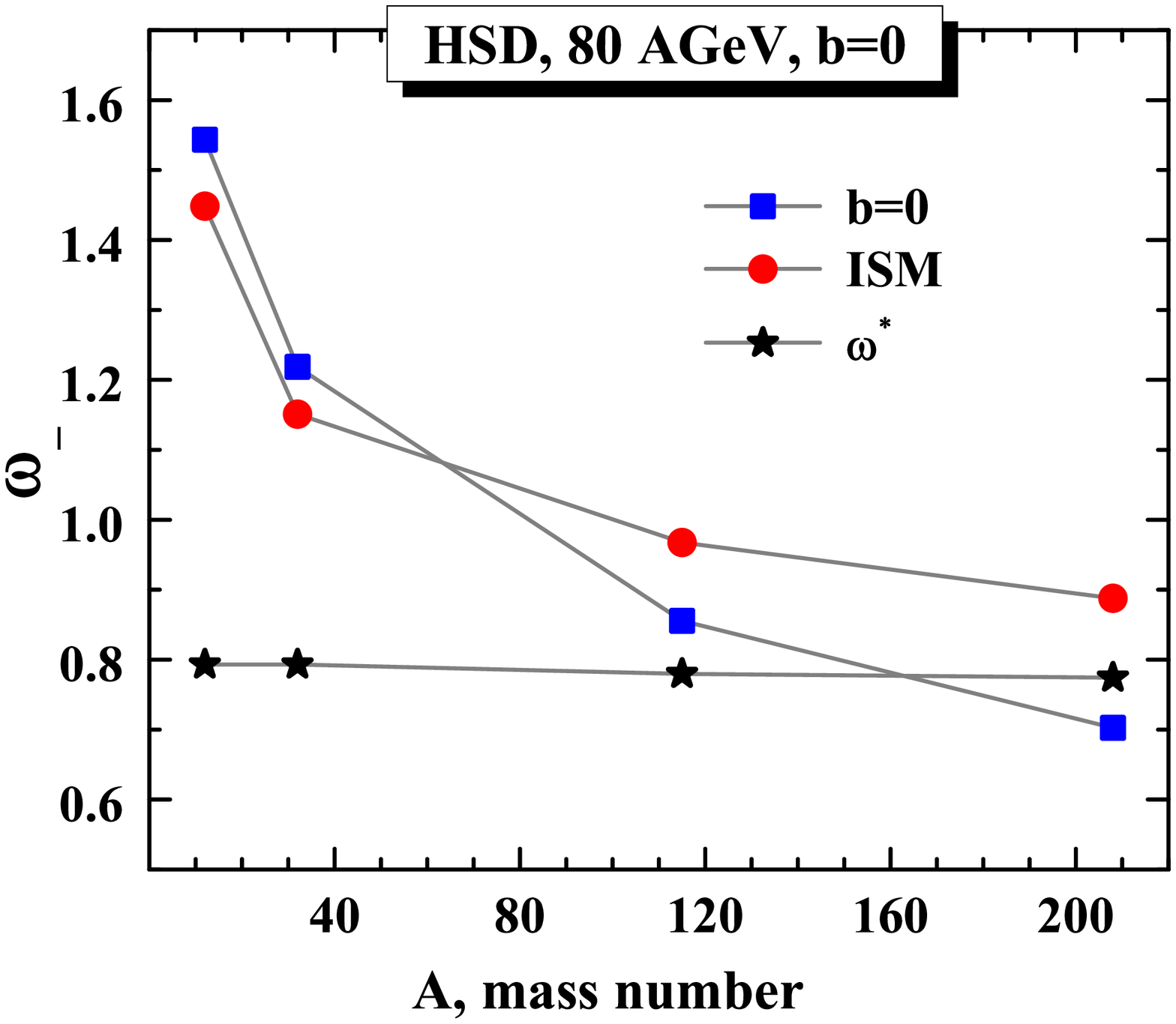,width=0.45\textwidth}
\end{center}
\caption{The {\it left} panel illustrates the energy dependence of
$\omega_-$ in S+S  collisions at $b=0$ in the full $4\pi$
acceptance, the {\it right} panel -- the $\omega_-$ dependence on
atomic mass  number at $E_{lab}$=80~AGeV. The HSD results are
shown by the squares while the circles correspond to the MIS
(\ref{omegai}). The stars show the first term, $\omega^*_-$, in
the r.h.s. of Eq.~(\ref{omegai}) -- the scaled variance for
negative hadrons in N+N collisions.  The values of $\omega_-^*$,
$n_i$, and $\omega_P$ are calculated within HSD.} \label{fig:MoIS}
\end{figure}

In Fig.~\ref{fig:MoIS} the HSD results for $\omega_i$ in $A+A$
collisions at $b=0$ are compared to the Eq.~(\ref{omegai}). One
concludes that the transport model results for the multiplicity
fluctuations are in qualitative agreement with MIS (\ref{omegai}).
Both $n_i$ and $\omega_i^*$ increase strongly with collision
energy as seen from Fig.~\ref{fig:flucP1}. This explains, due to
Eq.~(\ref{omegai}), the monotonous increase with energy of the
scaled variances $\omega_i$ in $A+A$ collisions at $b=0$ seen in
Fig.~\ref{fig:wi_b0_f}.
Note that $\omega_P$ at $b=0$ decreases with collision energy as
shown in Fig.~\ref{fig:Np_distr}, {\it right}. This, however, does
not compensate a strong increase of both $n_i$ and $\omega_i^*$.
The atomic mass  number dependence of the scaled variances
$\omega_i$ in $A+A$ collisions with $b=0$ follows from the
A-dependence of $\omega_P$. Fig.~\ref{fig:Np_distr} ({\it right})
demonstrates a strong increase of $\omega_P$ for light nuclei.
This, due to (\ref{omegai}), is transformed to the corresponding
behavior of $\omega_i$ seen in Fig.~\ref{fig:wi_b0_f}.

\subsection{Centrality Trigger with the Number of Projectile Participants}

In this subsection the centrality selection procedure by fixing the
number of projectile participants $N_P^{proj}$ is concidered.
This corresponds to the real situation of $A+A$ collisions in fixed target
experiments. As a first step one simulates in HSD and UrQMD the
minimal bias events
and calculate the event distribution over the number of
participants $N_{part}$. Then, one selects  1\% most central
collisions which correspond to the largest values of $N_P^{proj}$.
In such a sample of A+A collisions events with largest
$N_P^{proj}$ from different impact parameters can contribute.
After that one calculates the values of $\omega_P$ in these samples.
Note, that even for a fixed number of $N_P^{proj}$ the number of
target participants $N_P^{targ}$ fluctuates. Thus, the total
number of participants $N_P=N_P^{targ}+N_P^{proj}$ fluctuates,
too. In our 1\% sample, both $N_P^{targ}$ and $N_P^{proj}$
fluctuate. Besides there are correlations between $N_P^{targ}$
and $N_P^{proj}$.

\begin{figure}[ht!]
\begin{center}
\epsfig{file=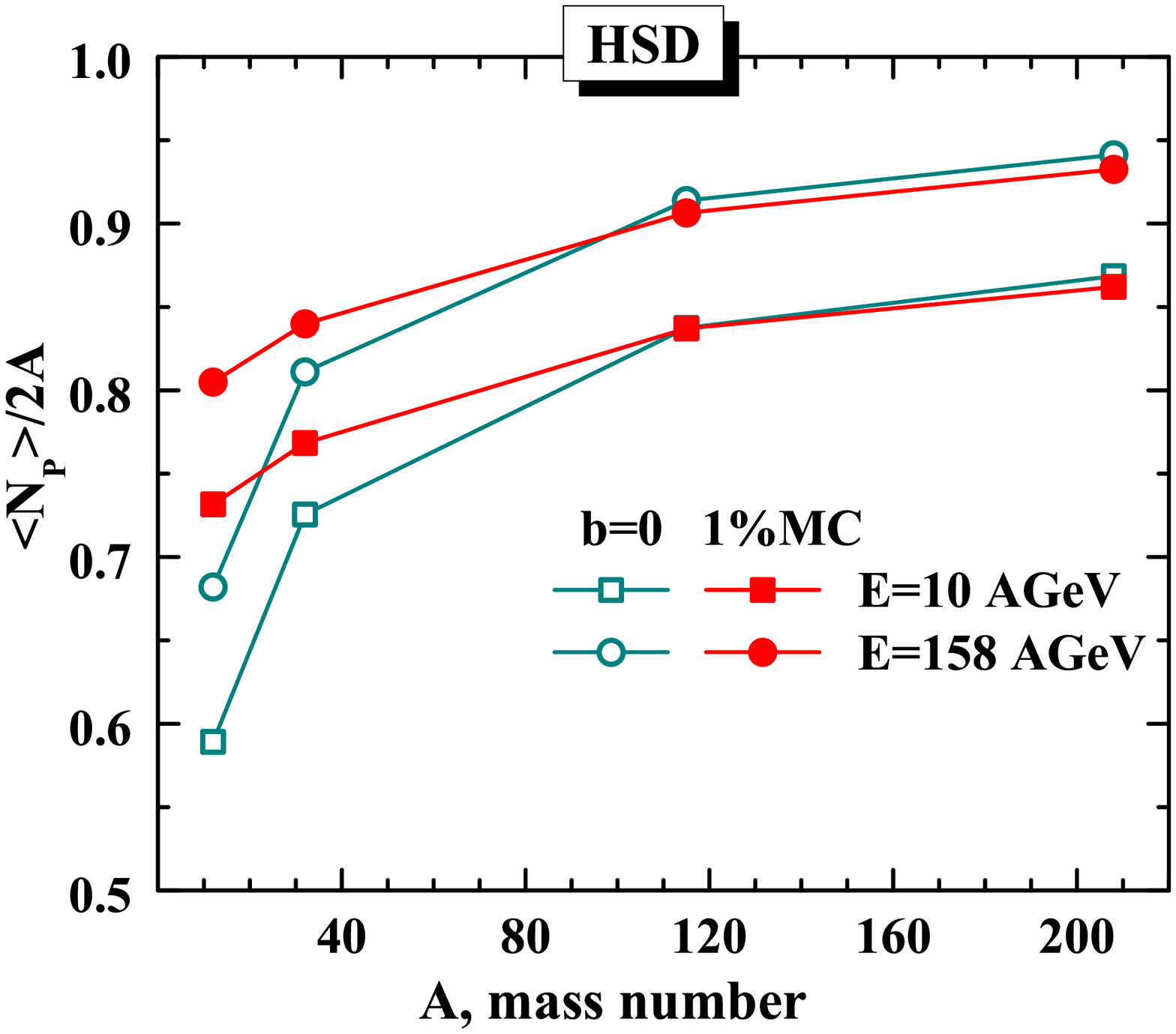,width=0.45\textwidth}
\epsfig{file=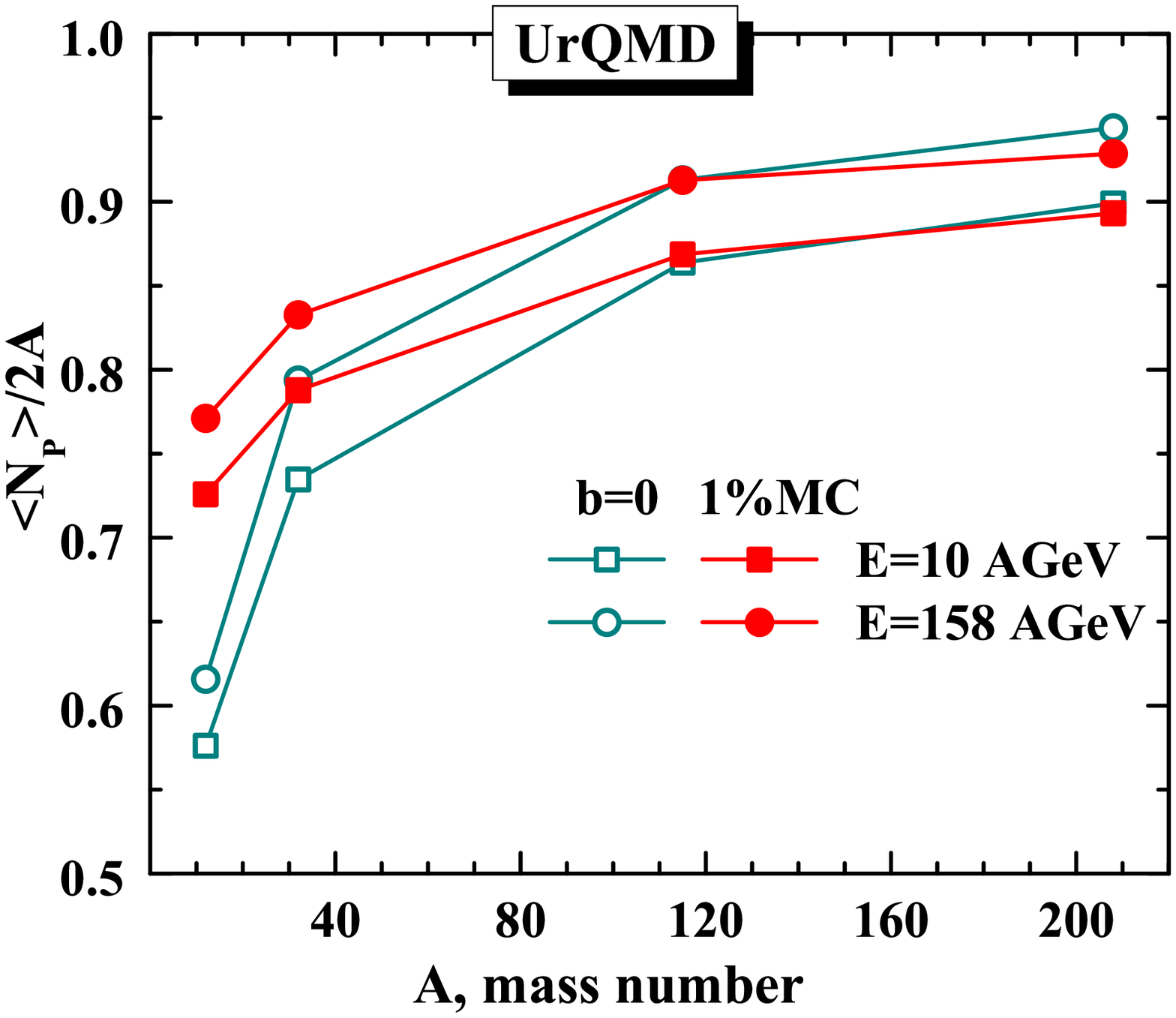,width=0.45\textwidth}
\epsfig{file=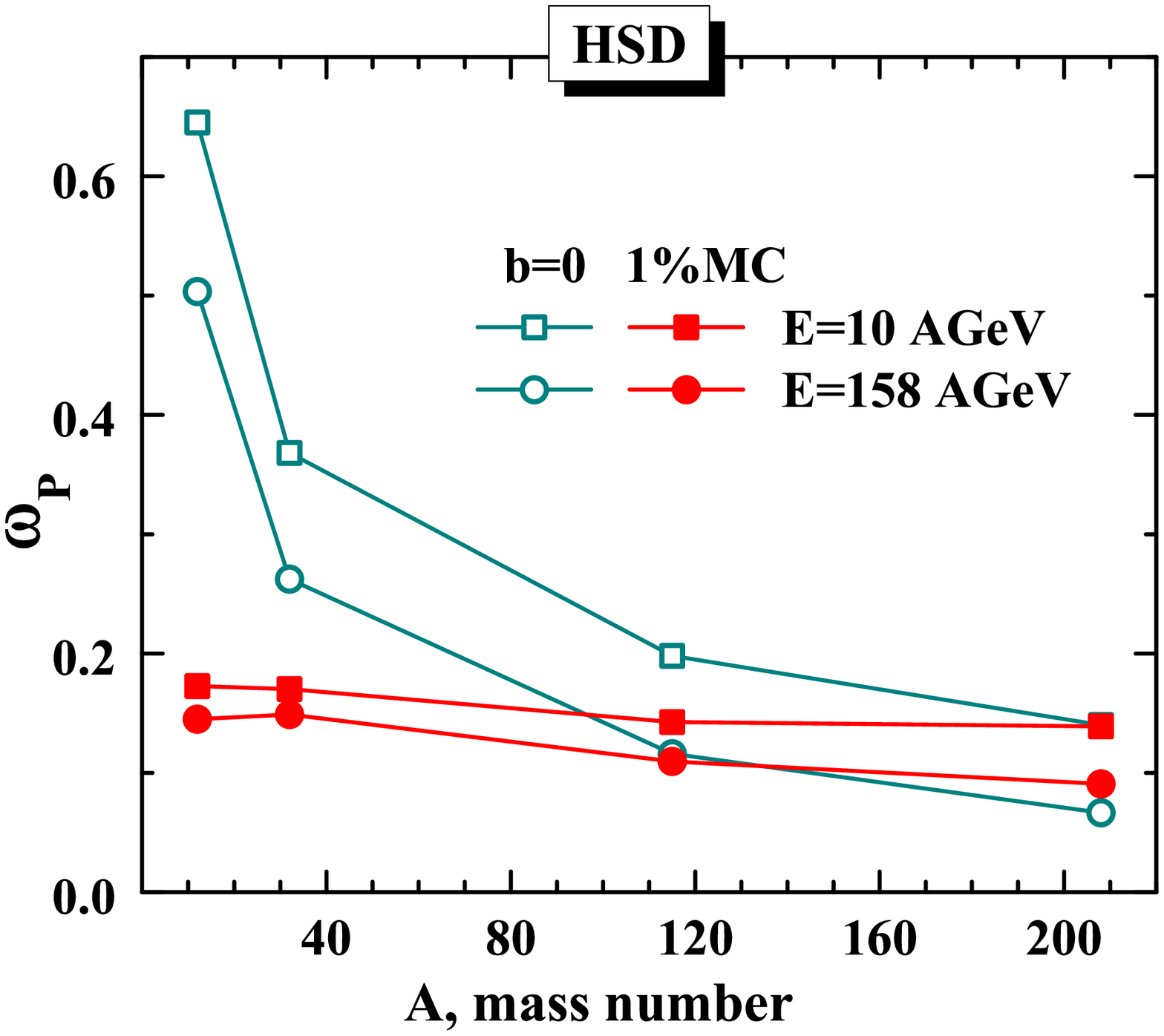,width=0.45\textwidth}
\epsfig{file=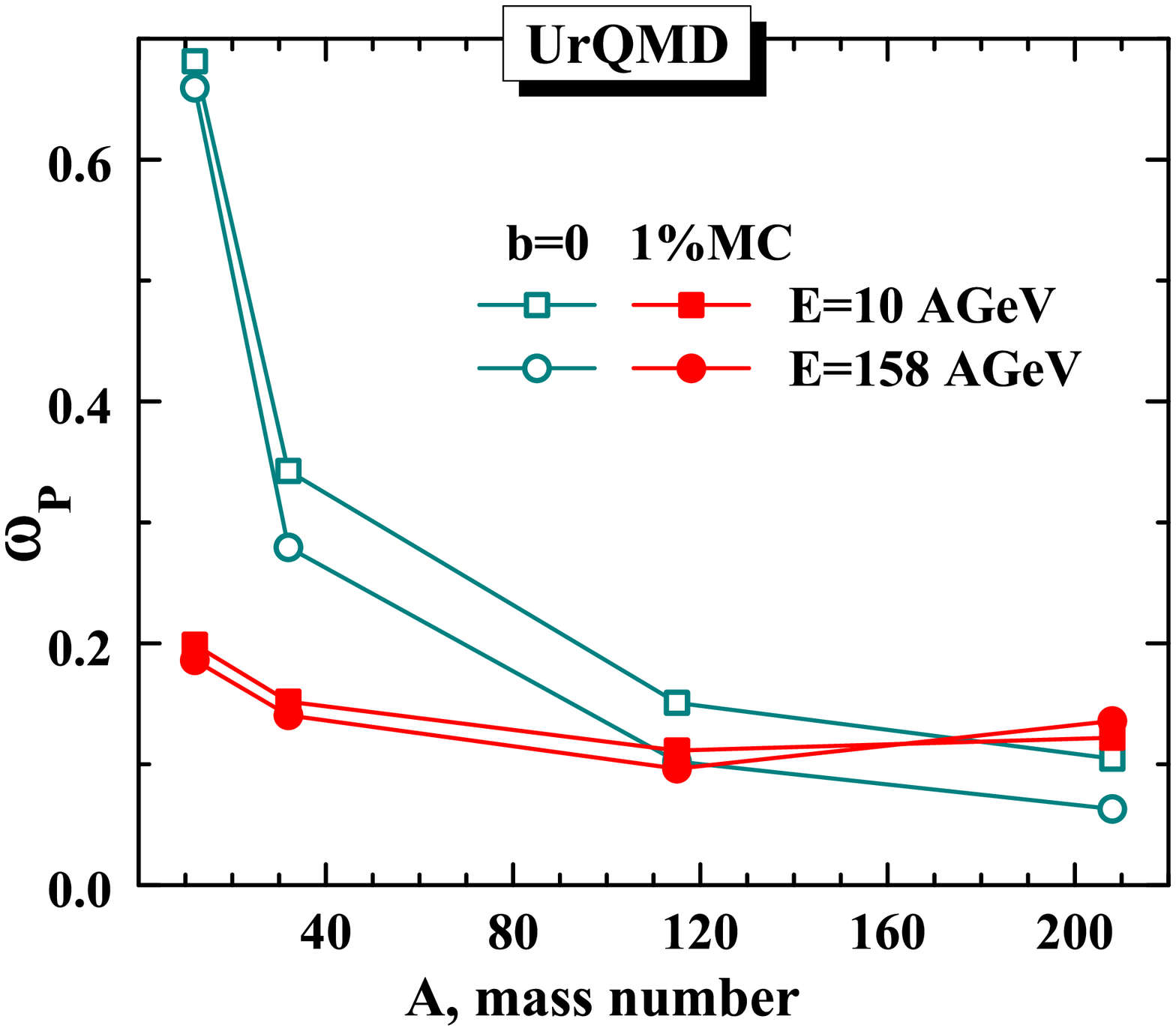,width=0.45\textwidth}
\end{center}
\caption{The HSD ({\it left}) and UrQMD ({\it right}) results for
the ratio $\langle N_P\rangle/2A$ (the upper panel) and the scaled
variance of the participant number fluctuations, $\omega_P$ (the
lower panel), for the 1\% most central collisions selected by the
largest values of $N_P^{proj}$ (full symbols), for different
nuclei at collision energies $E_{lab}$=10 and 158~AGeV. The open
symbols present the results of Fig.~\ref{fig:Np_distr} ({\it
right}) for $b=0$.} \label{fig:Np_distr2}
\end{figure}

Fig.~\ref{fig:Np_distr2} shows the ratio $\langle N_P\rangle/2A$ and
the scaled variance, $\omega_P$, for 1\% most central collisions
selected by the largest values of $N_P^{proj}$. These results are
compared with those for the $b=0$ centrality selection. For heavy
nuclei, like In and Pb, one finds no essential differences between
these two criteria of centrality selection. However, the 1\%
centrality trigger defined by the largest values of $N_P^{proj}$
looks much more rigid for light ions (S and C).  In this case the
ratio $\langle N_P\rangle/2A$ is larger, and $\omega_P$ is
essentially smaller than for the criterion  $b=0$. As a result the 1\%
centrality trigger by the largest values of $N_P^{proj}$  leads to
a rather weak $A$-dependence of $\omega_P$.

Some comments are appropriate at this point. Let's define the
centrality $c(N)$ as a percentage of events with a multiplicity
larger than $N$ (this can be the number of produced hadrons,
number of participants, etc.).  It was argued in Ref.~\cite{BrFl02}
that a selection of $c(N)$ of most central $A+A$ collisions is
equivalent to restricting the impact parameter, $b<b(N)$, with,
\eq{\label{centrality} b(N)~=~
\sqrt{\frac{\sigma_{inel}}{\pi}~c(N)}~, }
where $\sigma_{inel}$ is the total inelastic A+A cross section.
Thus, the centrality criterion by the multiplicity $N$ is
equivalent to the geometrical criterion by the impact parameter
$b$. Moreover, the result (\ref{centrality}) does not depend on
the specific observable $N$ used to define the $c$-percentage of
most central A+A collisions. The result (\ref{centrality}) should
be valid for any observable $N$ which is a monotonic function of
$b$. Therefore, the relation (\ref{centrality}) reduces any
centrality selection to the geometrical one. This result was
obtained in Ref.~\cite{BrFl02} by neglecting the fluctuations of
multiplicity $N$ at a given value of $b$. This is valid if $c$ is
not too small and the colliding nuclei are not too light. In the
sample of A+A events with 1\% of largest $N_P^{proj}$, the
relation (\ref{centrality}) can not be applied for S+S and C+C
collisions.

\begin{figure}[ht!]
\begin{center}
\epsfig{file=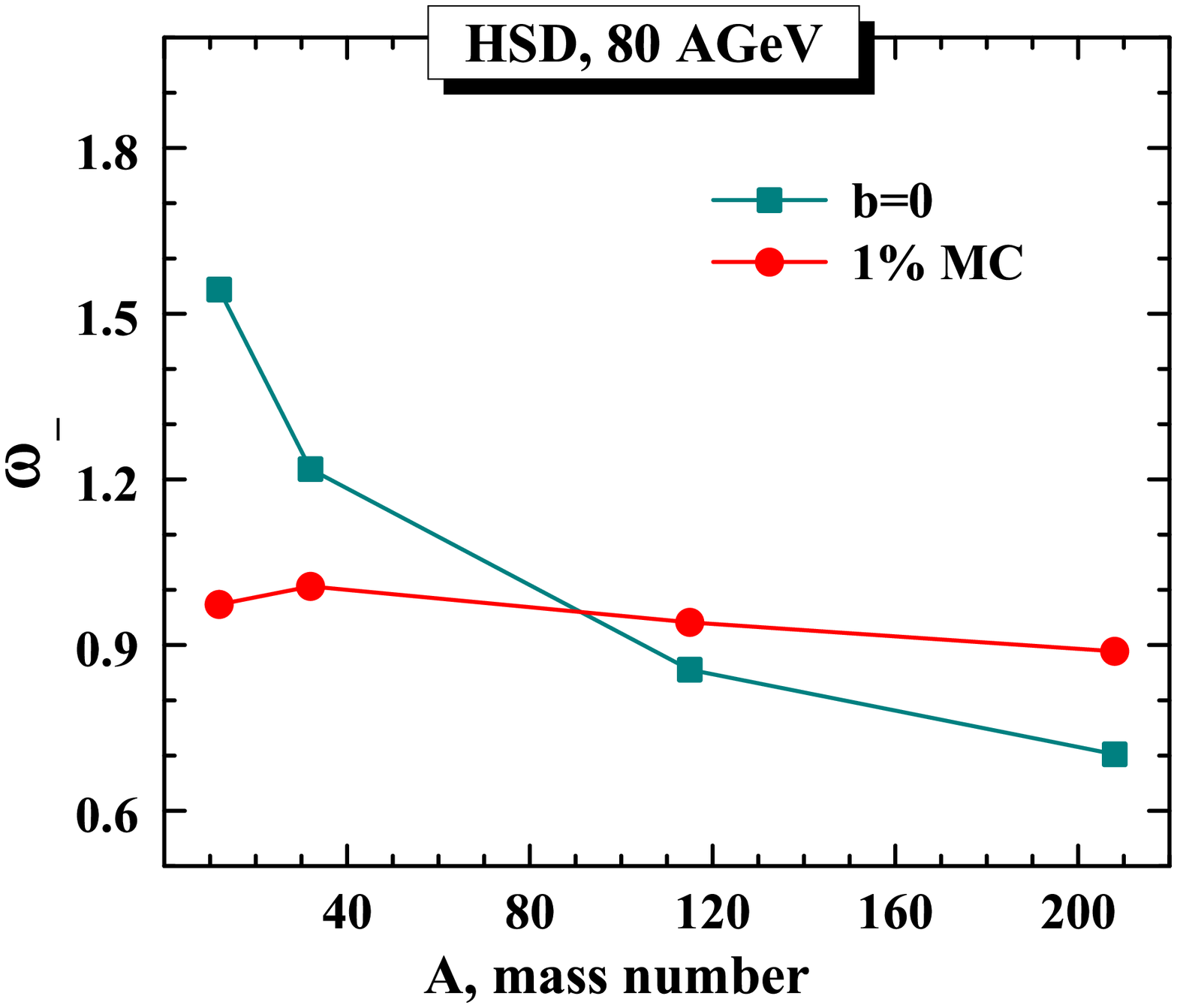,width=0.45\textwidth}
\epsfig{file=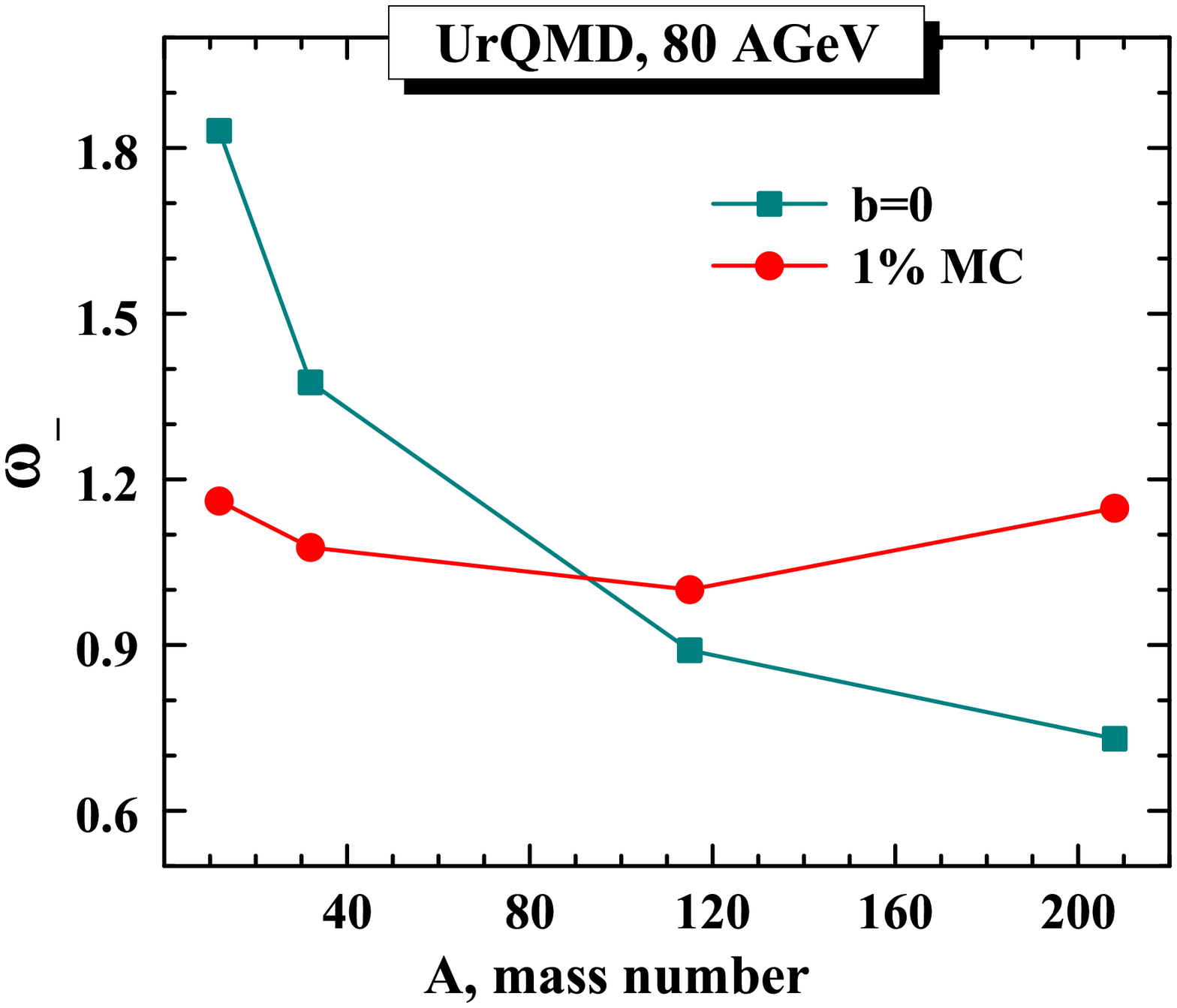,width=0.45\textwidth}
\end{center}
\caption{The  dependence of $\omega_-$  on atomic mass  number at
$E_{lab}$=80~AGeV for the HSD ({\it left}) and UrQMD ({\it left})
simulations. The squares correspond to $b=0$, and circles to 1\%
largest $N_P^{proj}$.} \label{fig:A-dep}
\end{figure}

Fig.~\ref{fig:A-dep} shows a comparison of the A-dependence of
$\omega_-$ in the transport models for two different samples of
the collision events: for $b=0$ and for the 1\% of events with
largest $N_P^{proj}$ values.  One can see that the multiplicity
fluctuations are rather different in these two samples.  Moreover,
these  differences are  in the opposite directions for heavy
nuclei and for light nuclei. For light nuclei, $\omega_-$ is
essentially smaller in the 1\% sample with largest $N_P^{proj}$
values, whereas for heavy nuclei the smaller fluctuations
correspond to $b=0$ events. Note that in the 1\% sample with
largest $N_P^{proj}$ values the A-dependence of multiplicity
fluctuations becomes much weaker. In this case a strong increase
of the multiplicity fluctuations for light nuclei, seen for $b=0$,
disappears.

\begin{figure}[ht!]
\begin{center}
\epsfig{file=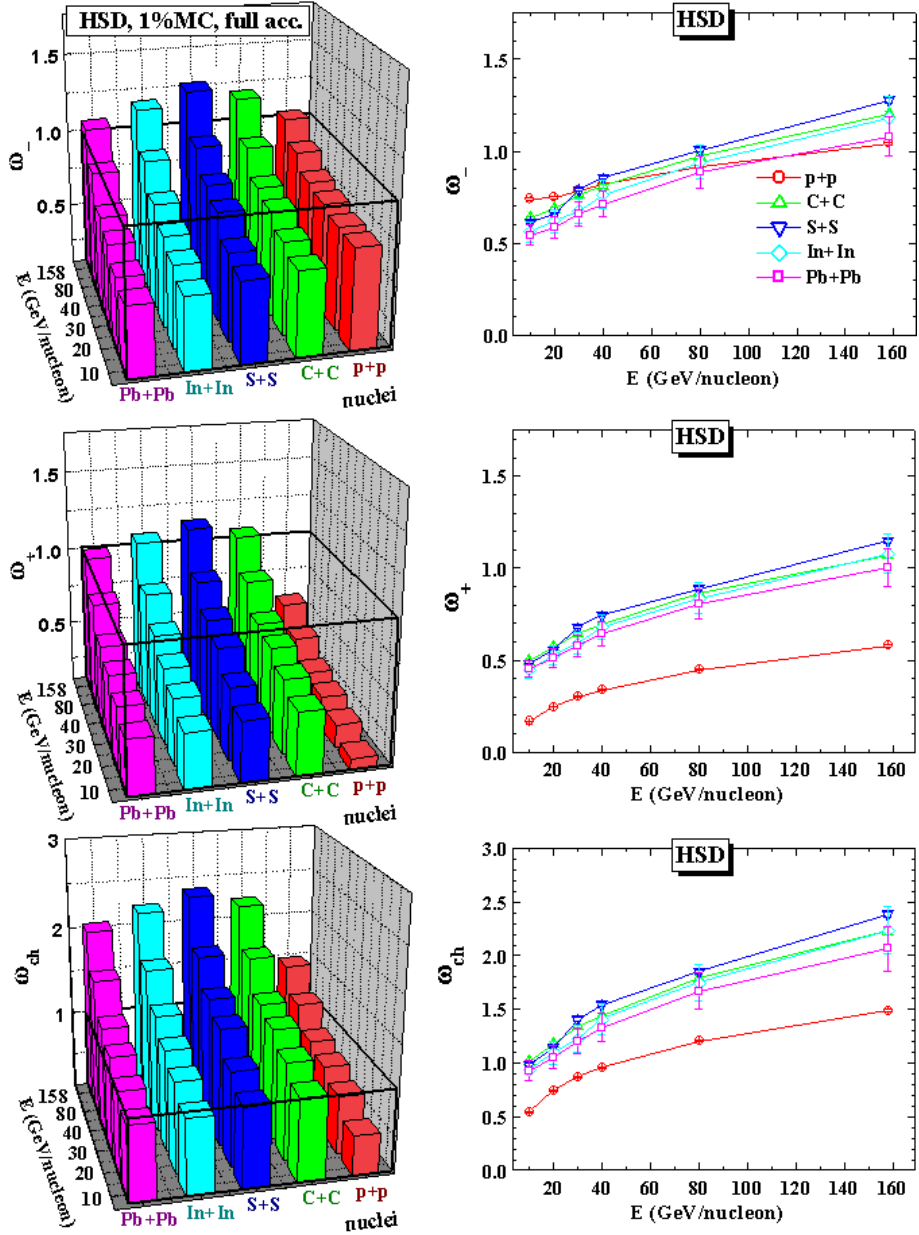,width=0.8\textwidth}
\end{center}
\caption{The HSD results for $\omega_-$ (upper panel), $\omega_+$
(middle panel), and $\omega_{ch}$ (lower panel) in $A+A$ and p+p
collisions for the full $4\pi$ acceptance in 3D ({\it left}) and
2D ({\it right}.) projection. The 1\%  most central C+C, S+S,
In+In, and Pb+Pb collisions are selected by choosing the largest
values of $N_P^{proj}$ at different collision energies
$E_{lab}$=10, 20, 30, 40, 80, 158~AGeV.  The errorbars indicate
the estimated uncertainties in the model calculations. The HSD
results from inelastic p+p collisions are the same as in
Fig.~\ref{fig:wi_b0_f}.} \label{fig:wi_1pc_f}
\end{figure}

\begin{figure}[ht!]
\begin{center}
\epsfig{file=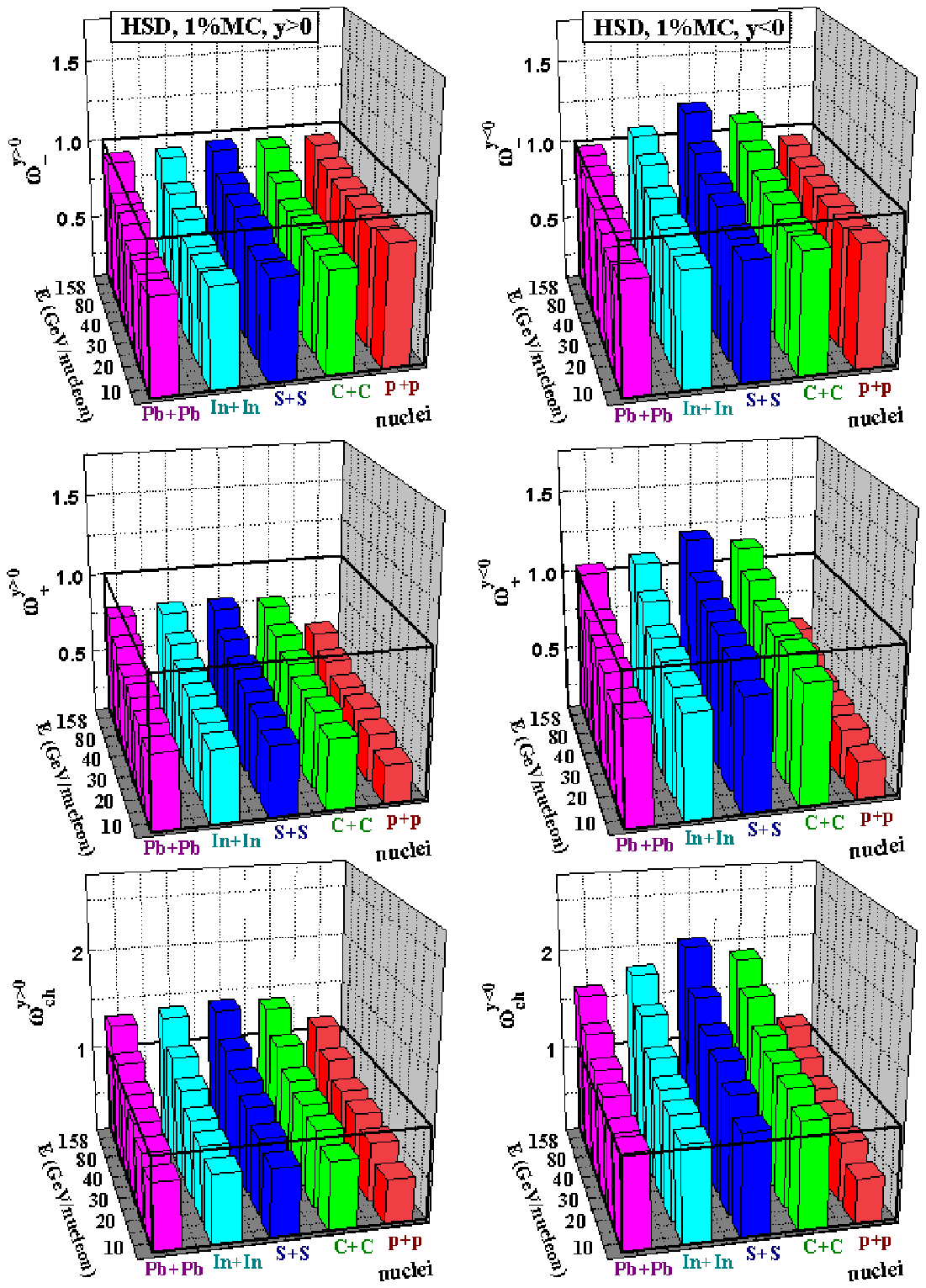,width=0.8\textwidth}
\end{center}
\caption{The same as in Fig.~\ref{fig:wi_1pc_f}, but for  final
hadrons accepted in the projectile hemisphere, $y>0$ ({\it left}),
and in the target hemisphere, $y<0$ ({\it right}).} 
\label{fig:wi_1pc_y}
\end{figure}

For the 1\% most central $A+A$ collision events - selected by the 
largest values of $N_P^{proj}$ - the HSD multiplicity fluctuations are 
shown in Fig.~\ref{fig:wi_1pc_f} and Fig.~\ref{fig:wi_1pc_y}.  For 
light nuclei (S and C) the multiplicity fluctuations in the samples of 
1\%  most central collisions are smaller than in the $b=0$ selection 
and the atomic mass  number dependencies become less pronounced 
(compare Fig.~\ref{fig:wi_1pc_f} and Fig.~\ref{fig:wi_b0_f}). This is 
because the participant number fluctuations $\omega_P$ have now 
essentially smaller A-dependence, as seen in Fig.~\ref{fig:Np_distr2}.

Fig.~\ref{fig:wi_b0_f} shows that both HSD and UrQMD predict a
monotonic dependence of the charge particle multiplicity with
energy. So, the hadronic `background' for the NA61 experiments is
expected to be a smooth monotonic function of beam energy.

The event-by-event observables show a higher sensitivity to the
initial nucleon density distribution than the standard single
particle observables \cite{TaDrKo07}.
A sensitivity of the A-dependence of $\omega_{ch}$ to these
details of the models indicates a necessity for further studies of
the initializations of the nuclei in transport model approaches.
This becomes important for the theoretical interpretation of
future experimental data on event-by-event fluctuations.

Note that the MIS and Eq.~(\ref{omegai}) work for the multiplicity
fluctuations simulated by the transport models in full $4\pi$
acceptance but not for the acceptance in a specific rapidity
region. The results for inelastic p+p collisions are identical in
the projectile and target hemispheres. This is not the case in the
sample of 1\% most central $A+A$ collisions selected by
$N_P^{proj}$. The total number of nucleons participating in $A+A$
collisions fluctuates.  These fluctuations are not symmetric in
forward-backward hemispheres: in the selected 1\% sample the
number of target participants $N_P^{targ}$ fluctuates essentially
stronger than that of $N_P^{proj}$.
The HSD results in Fig.~\ref{fig:wi_1pc_y} clearly demonstrate larger 
values for all scaled variances, $\omega_-$, $\omega_+$, and 
$\omega_{ch}$, for $y<0$ acceptance than those for $y>0$ one. This is 
due to stronger target participant fluctuations, 
$\omega_P^{targ}>\omega_P^{proj}$.


\section{ Fluctuations and Correlations in Au+Au Collisions at RHIC}
\label{ch:corr}

In this Section we discussed some features of the particle number
fluctuations \cite{KoGoBr07a} and forward-backward correlations
\cite{KoHaToGoBr08} in Au+Au collisions at
$\sqrt{s_{NN}}=200$~GeV.

\subsection{Multiplicity Fluctuations in Au+Au Collisions at RHIC}
\label{sec:mult2}

\begin{figure}[ht!]
\begin{center}
\epsfig{file=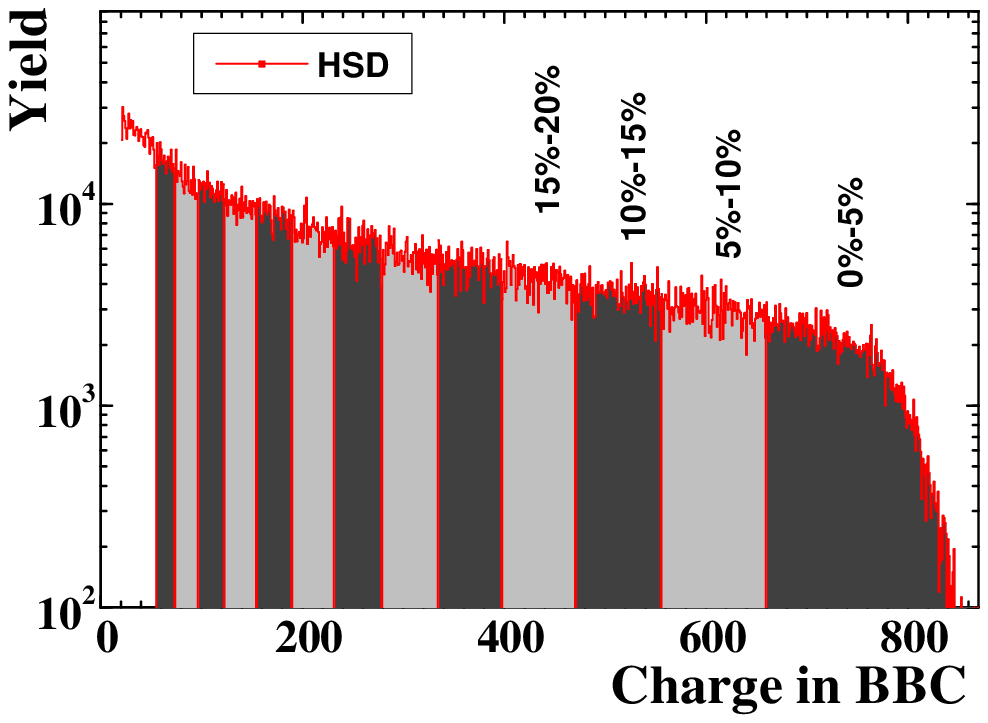,width=0.45\textwidth}
\epsfig{file=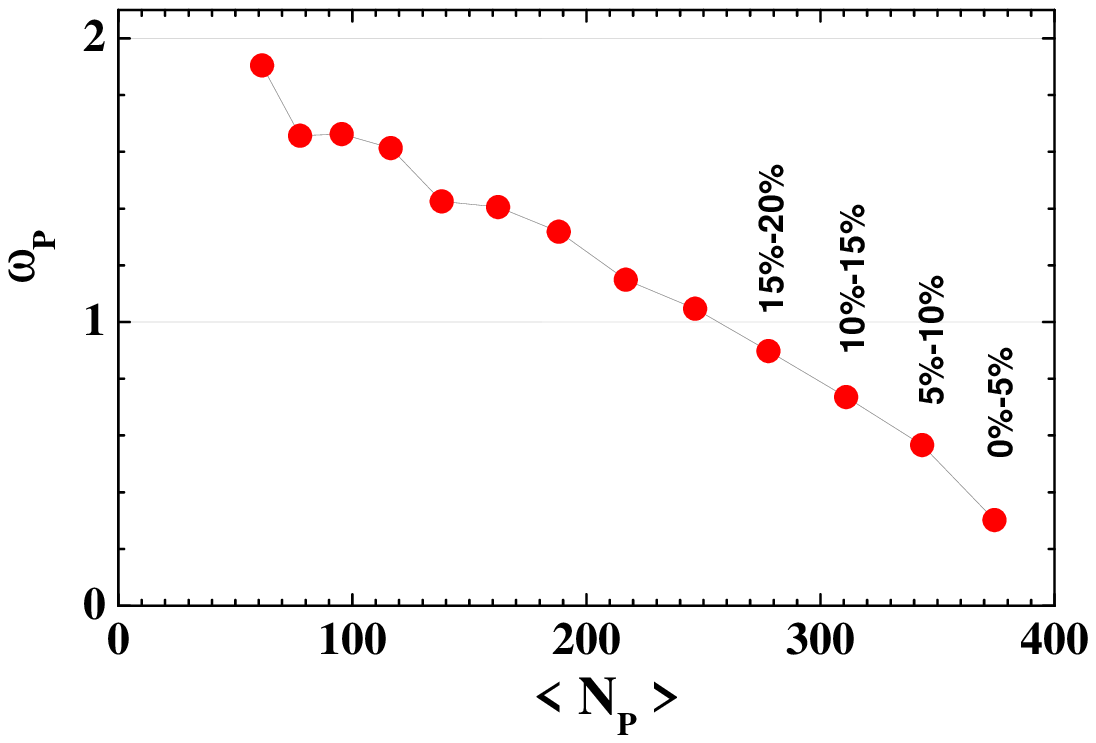,width=0.45\textwidth}
\end{center}
\caption{HSD model results for Au+Au collisions at
$\sqrt{s_{NN}}=200$~GeV. {\it Left:} Centrality classes defined
via the BBC distribution. {\it Right:} The average number of
participants, $\langle N_P\rangle$, and the scaled variance  of
the participant number fluctuations, $\omega_P$, calculated for
the 5\%  BBC centrality classes.} \label{fig:BBC}
\end{figure}

The charged multiplicity fluctuations in Au+Au collisions at
$\sqrt{s_{NN}}=200$~GeV have been measured recently by the PHENIX
Collaboration~\cite{Ad05a,Mi05}.
The centrality selection is an important aspect of fluctuation
studies in A+A collisions. As it was discussed in the previous
sections the samples of collisions with a fixed number of
projectile participants $N_P^{proj}$ can be selected to minimize
the participant number fluctuations at the fixed target
experiments. This selection is possible due to a measurement of
the number of nucleon spectators from the projectile,
$N_S^{proj}$, in each individual collision by a calorimeter which
covers the projectile fragmentation domain.
In a collider type experiments
another centrality trigger
should be used. For the multiplicity fluctuations data the PHENIX
Collaboration uses
two kinds of detectors which define the centrality of Au+Au
collision: Beam-Beam Counters (BBC) and Zero Degree Calorimeters (ZDC).
The BBC measure the charged particle multiplicity in the
pseudorapidity range $3.0<|\eta|<3.9$, and the ZDC -- the number
of neutrons with $|\eta|>6.0$ \cite{Ad05a,Mi05}. These neutrons
are part of the nucleon spectators. Due to technical reasons the
neutron spectators can be only detected by the ZDC (not protons
and nuclear fragments), but in both hemispheres. The BBC
distribution will be used in the HSD calculations to divide Au+Au
collision events into 5\% centrality samples. The HSD does not
specify different spectator groups -- neutrons, protons, and
nuclear fragments. Thus, one can not use the ZDC information. In
Fig.~\ref{fig:BBC} ({\it  left}) the HSD results for the BBC
distribution and centrality classes in Au+Au collisions at
$\sqrt{s_{NN}}=$200~GeV are shown. One finds a good agreement
between the HSD shape of the BBC distribution and the PHENIX data
\cite{Ad05a,Mi05}.
Note, however, that the HSD $\langle N_P\rangle$ numbers are not
exactly equal to the PHENIX values. It is also not obvious that
different definitions for the 5\% centrality classes give the same
values of the scaled variance $\omega_P$ for the participant number
fluctuations.

\begin{figure}[ht!]
\begin{center}
\epsfig{file=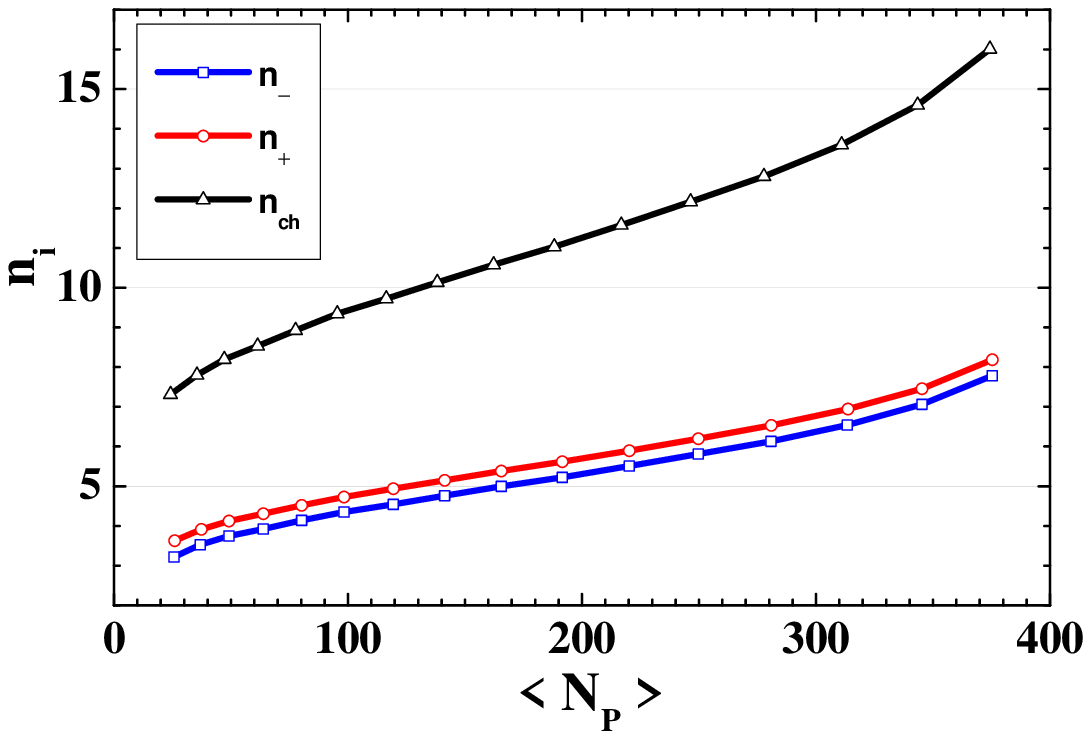,width=0.45\textwidth}
\epsfig{file=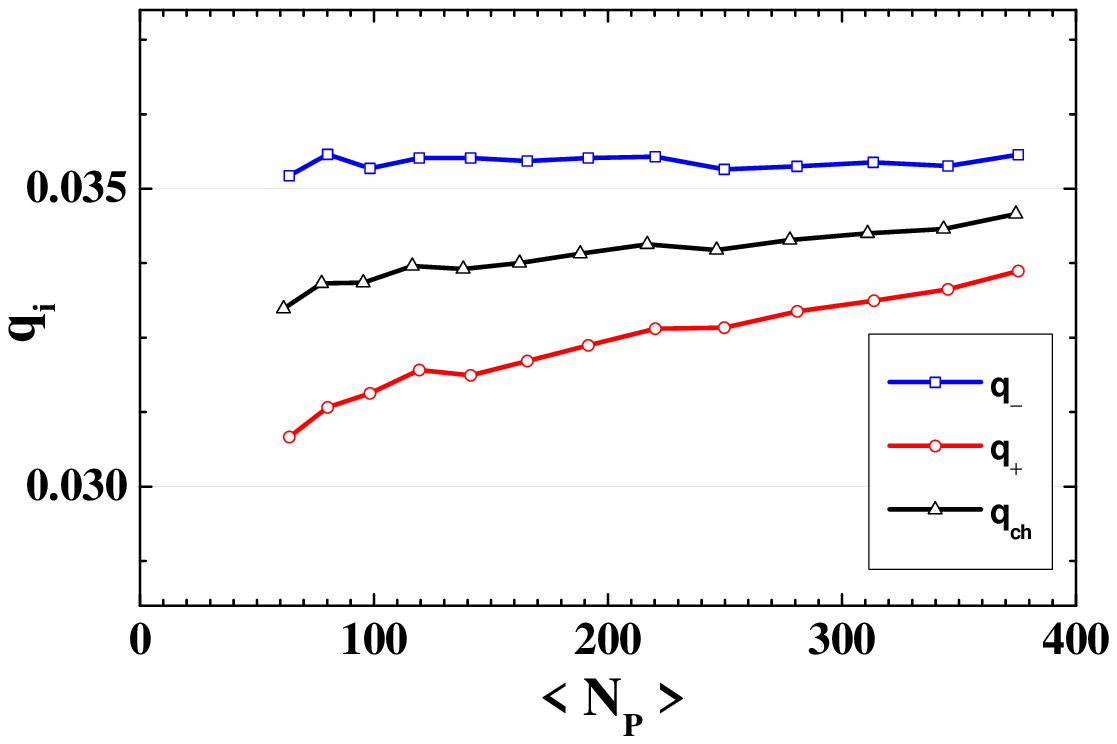,width=0.45\textwidth}
\end{center}
\caption{HSD results for different BBC centrality classes in Au+Au
collisions at $\sqrt{s_{NN}}=200$~GeV. {\it Left:} The mean number
of charged hadrons per participant, $n_{i}~=~\langle
N_{i}\rangle/\langle N_P\rangle$. {\it Right:} The fraction of
accepted particles, $q_i=\langle N^{acc}_{i}\rangle/\langle
N_{i}\rangle$.} \label{fig:omTeor}
\end{figure}

Defining the centrality selection via the HSD transport model
(which is similar to the BBC in the PHENIX experiment) we
calculate the mean number of nucleon participants, $\langle
N_P\rangle$, and the scaled variance of its fluctuations,
$\omega_P$,  in each 5\% centrality sample. The results are shown
in Fig.~\ref{fig:BBC}, right. The Fig.~\ref{fig:omTeor} ({\it
left}) shows the HSD results for the mean number of charged
hadrons per nucleon participant, $n_{i}=\langle
N_{i}\rangle/\langle N_P\rangle$, where  the index $i$ stands for
``$-$'', ``+'', and ``ch'' final hadrons.
Note that the centrality dependence of $n_i$ is opposite to that
of $\omega_P$.

The PHENIX detector accepts charged particles in a small region of
the phase space with pseudorapidity $|\eta|<0.26$, azimuthal
angle $\phi<245^o$, and the $p_T$ range from 0.2 to 2.0 GeV/c
\cite{Ad05a,Mi05}. The fraction of the accepted particles
$q_i=\langle N^{acc}_{i}\rangle/\langle N_{i}\rangle$ calculated
within the HSD model is shown in Fig.~\ref{fig:omTeor}, right.
According to the HSD results  only about
 3\% of charged particles are accepted by the mid-rapidity
PHENIX detector.

\begin{figure}[ht!]
\begin{center}
\epsfig{file=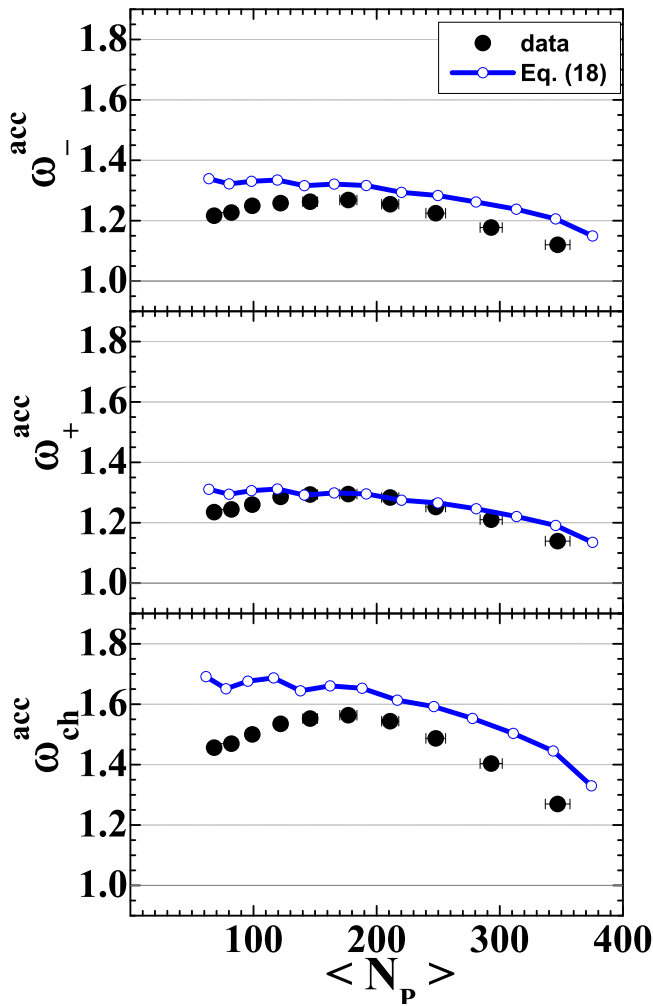,width=0.6\textwidth}
\end{center}
\caption{The scaled variance of charged particle fluctuations in
Au+Au collisions at $\sqrt{s_{NN}}=200$~GeV with the PHENIX
acceptance. The circles are the PHENIX data \cite{Ad05a,Mi05}
while the open points (connected by the solid line) correspond to
(\ref{omega-acc}) with the HSD results for $\omega_P$, $n_i$, and
$q_i$.} \label{fig:comp}
\end{figure}

To estimate the role of the participant number event-by-event
fluctuations the MIS relation (\ref{omegai}) has been used.
It has been
assumed that N+N collisions define the fluctuations
$\omega_i^*$ from a single source. To calculate the fluctuations
$\omega^{acc}_i$ in the PHENIX acceptance the acceptance scaling
formula (\ref{acc}) is used.
One finds,
\eq{
\omega_i^{acc}~=~1~-~q_i~+q_i~\omega_i^*~ +~ q_i~n_i~\omega_P~.
\label{omega-acc}
}

The HSD results for $\omega_P$ (Fig.~\ref{fig:BBC}, right), $n_i$
(Fig.~\ref{fig:omTeor}, {\it  left}), $q_i$
(Fig.~\ref{fig:omTeor}, right), together with  the HSD
nucleon-nucleon values, $\omega_-^*= 3.0$, $\omega_+^*=2.7$, and
$\omega_{ch}^*=5.7$ at $\sqrt{s_{NN}}=$~200~GeV, define completely
the results for $\omega_i^{acc}$ according to (\ref{omega-acc}).
We find a surprisingly  good agreement of the results given by
(\ref{omega-acc}) with the PHENIX data shown in
Fig.~\ref{fig:comp}. The centrality dependence of $\omega_i^{acc}$
stems from the product, $q_i n_i \omega_P$, in the last term of
the r.h.s. of (\ref{omega-acc}). Note that $\omega_P$ decreases
with $\langle N_P\rangle$, whereas both $n_i$ and $q_i$ increase.
It may lead to a nontrivial centrality dependence with a maximum
at intermediate values of $\langle N_P\rangle$ as seen in the
PHENIX data.

\subsection{Forward-Backward Correlations}

Correlations of particles between different regions of rapidity
have for a long time been considered to be a signature of new
physics.  A shortening in the correlation length in rapidity has
been thought to signal a transition to a quark-gluon plasma \cite{BiPe86,JeShBl06}.
Conversely, the appearance of long-range correlations has  been
associated with the onset of the percolation limit, also linked to
the QCD phase transition \cite{ArBrFePa96,BrPa00}. Recently, the
correlations across a large distance in rapidity have also been
suggested to arise from a color glass condensate
\cite{ArMcPa07,LaMc06}. The observation of
such  correlations in A+A collisions at RHIC energies by the
STAR Collaboration \cite{Ta07,TaScSr07} has
therefore elicited a lot of theoretical interest.

The purpose of this Section is to identify some {\em baseline}
contributions to the experimentally observed correlations,
contributions that do not depend on new physics
\cite{KoHaToGoBr08}. Two models that incorporate event-by-event
fluctuations in initial conditions have been used to illustrate
the effect of these contributions: the HSD transport model and a
`toy' wounded nucleon model.
Based on such a comparison one can argue that a study of the
dependence of correlations on the centrality bin definition as
well as the bin size may distinguish between `trivial'
correlations and correlations arising from `new physics'.


The statistical properties of a particular sample  of events can
be characterized by a set of moments or cumulants of some
observable. These properties depend upon a set of criteria which
are used to select this sample. Applied to the context of A+A
collisions this translates to the construction of centrality bins
of collision events from minimum-bias data. The charged hadron
multiplicities $N_A$ and $N_B$ will be considered in two symmetric
intervals $\Delta \eta$ of pseudo-rapidity. After construction of
the centrality bins, one can calculate the moments of a resulting
distribution $P^{\eta_{gap}}_{c}(N_A,N_B;\Delta\eta)$: \eq{
\langle N_A^k \cdot N_B^l \rangle^{\eta_{gap}}_c ~\equiv~
\sum_{N_A,N_B}~ N_A^k ~ N_B^l ~
P^{\eta_{gap}}_{c}(N_A,N_B;\Delta\eta)~. \label{eq1} } In
(\ref{eq1}) the subscript $c$ denotes a particular centrality bin,
while the superscript $\eta_{gap}$ denotes the separation of two
symmetric intervals $\Delta\eta$ in pseudo-rapidity space where
particle multiplicities $N_A$ and $N_B$ are measured. The
correlation coefficient
is defined by \eq{ \rho ~\equiv~ \frac{\langle \Delta N_A \cdot
\Delta N_B \rangle^{\eta_{gap}}_c}{\sqrt { \langle \left (\Delta
N_A \right)^2 \rangle^{\eta_{gap}}_c ~ \langle \left (\Delta N_B
\right)^2 \rangle^{\eta_{gap}}_c } } \label{eq2} } and measures
how strongly multiplicities $N_A$ and $N_B$ -- in a given
centrality bin $c$ for pseudo-rapidity separation $\eta_{gap}$ --
are correlated. In (\ref{eq2}), $\Delta N \equiv N-\langle N
\rangle^{\eta_{gap}}_c$ and $\langle \left (\Delta N_A \right)^2
\rangle^{\eta_{gap}}_c~ =~\langle \left (\Delta N_B \right)^2
\rangle^{\eta_{gap}}_c$ for symmetric intervals.

The recent preliminary data on the forward-backward correlation
coefficient (\ref{eq2}) of charged particles by the STAR
Collaboration
\cite{Ta07,TaScSr07} exhibit two striking
features:~ a)~an approximate independence on the width of the
pseudo-rapidity gap $\eta_{gap}$~,~ b) a strong increase of
$\rho$  with centrality.

\subsection{Glauber Monte Carlo Model}

The PHOBOS Glauber Monte Carlo code~\cite{AlBaLoSt08}
coupled to  a `toy' wounded nucleon model is used here, referred to as GMC.
The aim of this model is to emphasize two crucial aspects:
1)~an averaging over different system sizes
 within one centrality bin introduces correlations;
2)~the strength of these correlations depend on the criteria
 used for the centrality definition and on the size of the centrality bins.

\begin{figure}[ht!]
\begin{center}
\epsfig{file=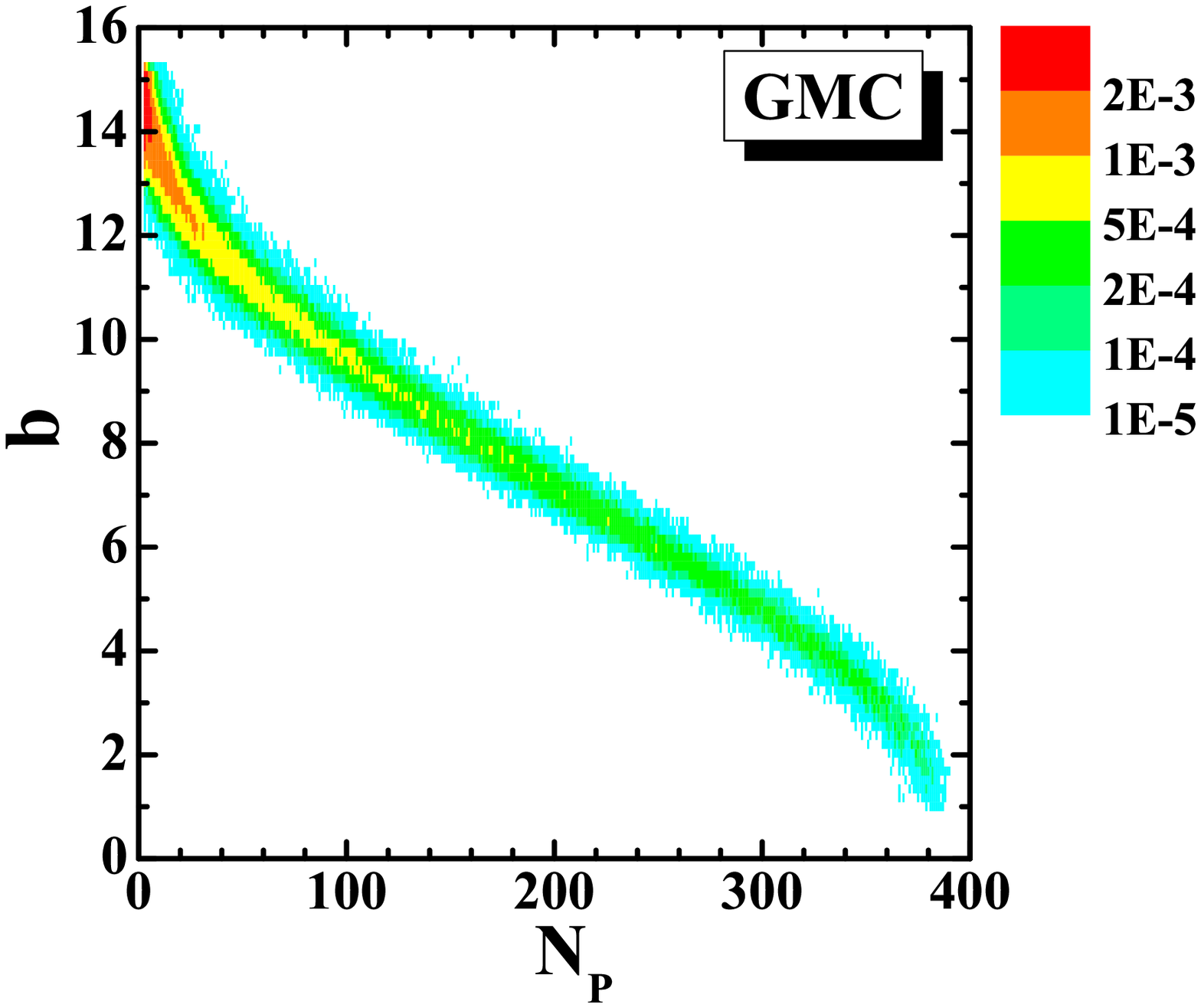,width=0.46\textwidth}
\epsfig{file=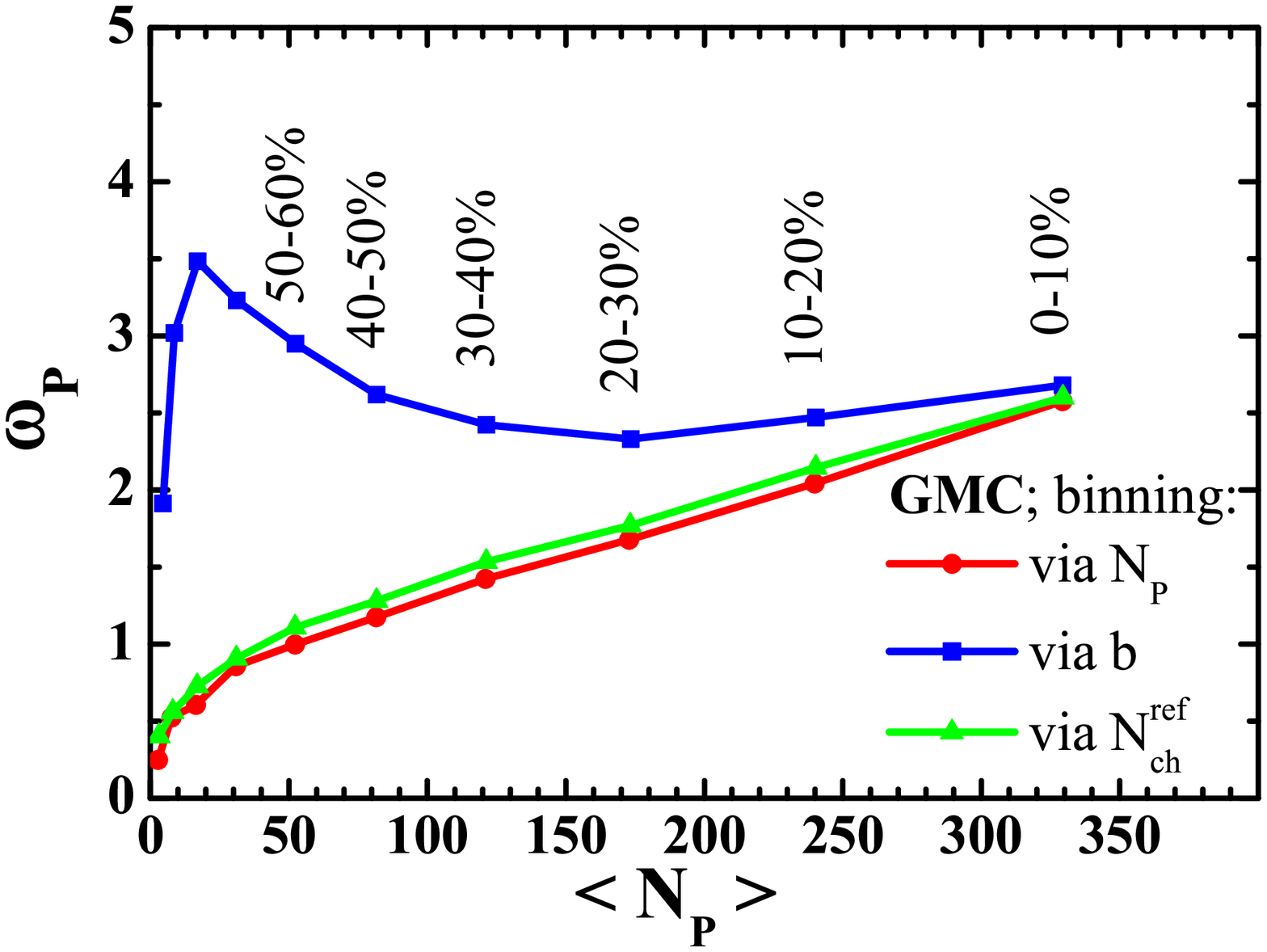,width=0.49\textwidth}
\end{center}
\caption{{\it Left}:  The histogram shows the distribution of
events with a fixed number of participating nucleons $N_{P}$ and
fixed impact parameter $b$ in Au+Au collisions at
$\sqrt{s_{NN}}=200$~GeV. {\it Right}: The scaled variance
$\omega_{P}$ of the distribution of participating nucleons in $10
\%$ bins as defined via $b$, $N_{P}$, and $N_{ch}^{ref}$.}
\label{fig:GMC_fluc_Npart}
\end{figure}

Employing the Glauber code the distribution of the number of
participating nucleons, $N_{P}$, is modelled in each A+A
collision for given impact parameter $b$ (cf. Fig.~\ref{fig:GMC_fluc_Npart}, {\it left}).
This is done for Au+Au with
standard Wood-Saxon profile and the N+N
cross section
of $\sigma_{NN}=42$~mb. The `event' construction proceeds then in a
two-step process.
Firstly, the total number of charged particles is randomly generated:
\eq{\label{GMC_nch}
N_{ch} ~=~ \sum_{i=1}^{N_{P}} ~ n_{ch}^i~ ,
}
where the number of charged particles $n_{ch}^i$ per participating
nucleon  are generated by independently sampling a Poisson
distribution with given mean value $\overline{n}_{ch}=10$.
Secondly, these charged particles are randomly distributed
according to a Gaussian in pseudo-rapidity space:
\eq{\label{GMC_dndy} \frac{dN_{ch}}{d\eta} ~\propto~ \exp \left(
-~ \frac{\eta^2}{2 \sigma_{\eta}} \right)~, } where $\sigma_{\eta}
= 3$ defines the width of the pseudo-rapidity distribution. Hence,
in each single event there are no correlations between the momenta
of any two particles. Note that numerical values of
$\overline{n}_{ch}$ and $\sigma_{\eta}$ are fixed in a way to have
a rough correspondence with the data on charged particle
production at $\sqrt{s_{NN}}=200$~GeV.

Fig.~\ref{fig:GMC_fluc_Npart} ({\it left}) shows the GMC event
distribution in the $(b,N_{P})$-plane. For each of these events we
randomly generate the number of charged particles $N_{ch}$ and
their $\eta$-distribution according to (\ref{GMC_nch}) and
(\ref{GMC_dndy}), respectively.  The construction of centrality
classes can now be done in several ways. The following criteria
are chosen: via impact parameter $b$, via the number of
participating (wounded) nucleons $N_{P}$, and via the charged
particle multiplicity $N_{ch}^{ref}$ in the midrapidity window
$|\eta| <1$.

In the case  of choosing the number of participating nucleons $N_P$
for centrality definition, one takes vertical cuts in
Fig.~\ref{fig:GMC_fluc_Npart} ({\it left}), while choosing the impact
parameter $b$, one takes horizontal cuts.  Hence, depending on the
centrality definition, one may assign a particular event
(characterized by $N_{P}$ and $b$) to two different centrality
bins.

Fig.~\ref{fig:GMC_fluc_Npart} ({\it right}) shows the resulting
scaled variance $\omega_{P}$,
\eq{
\omega_{P} ~\equiv~ \frac{ \langle \left( \Delta N_{P}
 \right)^2 \rangle_c}{\langle N_{P} \rangle_c}~,
} of the underlying distribution of the number of participating
nucleons $N_{P}$ in each centrality bin. Using the centrality
selection via impact parameter $b$, which is only the
theoretically available trigger,  one generally obtains a rather
wide distribution of participating nucleons in each bin. The lines
for centrality selections via $N_{ch}^{ref}$ and via $N_{P}$ are
similar due to the event construction by (\ref{GMC_nch}) and
(\ref{GMC_dndy}).

\begin{figure}[ht!]
\begin{center}
\epsfig{file=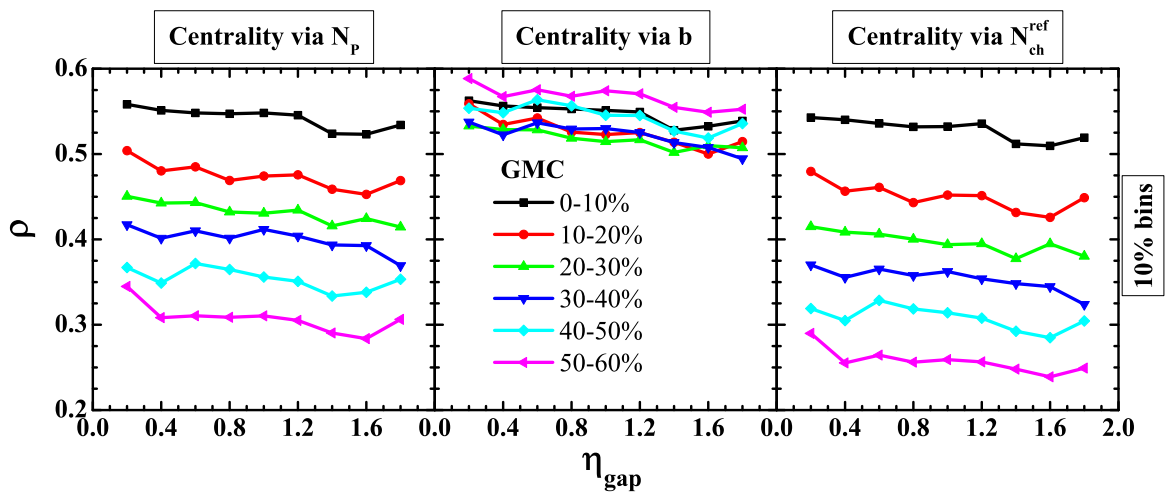,width=\textwidth}
\end{center}
\caption{The forward-backward correlation coefficient $\rho$ for $10\%$
centrality classes defined via $N_{P}$ ({\it left}), via the
impact parameter $b$ ({\it middle}), and via the multiplicity in
the central rapidity region  $N^{ref}_{ch}$ ({\it right}).}
\label{fig:GMC_FB_corr}
\end{figure}

The sensitivity of the forward-backward
correlation signal as a function of the separation $\eta_{gap}$ of
two narrow intervals ($\Delta \eta = 0.2$) on the centrality
definition is investigated now. This is done for the 10\% centrality defined via
$N_{P}$, via $b$, and via $N_{ch}^{ref}$. The results are shown in Fig.~\ref{fig:GMC_FB_corr}.
In the GMC one can identify the number of participating nucleons
$N_{P}$ with the system size, and $\omega_{P}$ as the measure for
system size fluctuations. Having a large system -- as measured by
$N_{P}$ -- implies a large number of charged particles $N_{ch}$. In
GMC they are distributed independently in pseudo-rapidity space.
Conversely, an event with small $N_{P}$ contains only few
charged particles. By grouping the collision events into 10\%
centrality bins one finds rather large $N_P$-fluctuations in one
specific bin. The averaging over different states in the
centrality bin introduces correlations between any two regions
of pseudo-rapidity. Small systems will have few particles `on the
left' and few particles `on the right' with respect to
midrapidity. Large systems will have many particles `on the left'
and many particles `on the right'.  But this just means a non-zero
forward-backward correlation. From the definition (\ref{eq2}) one
finds a positive correlation coefficient~$\rho$ due to averaging
over system sizes.

Note that centrality selections via $N_P$ and via $N_{ch}$ give
essentially the same results for $\rho$ in the GMC (cf. {\it left}
and {\it right} panels of Fig.~\ref{fig:GMC_FB_corr}). Using the
impact parameter $b$ for the centrality definition generates
centrality bins with almost constant $\rho$ as seen in
Fig.~\ref{fig:GMC_FB_corr} ({\it middle}). This is due to a rather
flat dependence of $\omega_{P}$ on the centrality defined via $b$
as shown in Fig.~\ref{fig:GMC_fluc_Npart} ({\it right}). In the GMC
model the apparent ordering of $\rho$ values with respect to
centrality bins originates from the width of the underlying
distribution in the number of wounded nucleons in each bin, i.e.
from the values of $\omega_P$ .

The measured and apparently strong forward-backward correlations
can be accounted for by a `toy' model such as the GMC, provided it
produces particles over the whole rapidity range and includes
strong enough event-by-event fluctuations of $N_P$. The next
section will show that an introduction of dynamics and hadron
re-interactions within HSD does not alter these conclusions
significantly.

\subsection{HSD Transport Model Simulations}

A physically more reasonable scenario which, however, also does not
include any `new physics' (such as color glass condensate,
quark-gluon plasma, etc.) can be obtained in the
HSD transport approach. 
As before within the GMC, the HSD events are generated according to a
uniform distribution, $N_{ev}(b) \sim b$.  The resulting
distribution of events in the $(N_{P},b)$-plane is similar to the
GMC result depicted in Fig.~\ref{fig:GMC_fluc_Npart} ({\it left}).
Note, that the peripheral part of the distribution determines also
the centrality binning and the real bin widths. This is crucial
for most central collisions where the number of events is small.
Slight uncertainties in the peripheral ``tail'' of the
distribution leads to large errors in the sizes of most central
bins and hence to large changes in results for fluctuations and
correlations.

In contrast to the STAR data,
the charged particle reference multiplicity $N_{ch}^{ref}$
in the same pseudo-rapidity range $|\eta|<1$ for all values of $\eta_{gap}$
is used in the HSD simulations.
This procedure introduces a systematic bias, since the
pseudo-rapidity regions for the measured multiplicity in a small
$\Delta \eta$ window (signal) and for the reference multiplicity
partially overlap. This bias, however, is small and does not
affect any of the conclusions.

\begin{figure}[ht!]
\begin{center}
\epsfig{file=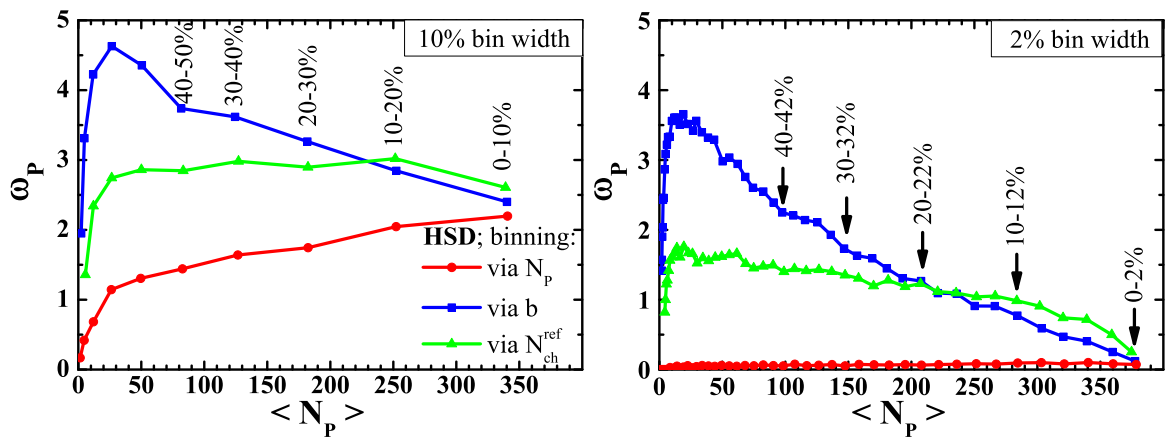,width=\textwidth}
\end{center}
\caption{The HSD results for the fluctuations $\omega_{part}$ as a function
of the mean value $\langle N_{P}\rangle$ of the participating
nucleons within bins as defined via $b$, $N_{P}$, and
$N_{ch}^{ref}$.  The {\it left} panel corresponds to a 10\% and
the {\it right} to a 2\% bin width.}
\label{fig:HSD_fluc_Npart}
\end{figure}

\begin{figure}[ht!]
\begin{center}
\epsfig{file=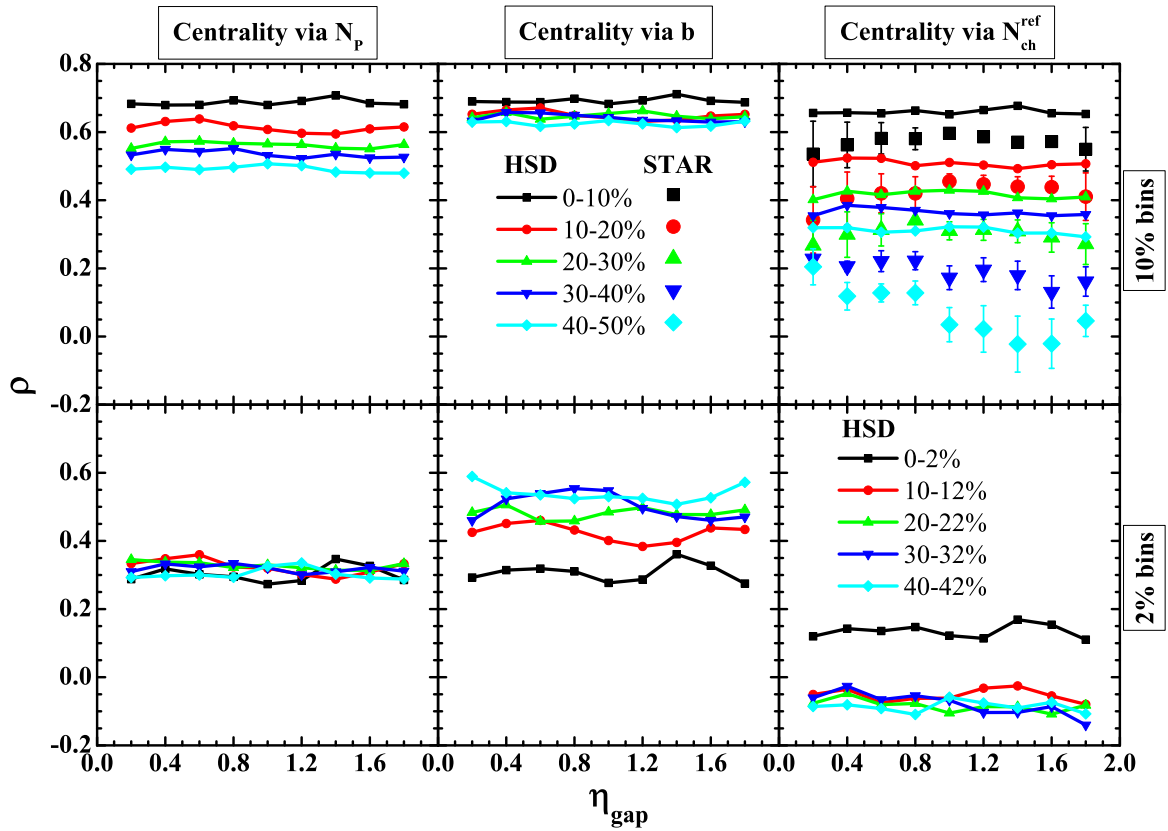,width=\textwidth}
\end{center}
\caption{The HSD results for the forward-backward correlation
coefficient $\rho$ for 10\% ({\it top}) and 2\% ({\it bottom})
centrality classes defined via $N_{P}$ ({\it left}), via impact
parameter $b$ ({\it center}), and via the reference multiplicity
$N_{ch}^{ref}$ ({\it right}). The symbols in the {\it top right}
panel present the STAR data in Au+Au collisions at
$\sqrt{s_{NN}}=200$~GeV \cite{Ta07,TaScSr07}.}
\label{fig:HSD_FB_corr}
\end{figure}

Fig.~\ref{fig:HSD_fluc_Npart} shows the scaled variance of the
underlying $N_{P}$ distribution for 10\% ({\it left}) and 2\%
({\it right}) centrality bins defined via different centrality
triggers within HSD. The results for 10\% bins can be compared
with the scaled variance $\omega_P$ in the GMC model in
Fig.~\ref{fig:GMC_fluc_Npart} ({\it right}). Fluctuations of the
number of participants, as well as their average values, are
similar in both HSD and GMC models when the centrality bins are
defined via $N_{P}$. These quantities are completely defined by
the $N_{P}$ distribution, which is similar in both models.
Binning via the impact parameter $b$
in HSD, as well as in GMC, gives decreasing fluctuations in the
participant number with increasing collision centrality.  The
results for  10\% bins defined via the reference multiplicity are
rather different in the GMC and HSD models. In GMC the charged
multiplicity distribution is implemented according to
(\ref{GMC_nch}) and (\ref{GMC_dndy}). Hence, the results obtained
by binning via the reference multiplicity follow the line obtained
by binning via $N_{P}$. In contrast to the GMC, in the HSD
simulations the average number of charged particles
$\overline{n}_{ch}$ per participating nucleon is not a constant,
but increases with $N_P$. Additionally, the shape of rapidity
distribution is also different in different centrality bins. These
two effects lead to different values of $\omega_P$ in the
centrality bins defined via $N_{ch}^{ref}$ in the GMC and HSD
models.
An interesting feature seen in Figs.~\ref{fig:GMC_fluc_Npart}
and \ref{fig:HSD_FB_corr}
is that $\omega_{P}$ is large for 10\% most central bins.
The behavior for 2\% centrality bins
is rather different: $\omega_P$ decreases with centrality as seen
from Fig.~\ref{fig:HSD_fluc_Npart} right. This becomes
similar to the dependence of $\omega_P$ on $N_P^{proj}$
considered in Sec.~\ref{ch:mult} for the fixed target experiments.

One comment is appropriate here. It was argued in
Ref.~\cite{BrFl02} that any centrality selection in A+A collisions
is equivalent to the geometrical one via impact parameter $b$.
This was already discussed in Sec.~\ref{ch:ppAA}.
Different centrality selection criterions give indeed the same
{\it average} values of physical observables. However, they may
lead to rather different fluctuations of these observables in the
corresponding centrality bins, cf. equal values of $\langle
N_P\rangle$ and different values of $\omega_P$ for different
centrality selections presented in Fig.~\ref{fig:HSD_fluc_Npart}.
When considering smaller centrality bins (2\% in
Fig.~\ref{fig:HSD_fluc_Npart}, {\it right}) the fluctuations in
the participant number become smaller but more strongly dependent
on the definition of the binning.

Fig.~\ref{fig:HSD_FB_corr} summarizes the dependence of
forward-backward correlation coefficient $\rho$ as a function of
$\eta_{gap}$ on the bin size and centrality definition within the
HSD model. The dependence of $\rho$ on $\eta_{gap}$ is almost
flat, reflecting a boost-invariant distribution of particles
created by string breaking in the HSD.
The {\it right top} panel of Fig.~\ref{fig:HSD_FB_corr} demonstrates
also a comparison of the HSD results with the STAR data
\cite{Ta07,TaScSr07}. One observes that the
HSD results exceed systematically the STAR data. However, the main
qualitative features of the STAR data -- an approximate
independence of the width of the pseudo-rapidity gap $\eta_{gap}$
and a strong increase of $\rho$  with centrality -- are fully
reproduced by the HSD simulations.

\begin{figure}[ht!]
\begin{center}
\epsfig{file=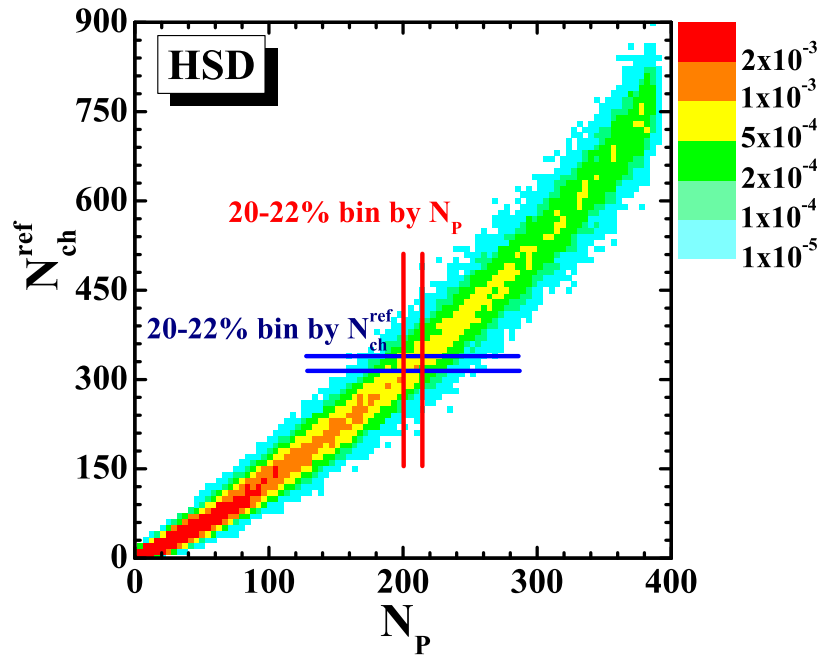,width=0.7\textwidth}
\end{center}
\caption{The histogram shows the distribution of HSD events with
fixed number of participating nucleons $N_{P}$ and fixed reference
charge particle multiplicity $N_{ch}^{ref}$.  The same centrality
class (20-22\% as an example) defined in various ways contains
different events.} \label{fig:HSD_Npart_Nref}
\end{figure}

The correlation coefficient $\rho$ largely follows the trend of
the participant number  fluctuations $\omega_P$ as a function of
centrality. The actual results, however, strongly depend on the
way of defining the centrality bins. For instance,  choosing
smaller centrality bins leads to weaker forward-backward
correlations, a less pronounced centrality dependence, and a
stronger dependence on the bin definition.
The physical origin for this is demonstrated in
Fig.~\ref{fig:HSD_Npart_Nref}. As the bin size becomes comparable to
the width of the correlation band between $N_{P}$ and
$N_{ch}^{ref}$, the systematic deviations of different centrality
selections  become dominant: the same centrality bins defined by
$N_P$ and by $N_{ch}^{ref}$ contain different events and may give
rather different values for the forward-backward correlation
coefficient $\rho$.

It should be underlined that these properties are specific to the
{\em geometric} nature of the correlations analyzed here. If the
observed fluctuations are of {\em dynamical} origin (for example,
arising from the quantum fluctuations of coherent fields created
in the first $fm/c$ of the system's lifetime as in Refs.
\cite{ArMcPa07,LaMc06}), there are no evident reasons why they
should strongly  depend on the centrality bin definitions and bin
sizes. Thus, the experimental analysis for different bin sizes and
centrality definitions -- as performed here -- may serve as a
diagnostic tool for an origin of the observed correlations. A
strong specific dependence of the correlations on  bin size and
centrality definition would signify their geometrical origin.

\section{ Fluctuations of Conserved Charges in Nucleus-Nucleus Collisions}
\label{ch:QB}

The aim of the present Section is to study event-by-event fluctuations
of the conserved charges -- net baryon number and electric charge -- in Pb-Pb collisions
at 158 AGeV energies within the HSD transport approach \cite{KoGoBrSt06}.

\subsection{Net Baryon Number Fluctuations}

Let's start with a  quantitative discussion by first considering the
fluctuations of the net baryon number in different regions
of the participant domain in collisions of two identical nuclei.
These fluctuations are most closely related to the fluctuations of
the number of participant nucleons because of baryon number
conservation.

\begin{figure}[ht!]
\begin{center}
\epsfig{file=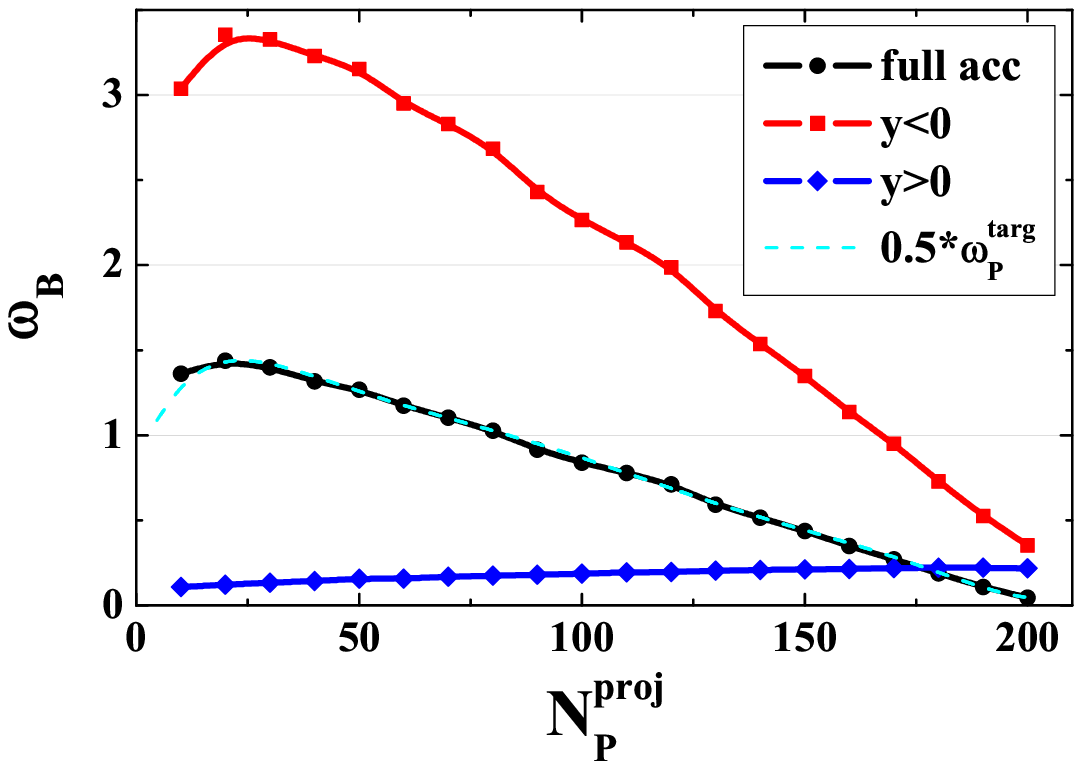,width=0.45\textwidth}
\epsfig{file=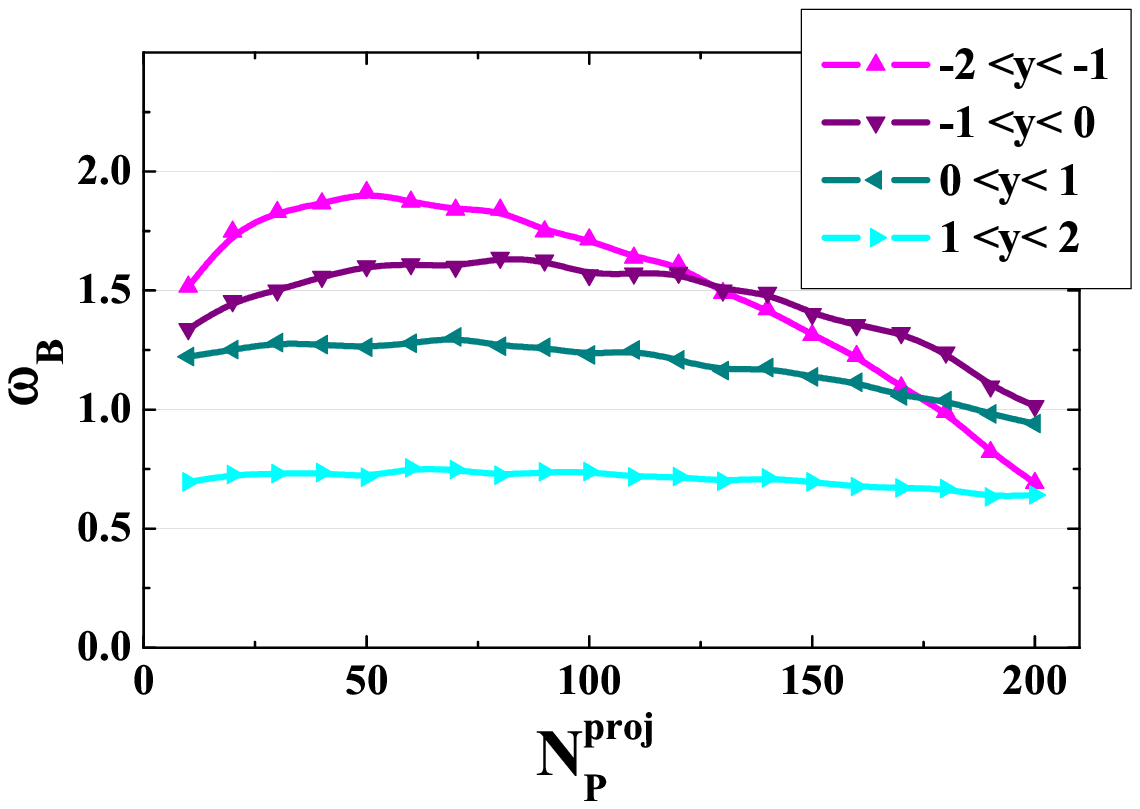,width=0.45\textwidth}
\end{center}
\caption{The HSD simulations for Pb+Pb collisions  at 158~AGeV for
fixed values of $N_P^{proj}$. {\it Left:} The baryon number
fluctuations in full acceptance, $\omega_B$, in projectile
hemisphere, $\omega_B^p$ (lower curve), and in target hemisphere,
$\omega_B^t$ (upper curve).
{\it Right:} The scaled variances of the baryon number fluctuations in different rapidity intervals.}
\label{fig:wB1}
\end{figure}

The HSD results for $\omega_B$ in Pb+Pb at 158~AGeV are presented in Fig.~\ref{fig:wB1}.
In each event the nucleon spectators when counting the number of baryons are subtracted.
Otherwise there would be no fluctuations of the net baryon number
in the full acceptance and $\omega_B$ equals to zero.
The net baryon number in the full phase space,
$B\equiv N_B-N_{\overline{B}}$, equals to the total number of
participants $N_P=N_P^{targ}+N_P^{proj}$. At fixed $N_P^{proj}$ the
$N_P$ number fluctuates due to  fluctuations of $N_P^{targ}$.
These fluctuations correspond to an average value, $\langle
N_P^{targ}\rangle \simeq N_P^{proj}$, and a scaled variance,
$\omega_P^{targ}$ (see Fig.~\ref{fig:wNtarg}). Thus, for the net baryon
number fluctuations in the full phase space one finds,
$\omega_B\cong 0.5\omega_B^{targ}$.
A factor $1/2$ appears because only a half of the total number of 
participants fluctuates.

One introduces $\omega_B^p$ and $\omega_B^t$, where the
superscripts $p$ and $t$ mark quantities measured in the
projectile and target momentum hemispheres, respectively.
Fig.~\ref{fig:wB1} demonstrates that $\omega_B^t>\omega_B^p$, both
in the whole projectile-target hemispheres and in the symmetric
rapidity intervals. On the other hand one observes that
$\omega_B^p \approx \omega_B^t$ in most central collisions. This
is because the fluctuations of the target participants become
negligible in this case, i.e. $\omega_P^{targ}\cong 0$
(Fig.~\ref{fig:wNtarg}, {\it  right}). As a consequence the
fluctuations of any observable in the symmetric rapidity intervals
become identical in most central collisions. Note also that
transparency-mixing effects are  different at different
rapidities. From Fig.~\ref{fig:wB1} ({\it  right}) it follows that
$\omega_B^p$ in the target rapidity interval $[-2,-1]$ is much
larger than $\omega_B^t$ in the symmetric projectile rapidity
interval $[1,2]$. This fact reveals the strong transparency
effects. On the other hand, the behavior is different in symmetric
rapidity intervals near the midrapidity. From Fig.~\ref{fig:wB1}
({\it  right}) one observes that $\omega_B^p$ in the target
rapidity interval $[-1,0]$ is already much closer to $\omega_B^t$
in the symmetric projectile rapidity interval $[0,1]$. This gives
a rough estimate of the width, $\Delta y \approx 1$, for the
region in rapidity space where projectile and target nucleons
communicate to each others.

\subsection{Net Electric Charge Fluctuations}

\begin{figure}[ht!]
\begin{center}
\epsfig{file=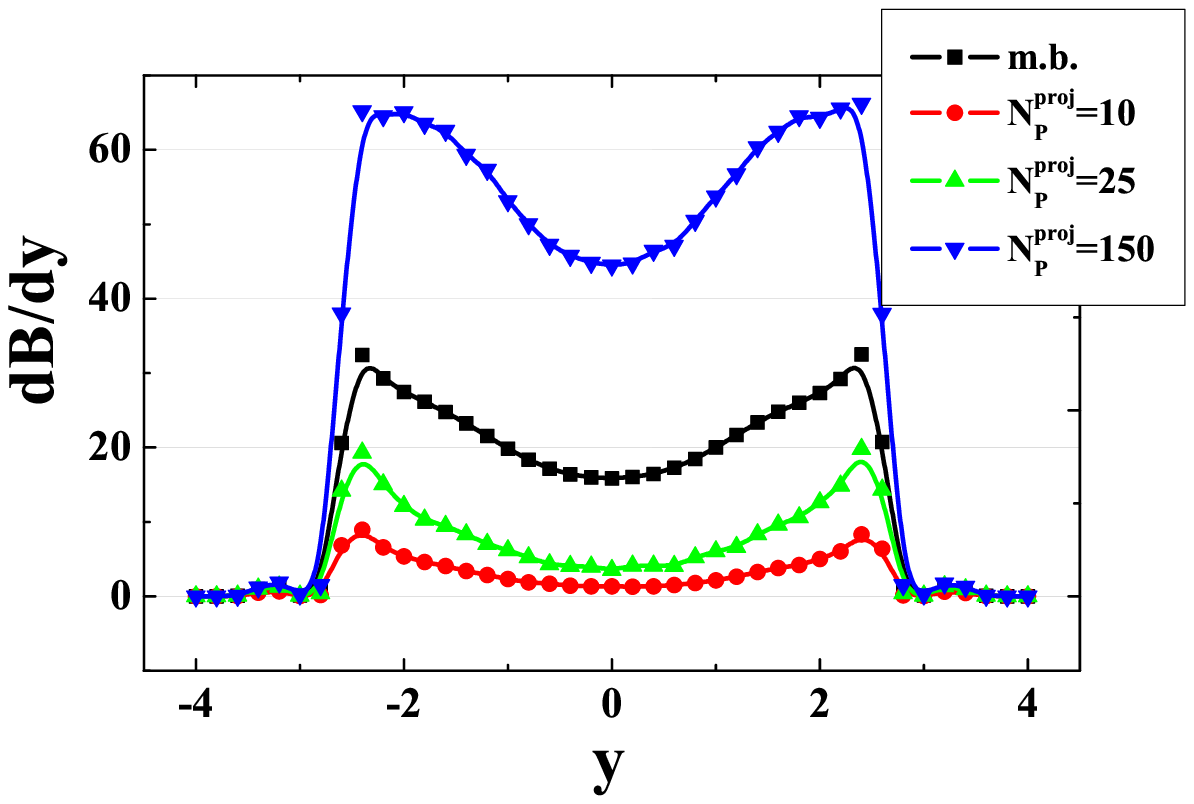,width=0.45\textwidth}
\epsfig{file=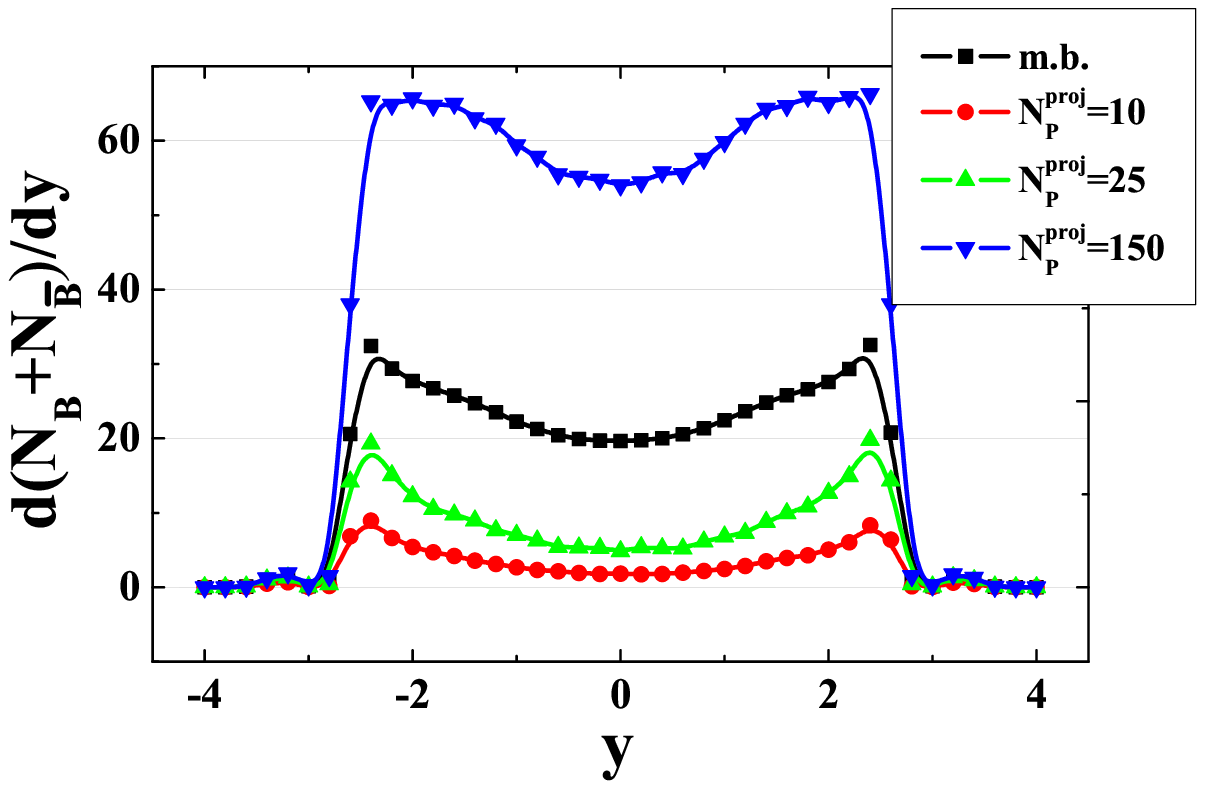,width=0.45\textwidth}
\end{center}
\caption{The HSD rapidity distributions in Pb+Pb collisions at
158~AGeV for the net baryon number, $B=N_B-N_{\overline{B}}$ ({\it
left}), and total number of baryons, $N_B+N_{\overline{B}}$ ({\it
right}), at different $N_P^{proj}$ and in the minimum bias (m.b.)
sample.} \label{fig:Bt}
\end{figure}

It is difficult to study
experimentally the baryon number fluctuations since an identification of
neutrons in a large acceptance  in a single event is difficult.
In this subsection we consider the HSD results for the net
electric charge, $Q$, fluctuations. As $Q\cong 0.4 B$ in the
initial heavy nuclei one can naively expect that $Q$ fluctuations
are quite similar to $B$ fluctuations. However, there is a
principal difference between $Q$ and $B$ in relativistic A+A
collisions. Fig.~\ref{fig:Bt} demonstrates the rapidity
distributions of the net baryon number, $B=N_B-N_{\overline{B}}$~
({\it  left}), and {\it total} number of baryons,
$N_B+N_{\overline{B}}$~ ({\it  right}), for different centralities
in Pb+Pb collisions at 158~AGeV. One observes that both quantities
are very close to each other; the $y$-dependence and absolute
values are very close for $B$ and $N_B+N_{\overline{B}}$
distributions. This is, of course, because the number of
antibaryons is rather small, $N_{\overline{B}}\ll N_B$.

\begin{figure}[ht!]
\begin{center}
\epsfig{file=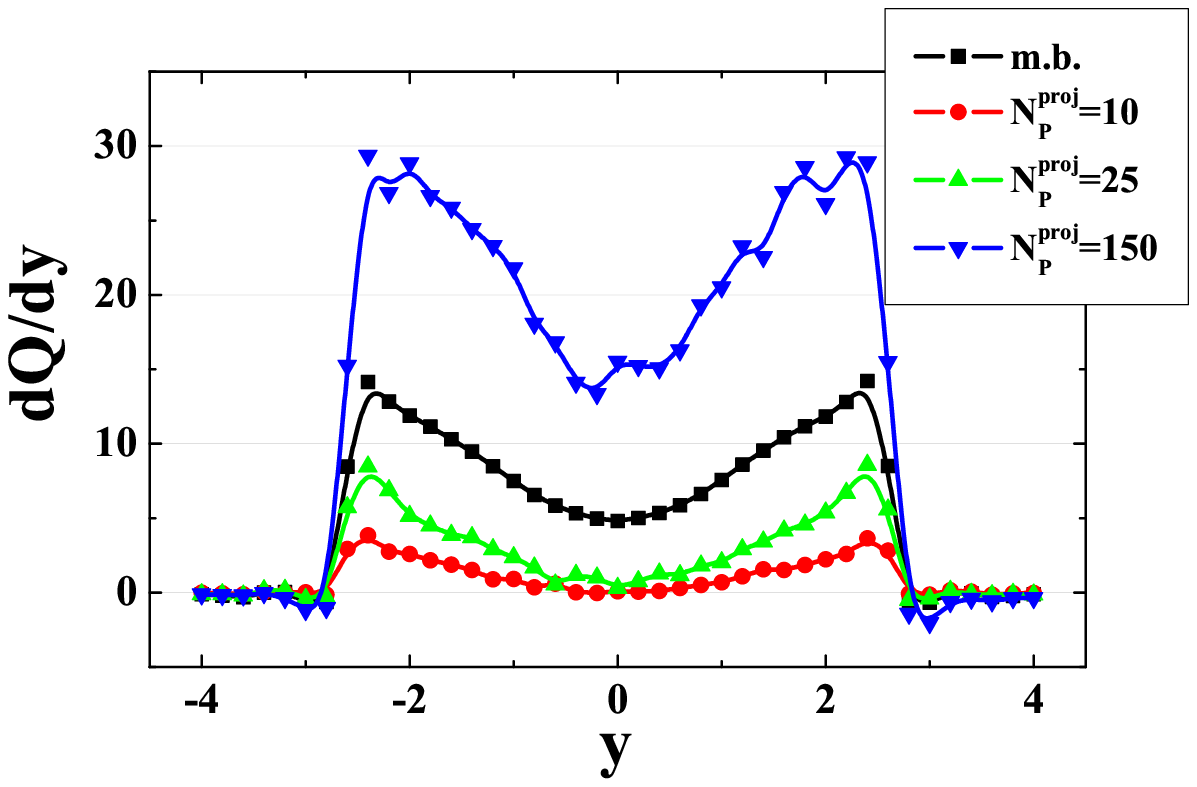,width=0.45\textwidth}
\epsfig{file=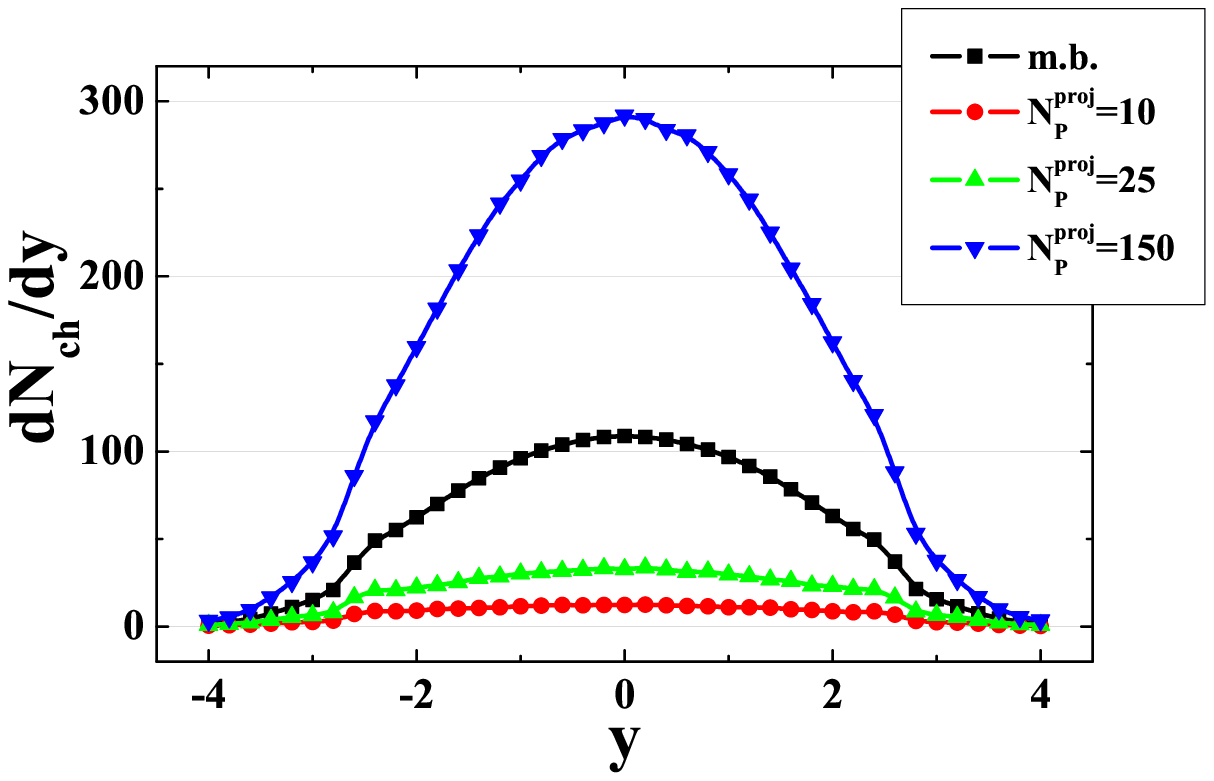,width=0.45\textwidth}
\end{center}
\caption{The same as in Fig.~\ref{fig:Bt} but for the electric
charge  $Q=N_+ -N_{-}$~ ({\it left}), and total number of charged
particles, $N_{ch}\equiv N_++N_{-}$~ ({\it right}).}
\label{fig:Qt}
\end{figure}

\begin{figure}[ht!]
\begin{center}
\epsfig{file=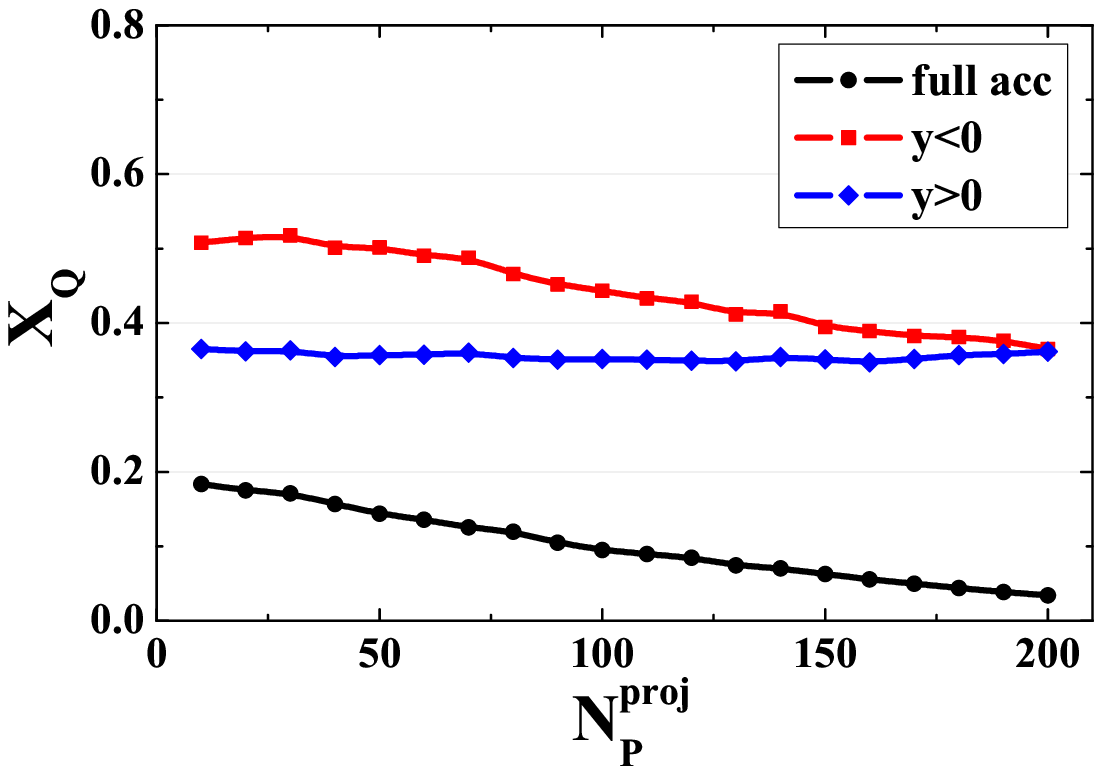,width=0.45\textwidth}
\epsfig{file=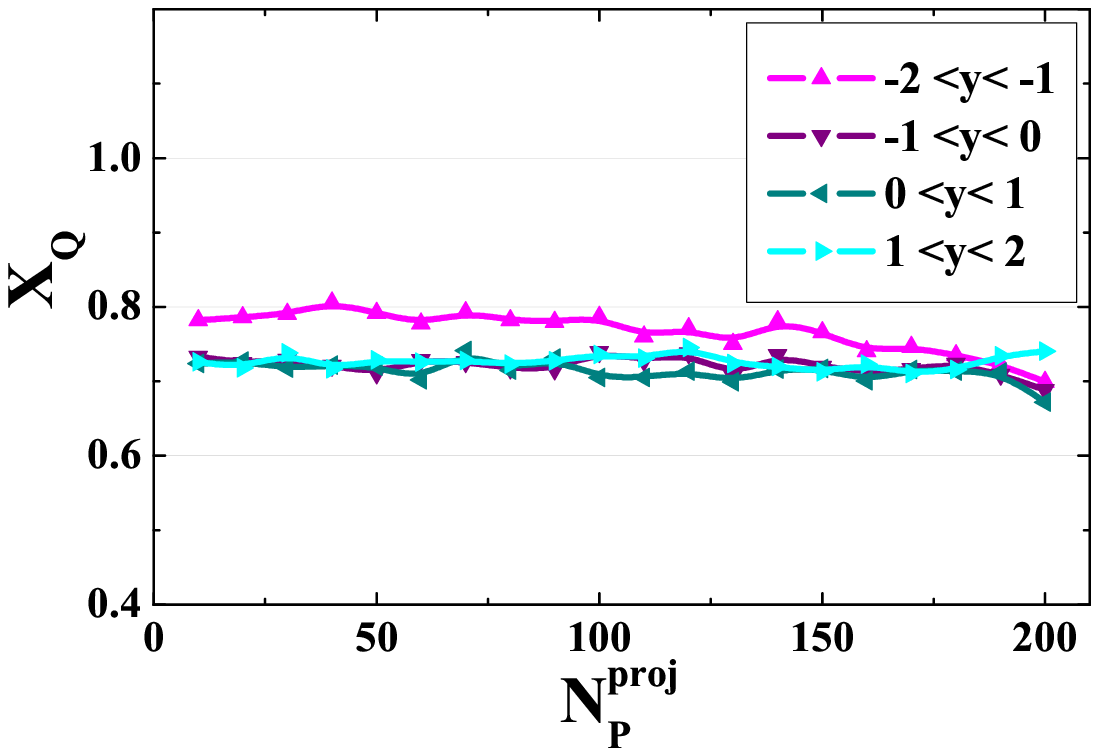,width=0.45\textwidth}
\end{center}
\caption{{\it Left}: The HSD simulations in Pb+Pb collisions at
158~AGeV for $X_Q$ at different values of $N_P^{proj}$ in the full
acceptance (lower curve), for the projectile (middle curve) and
target (upper curve) hemispheres. {\it Right:} The same, but for
the fixed rapidity intervals in the c.m.s. } \label{fig:xQ1}
\end{figure}

Fig.~\ref{fig:Qt} shows the same as Fig.~\ref{fig:Bt} but for the
electric charge  $Q=N_+ -N_{-}$~ ({\it  left}), and {\it total}
number of charged particles, $N_{ch}\equiv N_++N_{-}$~ ({\it
right}).
The $y$-dependence of $dQ/dy$ and $dN_{ch}/dy$ is quite different.
Besides, the absolute values of $N_{ch}$ are about 10 times larger
than those of $Q$.  This implies that $Q\ll N_+\approx N_-$.
In the previous subsection the scaled variance $\omega_B$ to
quantify the measure of the net baryon fluctuations has been used.
It appears to be a useful variable as $\omega_B$ is
straightforwardly connected to $\omega_P^{targ}$ and due to the
relatively small number of antibaryons. Fig.~\ref{fig:Qt} tells
that $\omega_Q$ is a bad  measure of the electric charge
fluctuations in high energy A+A collisions. One observes that
$\omega_Q\equiv Var(Q)/\langle Q\rangle$ is much larger than 1
simply due to the small value of $\langle Q\rangle$ in a
comparison with $N_+$ and $N_-$. If the A+A collision energy
increases, it follows, $\langle N_+\rangle\cong \langle
N_-\rangle$
increases with collision energy, whereas  $\langle Q\rangle \cong 0.4B$
remains constant. This leads to very large values of
$\omega_Q\rightarrow$. The same would happen with $\omega_B$,
too, but at much larger collision energies. A useful measure of the
net electric charge fluctuations is the quantity (see, e.g.,
\cite{JeKo03}):
\eq{
X_Q~\equiv~\frac{Var(Q)}{\langle N_{ch}\rangle}~. \label{XQ}
}
%
A value of $X_Q$ can be easily calculated for the Boltzmann ideal
gas in the grand canonical ensemble. In this case the number of
negative and positive particles fluctuates according to the
Poisson distribution (i.e. $\omega_-=\omega_+=1$), and the
correlation between $N_+$ and $N_-$ are absent (i.e. $\langle
N_+N_-\rangle = \langle N_+\rangle \langle N_-\rangle$), so that
$X_Q=1$. On the other hand, the canonical ensemble formulation
(i.e. when $Q=const$ fixed exactly for all microscopic states of
the system) leads to $X_Q=0$. Fig.~\ref{fig:xQ1} shows the results
of the HSD simulations for the full acceptance, for the projectile
and target hemispheres ({\it  left}), and also for symmetric
rapidity intervals in the c.m.s. ({\it  right}).

The $Q$ fluctuation in the full acceptance is due to $N_P^{targ}$
fluctuations. As $Q\cong 0.4 B$ in colliding (heavy) nuclei, one
may expect $Var(Q)\cong 0.16~ Var(B)$. In addition, $\langle
N_{ch}\rangle \cong 4 \langle N_P \rangle$ at 158~AGeV, so that
one estimates $X_Q \cong 0.04~ \omega_B$ for the fluctuations in
the full phase space. The actual values of $X_Q$ presented in
Fig.~\ref{fig:xQ1} ({\it  left}) are about 3 times larger. This is
because of $Q$ fluctuations  due to different event-by-event
values of proton and neutron participants even in a sample with
fixed values of $N_P^{proj}$ and $N_P^{targ}$.

From Fig.~\ref{fig:xQ1} ({\it  right}) one sees only a tiny
difference between the $X_Q$ values in the symmetric rapidity
intervals in the projectile and target hemispheres, and slightly
stronger effects for the whole projectile and target hemispheres
(Fig.~\ref{fig:xQ1}, {\it  left}). In fact, the fluctuations of
$N_+$ and $N_-$ are very different in the projectile and target
hemispheres, and the scaled variances $\omega_{+}^t$ and
$\omega_-^t$ have a very strong $N_P^{proj}$-dependence. This is
shown on top-{\it  left} and middle-{\it  left} panels of
Fig.~\ref{fig:proj-targ}.

 The $X_Q$ can be presented in two equivalent forms
\eq{
X_Q~&=~ \omega_+~\frac{\langle N_+\rangle }{\langle N_{ch}\rangle}~
     +~\omega_-~\frac{\langle N_-\rangle}{\langle N_{ch}\rangle}~
     -~2~\frac{\Delta(N_+,N_-)}{\langle N_{ch}\rangle}~~ \nonumber
\\  &=~2~\omega_+~\frac{\langle N_+\rangle}{\langle N_{ch}\rangle}~
     +~2~\omega_-~\frac{\langle N_-\rangle}{\langle N_{ch}\rangle}~
     -~\omega_{ch}~.\label{XQ-1}
}
The Eq.~(\ref{XQ-1}) is valid for any region of the phase space:
full phase space, projectile or target hemisphere, etc. As seen
from Fig.~\ref{fig:proj-targ}, both $\omega_+^t$ and $\omega_-^t$
are large and strongly $N_P^{proj}$-dependent. This is not seen in
$X_Q^t$ because of strong correlations between $N_+^t$ and
$N_-^t$, i.e. the term $2~\Delta(N_+,N_-)/\langle N_{ch}\rangle $
compensates $\omega_+$ and $\omega_-$ terms in (\ref{XQ-1}).

\begin{figure}[ht!]
\begin{center}
\epsfig{file=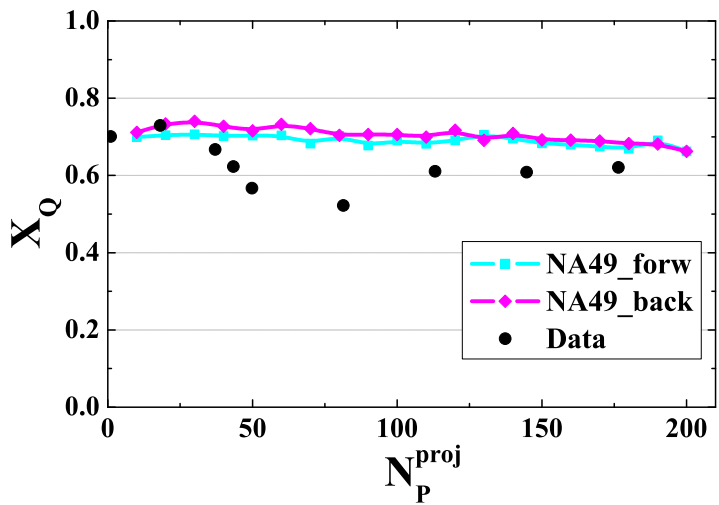,width=0.6\textwidth}
\end{center}
\caption{The HSD results for $X_Q$ for Pb+Pb collisions at
158~AGeV for the forward rapidity interval $1.1<y<2.6$ inside the
projectile hemisphere. The solid dots are the estimates obtained
from  Eq.~(\ref{XQ-1}) using the NA49 experimental data
\protect\cite{Ry05,An04,Di05} (the error bars are not indicated
here). For  illustration, the HSD results in the symmetric
backward rapidity interval $-2.6<y<-1.1$ (target hemisphere) are
also presented.} \label{fig:xQdata}
\end{figure}

Fig.~\ref{fig:xQdata} shows a comparison of the HSD results for $X_Q$
with NA49 data in Pb+Pb collisions at 158~AGeV for the forward
rapidity interval $1.1<y<2.6$ inside the projectile hemisphere with
additional $p_T$-filter imposed.

As an illustration, the HSD results in the symmetric backward
rapidity interval $-2.6<y<-1.1$ (target hemisphere) are also
included. One observes no difference between the $X_Q$ results for
the NA49 acceptance in the projectile and target hemispheres. The
HSD values for $\omega_+$, $\omega_-$, and $\omega_{ch}$ are
rather different in the projectile and target hemispheres for the
NA49 acceptance (see Fig.~\ref{fig:proj-targ}). This is not seen
in Fig.~\ref{fig:xQdata} for $X_Q$. As explained above a
cancellation between $\omega_+$, $\omega_-$ and $\omega_{ch}$
terms take place in (\ref{XQ-1}). In fact, NA49 did not perform
the $X_Q$ measurements. The $X_Q$-data (solid dots) presented in
Fig.~\ref{fig:xQdata} are obtained from Eq.~(\ref{XQ-1}) using the
NA49 data for $\omega_+$, $\omega_-$, and $\omega_{ch}$ as well as
$\langle N_+\rangle$, $\langle N_-\rangle$, and $\langle
N_{ch}\rangle$ \cite{Ry05,An04,Di05}. Such a procedure leads,
however, to very large errors  for $X_Q$ (not indicated in
Fig.~\ref{fig:xQdata}) which excludes any conclusion about the
(dis)agreement of HSD results with NA49 data.

\begin{figure}[ht!]
\begin{center}
\epsfig{file=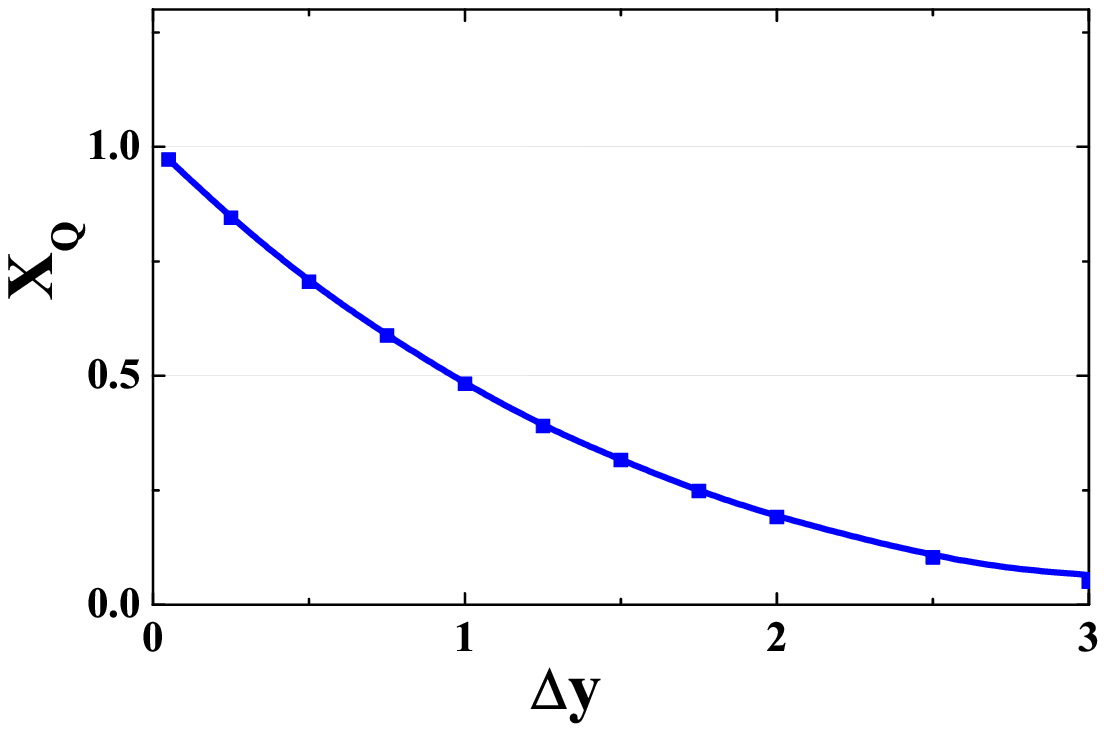,width=0.45\textwidth}
\epsfig{file=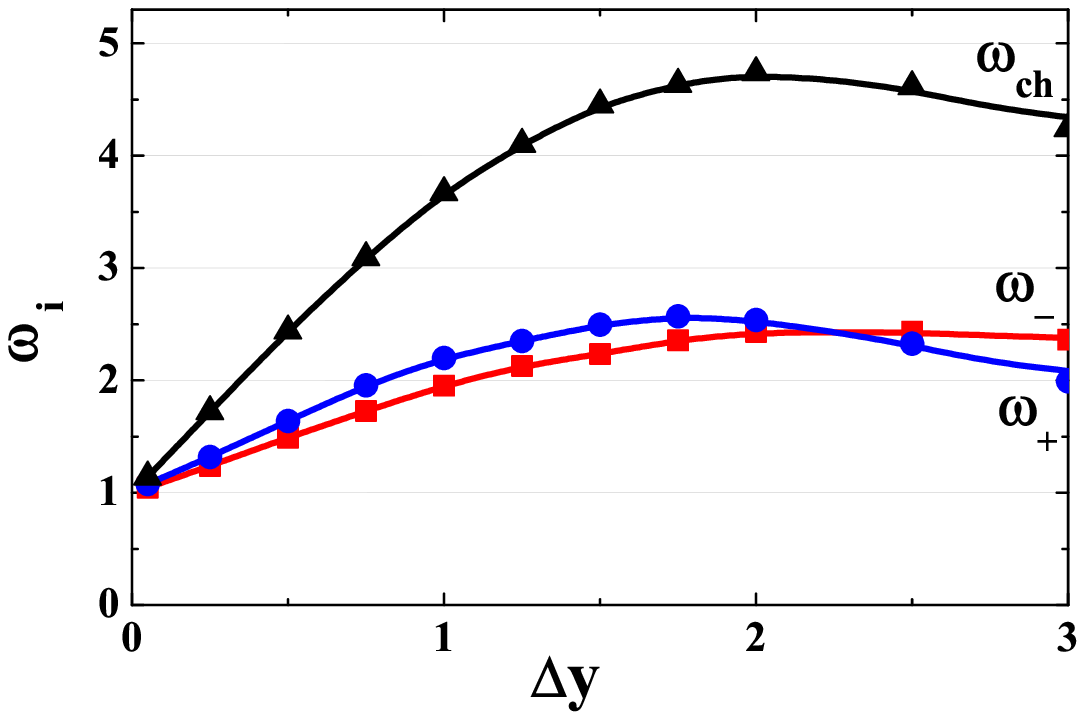,width=0.45\textwidth}
\end{center}
\caption{The HSD results for electric charge fluctuations in 2\%
most central Pb+Pb collisions at 158~AGeV in the symmetric
rapidity interval, $\Delta Y = [-y, y]$ as a function of  $\Delta
y= \Delta Y/2$ in the c.m.s. A {\it left} panel shows the behavior
of $X_Q$, and a {\it right} one demonstrates separately
$\omega_+$, $\omega_-$, and $\omega_{ch}$.} \label{fig:xQ-dY}
\end{figure}

It is instructive to consider the electric charge fluctuations in the 
symmetric rapidity interval $[-y, y]$ in the c.m.s.  for the most 
central Pb+Pb events.  The sample of most central events is chosen by 
restricting the impact parameter to $b<2$~fm. It gives about 2\%  most 
central Pb+Pb collisions from the whole minimum bias sample. 
Fig.~\ref{fig:xQ-dY} shows the HSD results for electric charge 
fluctuations in 2\% most central Pb+Pb collisions  for the symmetric 
rapidity interval $\Delta Y = [-y, y]$ in the c.m.s. as the function of 
$\Delta y =\Delta Y/2$.

For $\Delta Y\rightarrow 0$ one finds $X_Q\rightarrow 1$. This can
be understood as follows: For $\Delta Y\rightarrow 0$ the
fluctuations of negatively, positively and all charged particles
behave as for the Poisson distribution: $\omega_+\cong
\omega_-\cong \omega_{ch}\cong 1$. Then from Eq.~(\ref{XQ-1}) it
follows that $X_Q\cong 1$, too. From Fig.~\ref{fig:xQ-dY} ({\it
right}) one observes that $\omega_+$, $\omega_-$, and
$\omega_{ch}$ all increase with increasing interval $\Delta Y$.
However, $X_Q$ decreases with $\Delta Y$ and -- because of global
$Q$ conservation -- it goes approximately to zero when all final
particles are accepted.


\section{Fluctuations of Ratios in Nucleus-Nucleus Collisions}
\label{ch:ratio}

The measurement of the fluctuations in the kaon to pion
ratio by the NA49 Collaboration \cite{Af01} was the first
event-by-event measurement in nucleus-nucleus collisions.
It was suggested that this ratio might allow to distinguish events with
enhanced strangeness production attributed to the QGP phase.
Nowadays, the excitation function for this observable is available in a wide
range of energies: from the NA49 collaboration \cite{Al08} in Pb+Pb
collisions at the CERN SPS and from the STAR collaboration \cite{Da06,Ab09} in
Au+Au collisions at RHIC.
First statistical model estimates of the $K/\pi$ fluctuations have been
reported in Refs.~\cite{BaHe99,JeKo99}.
The results from the transport model UrQMD have been presented in Ref.~\cite{Ro04}
for the SPS energy range.

This Section presents the results of a systematic study of
$K/\pi$, $K/p$ and $p/\pi$ ratio fluctuations
($K=K^++K^-$, $\pi=\pi^++\pi^-$, and $p$ means $p+\overline{p}$)
based on the HSD transport model.
These results will be copmared to statistical model results in
different ensembles~\cite{GoHaKoBr09,KoHaGoBr09}
and available experimental data.

\subsection{Notations and Approximations}

The fluctuations of the ratio $R_{AB}\equiv N_A/N_B$ will be
characterized by \cite{BaHe99,JeKo99}
\eq{\label{sigma-def}
\sigma^2~\equiv~\frac{\langle \left(\Delta R_{AB}\right)^2\rangle}
{\langle R_{AB}\rangle^2 } ~.
}
Using the expansion,
\eq{
\frac{N_A}{N_B}~=~&\frac{\langle N_A\rangle +\Delta N_A} {\langle
N_B \rangle +\Delta N_B}~ = ~ \frac{\langle N_A\rangle +\Delta
N_A} {\langle N_B \rangle }~ \times ~~ \nonumber
\\ &\left[1~-~\frac{\Delta N_B}{
\langle N_B \rangle } ~+~ \left(\frac{\Delta N_B}{ \langle N_B
\rangle}\right)^2~ -~\cdots~\right]~,\label{expand}
}
one finds to second order in $\Delta N_A/\langle N_A \rangle$ and
$\Delta N_B/\langle N_B \rangle$
the average value and fluctuations of the $A$ to $B$ ratio:
\eq{\label{NANBav}
\langle R_{AB}\rangle~& \cong ~ \frac{\langle N_A\rangle}{\langle
N_B \rangle}~\left[1~+~ \frac{\omega_{B}}{\langle N_B
\rangle}~-~\frac{
\Delta\left(N_A,N_B\right)}{\langle N_A\rangle \langle
N_B\rangle}\right]~,
\\
\label{sigma}
 \sigma^2~&
\cong~\frac{\Delta \left(N_A,N_A\right)}{\langle
N_A\rangle^2}~+~\frac{\Delta \left(N_B,N_B\right)}{\langle
N_B\rangle^2}~-~2~\frac{\Delta \left(N_A,N_B\right)}{\langle
N_A\rangle\langle N_B\rangle}~
\nonumber \\
 &=~\frac{\omega_A}{\langle N_A\rangle}~+~\frac{\omega_B }{\langle
N_B\rangle}~-~2\rho_{AB}~\left[\frac{\omega_A \omega_B}{\langle
N_A\rangle\langle N_B\rangle}\right]^{1/2}~.
}

If species $A$ and $B$ fluctuate independently according to the
Poisson distributions (this takes place, for example, in the GCE
for an ideal Boltzmann gas) one finds, $\omega_A=\omega_B=1$ and
$\rho_{AB}=0$. The Eq.~(\ref{sigma}) then reads
\eq{\label{Boltz}
\sigma^2~=~\frac{1}{\langle N_A\rangle}~+~\frac{1}{\langle
N_B\rangle}~.
}
In a thermal gas, the average multiplicities are proportional to
the system volume $V$. The Eq.~(\ref{Boltz}) demonstrates then a
simple dependence $\sigma^2\propto 1/V$ on the system volume.

A few examples concerning to Eq.~(\ref{sigma}) are appropriate
here. When $\langle N_B\rangle \gg \langle N_A\rangle$, e.g.,
$A=K^++K^-$ and $B=\pi^++\pi^-$, the $\sigma^2$ (\ref{sigma}) is
dominated by the less abundant particles and the resonances
decaying into it. When $\langle N_A\rangle \cong \langle
N_B\rangle$, e.g., $A=\pi^+$ and $B=\pi^-$, the correlation term
in Eq.~(\ref{sigma}) may become especially important. A resonance
decaying always into a $\pi^+\pi^-$-pair does not contribute to
$\sigma^2$ (\ref{sigma}), but contributes to $\pi^+$ and $\pi^-$
average multiplicities. This leads \cite{JeKo99} to a suppression
of $\sigma^2$ (\ref{sigma}) in comparison to its value given by
Eq.~(\ref{Boltz}). For example, if all $\pi^+$ and $\pi^-$
particles come by pairs from decay of resonances, one finds the
correlation coefficient $\rho_{\pi^+\pi^-}=1$ in
Eq.~(\ref{sigma}), and thus $\sigma^2=0$. In this case, the
numbers of $\pi^+$ and $\pi^-$ fluctuate as the number of
resonances, but the ratio $\pi^+/\pi^-$ does not fluctuate.

\subsection{Mixed Events Procedure}
The experimental data for
$N_A/N_B$ fluctuations are usually presented in terms of the so called
dynamical fluctuations \cite{VoKoRi99}\footnote{Other dynamical measures, $\Phi$ \cite{GaMr92,Mr98}
and $F$ 
\cite{JeKo99}, can be also used. See the discussion of this point
in the recent paper \cite{KoSc09}.}
\eq{\label{sigmadyn}
\sigma_{dyn}~\equiv~\texttt{sign}\left(\sigma^2~-~\sigma^2_{mix}\right)\left|\sigma^2~-~\sigma^2_{mix}\right|^{1/2}~,
}
where $\sigma^2$ is defined by Eq.~(\ref{sigma}), and
$\sigma^2_{mix}$ corresponds to the following {\it mixed events}
procedure\footnote{ The idealized mixed events procedure
appropriate for model analysis is described here. The real
experimental mixed events procedure is more complicated and
includes experimental uncertainties, such as particle
identification etc.}. One takes a large number of nucleus-nucleus
collision events, and measures the numbers of $N_A$ and $N_B$ in
each event. Then all $A$ and $B$ particles from all events are
combined into one {\it set}. A construction of {\it mixed events}
is done like the following: One fixes a random number $N=N_A+N_B$
according to the experimental probability distribution $P(N)$,
takes randomly $N$ particles ($A$ and/or $B$) from the {\it whole
set}, fixes the values of $N_A$ and $N_B$, and returns these $N$
particles into the {\it set}. This is the mixed event number one.
Then one constructs event number 2, number 3, etc.

Note that the number of events is much larger than the number
of hadrons, $N$, in any single event. Therefore, the probabilities
$p_{A}$ 
and $p_B 
= 1-p_{A}$ to take the
$A$ and $B$ species from the whole {\it set} can be considered as
constant values during the event construction.
Another consequence of a large number of events is the fact that
all $A$ and $B$ particles in any constructed {\it mixed event}
most probably belong to different {\it physical events} of nucleus-nucleus
collisions. Therefore, the correlations between $N_B$ and $N_A$
numbers in a physical event are expected to be destroyed in a
mixed event. This is the main purpose of the mixed events
construction.
For any function $f$ of $N_A$ and $N_B$ the mixed events averaging is
then defined as,

\eq{
\langle f(N_A,N_B) \rangle_{mix} ~=~& \sum_N P(N)\sum_{N_A,N_B}f(N_A,N_B)~\times~ \nonumber
\\ & \delta(N-N_A-N_B)~\frac{(N_A+N_B)!}{N_A! N_B!} p_{A}^{N_A} p_B^{N_B}~.
\label{mix1}
}
The straightforward calculation of mixed averages (\ref{mix1})
can be simplified by introducing the generating function $Z(x,y)$,
\eq{\label{mix2}\fl\hskip3em
 Z(x,y) ~\equiv~ \sum_N P(N) \sum_{N_A,N_B} \delta(N-N_A-N_B)
 \frac{(N_A +N_B)!}{N_A!~ N_B!}~ (xp_{A})^{N_A}~
 (yp_B)^{N_B}~
 \nonumber \\
 =~ \sum_N P(N) \left(xp_{A}+yp_{B}\right)^N~.
}
 The averages (\ref{mix1}) are then expressed as $x$- and
$y$-derivatives of $Z(x,y)$ at $x=y=1$.
One finds:
\eq{
   &\langle N_A \rangle_{mix} ~ = ~
       \left(\frac{\partial Z}{\partial x}\right)_{x=y=1}~ =~
       p_{A} ~\langle N \rangle ~,~ \nonumber
\\ &\langle N_B \rangle_{mix} ~ = ~
      \left(\frac{\partial Z}{\partial y}\right)_{x=y=1}~ = ~
       p_{B} ~ \langle N \rangle ~, \label{mix-av}
\\ &\langle N_A(N_A-1) \rangle_{mix} ~ =~
 \left(\frac{\partial^2 Z}{\partial^2 x}\right)_{x=y=1} ~ =~
 p_{A}^2 ~\langle N (N-1)\rangle~,\label{mix-flA}\\~~~
& \langle N_B(N_B-1) \rangle_{mix} ~ =~
 \left(\frac{\partial^2 Z}{\partial^2 y}\right)_{x=y=1} ~ =~
 p_{B}^2 ~\langle N (N-1)\rangle~, \label{mix-flB}\\
&\langle N_AN_B \rangle_{mix} ~ - ~\langle N_A\rangle_{mix}
\langle N_B\rangle_{mix}~ = ~ \left(\frac{\partial^2 Z} {\partial
x\partial y}\right)_{x=y=1}
~ \nonumber
\\ &~=~ p_A p_B~\omega_N~
\langle N \rangle~.\label{mix-cor}
 }

Calculating the $N_A/N_B$ fluctuations for mixed events according
to (\ref{sigma}) one
gets:
\eq{ & \sigma^2_{mix}~ \equiv~\frac{\Delta_{mix}
\left(N_A,N_A\right)}{\langle N_A\rangle^2}~+~\frac{\Delta_{mix}
\left(N_B,N_B\right)}{\langle N_B\rangle^2}~-~2~\frac{\Delta_{mix}
\left(N_A,N_B\right)}{\langle N_A\rangle\langle
N_B\rangle}~\nonumber \\
&=~\left[\frac{1}{\langle N_A \rangle }~+~ \frac{\omega_N-
1}{\langle N \rangle}\right] ~+~\left[\frac{1}{\langle N_B \rangle}~+~
\frac{\omega_N-1}{\langle N \rangle}\right]~-~
2~\frac{\omega_N-1}{\langle N \rangle}~\nonumber \\
&=~ \frac{1}{\langle N_A \rangle }~+~\frac{1}{\langle N_B \rangle
} ~. \label{Dmix}
}
A comparison of the final result in Eq.~(\ref{Dmix}) with
Eq.~(\ref{Boltz}) shows that the mixed event procedure gives the
same $\sigma^2$ for $N_A/N_B$ fluctuations as in the GCE
formulation for an ideal Boltzmann gas, i.e. $\omega_A=\omega_B=1$
and $\rho_{AB}=0$. If $\omega_N=1$
(e.g., for the Poisson distribution
$P(N)$), one indeed finds $\omega_A^{mix}=\omega_B^{mix}=1$ and
$\rho_{AB}^{mix}=0$. Otherwise,
if
$\omega_N\neq 1$, the mixed events procedure leads to
$\omega_A^{mix}\neq 1$, $\omega_B^{mix}\neq 1$, and to non-zero
$N_AN_B$ correlations, as seen from the second line of
(\ref{Dmix}). However, the final result for $\sigma^2_{mix}$
(\ref{Dmix}) is still the same simple. It does not depend on the
specific form of $P(N)$. Non-trivial ($\omega_{A,B}^{mix}\neq 1$)
fluctuations of $N_A$ and $N_B$ as well as non-zero
$\rho_{AB}^{mix}$ correlations may exist in the mixed events
procedure, but they are cancelled out exactly in $\sigma_{mix}^2$.

\subsection{Hadron Scaled Variances and Correlation Coefficients}

\begin{figure}[ht!]
\begin{center}
\epsfig{file=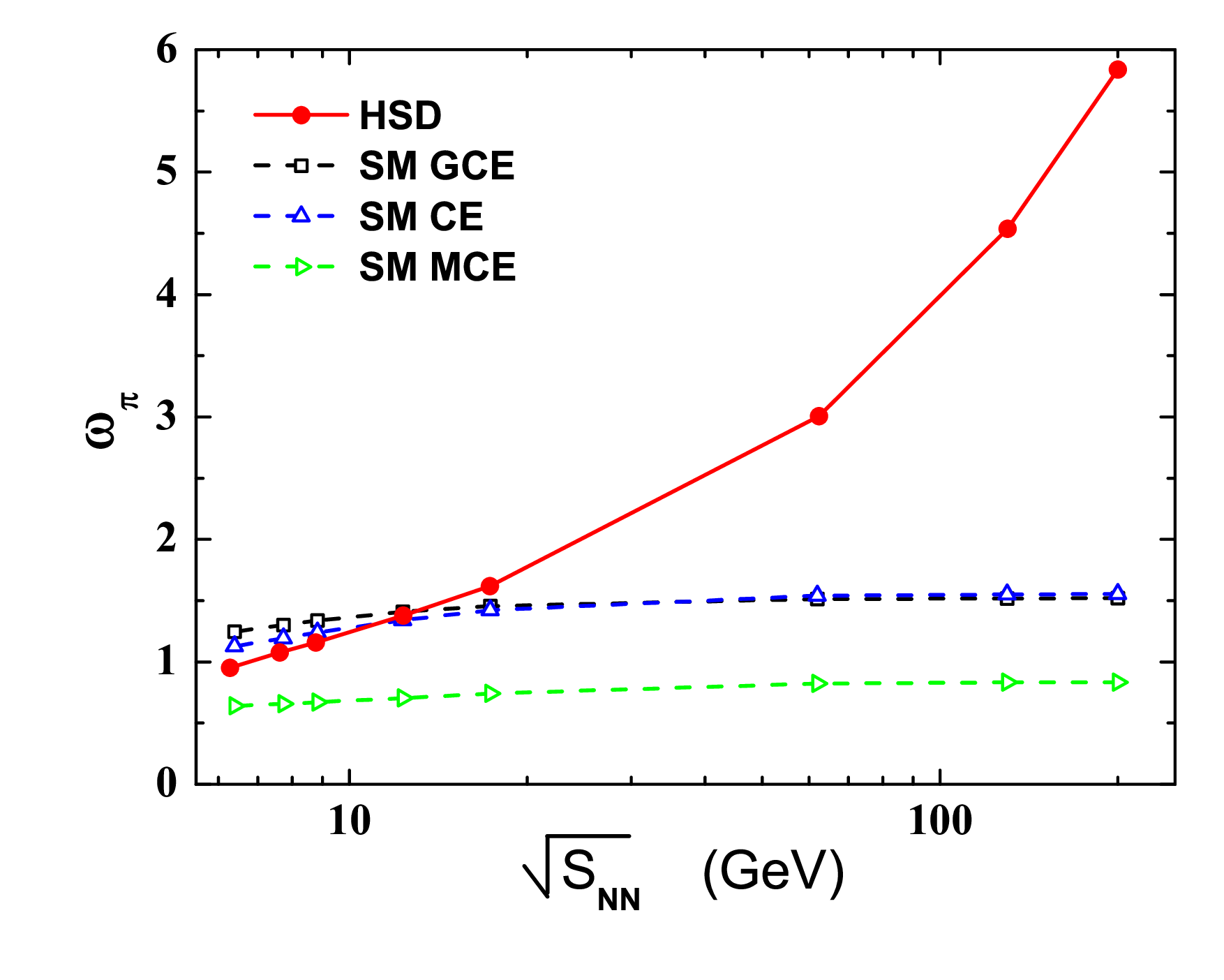,width=8.5cm}
\epsfig{file=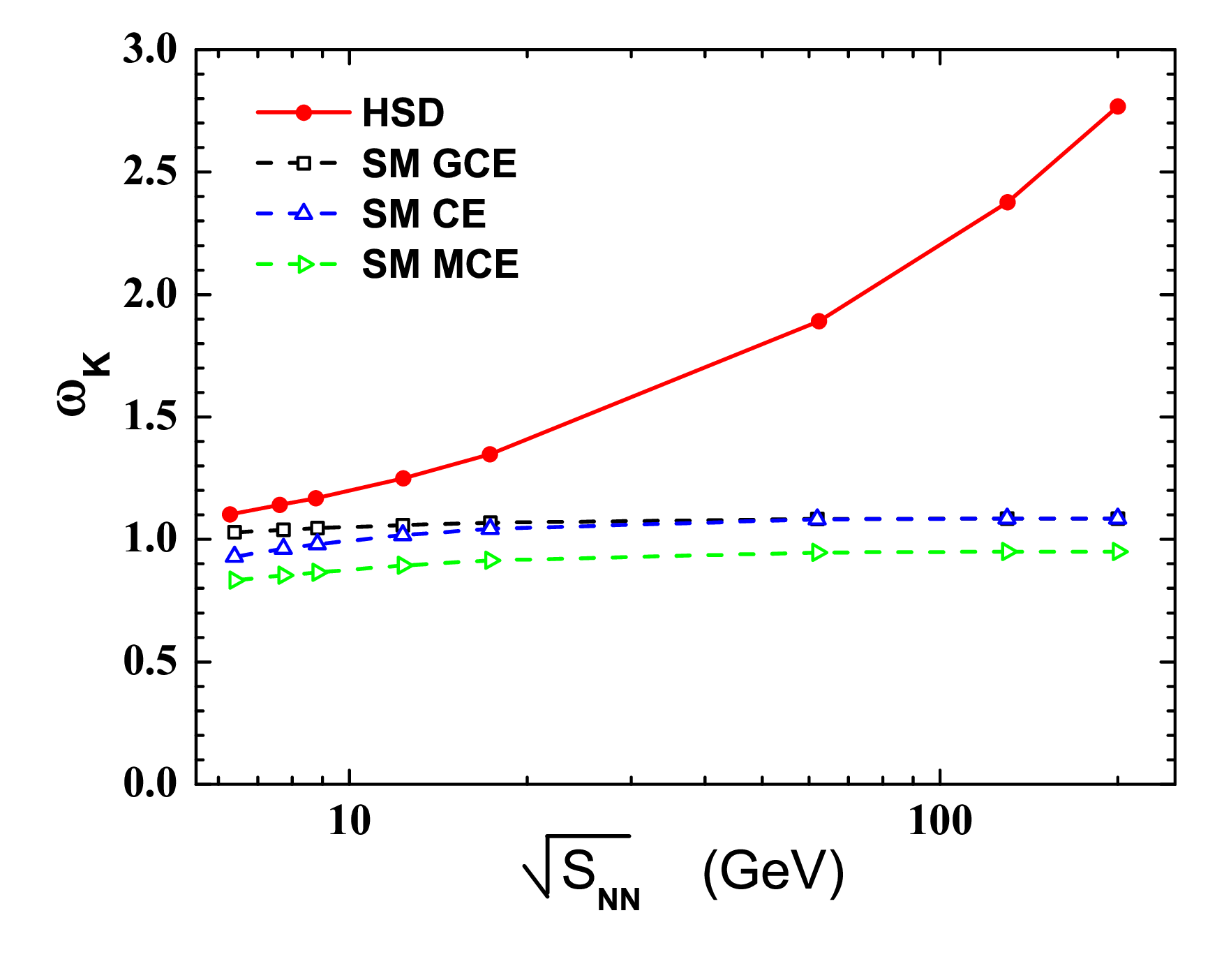,width=8.5cm}
\epsfig{file=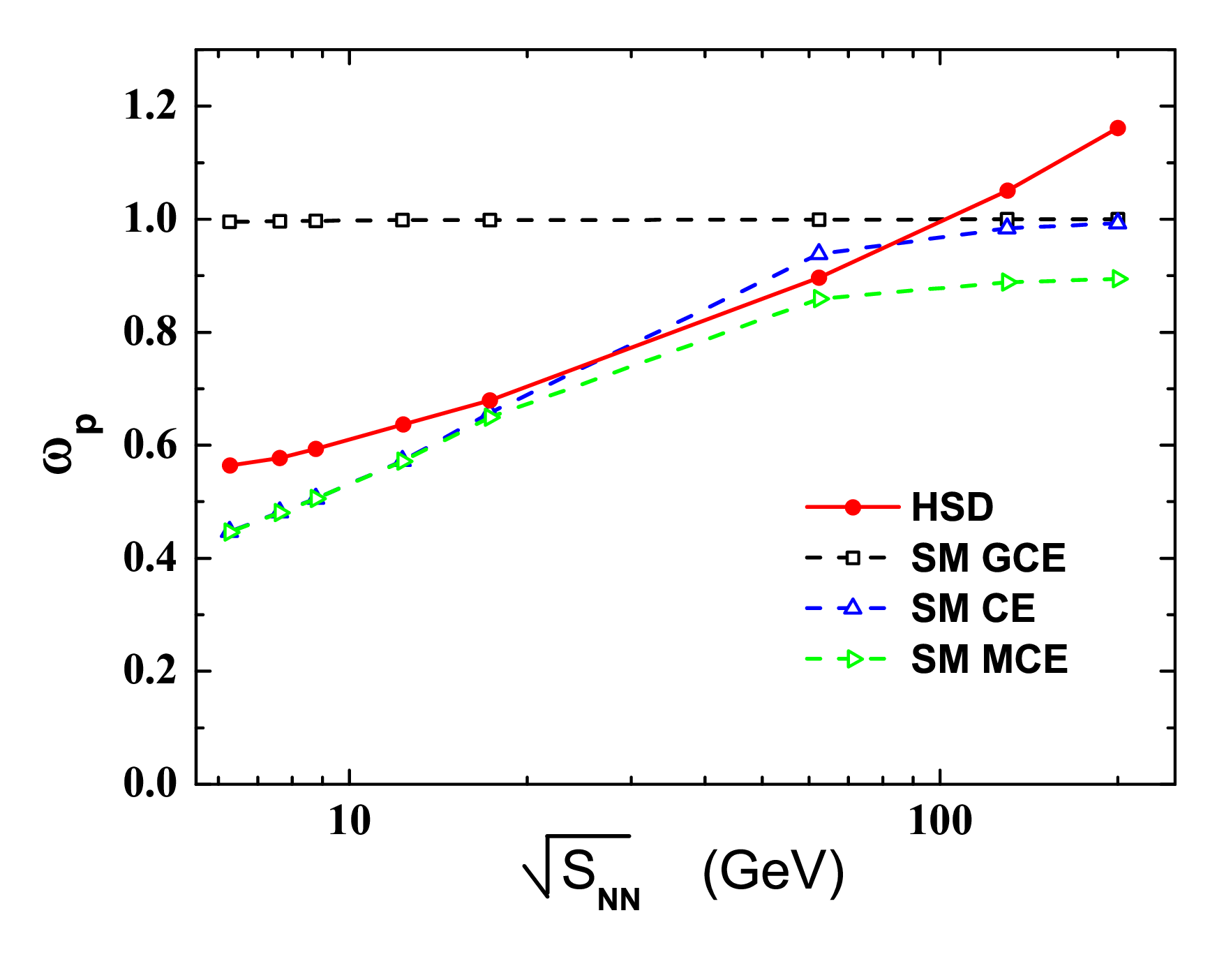,width=8.5cm}
\end{center}
\caption{The SM results  in the GCE, CE, and MCE ensembles and the
HSD results (impact parameter $b=0$) are presented for the scaled
variances $\omega_{\pi}$, $\omega_K$, $\omega_p$ for Pb+Pb (Au+Au)
collisions at different c.m. energies $\sqrt{s_{NN}}$.}
\label{fig-omegaAB}
\end{figure}

\begin{figure}[ht!]
\begin{center}
\epsfig{file=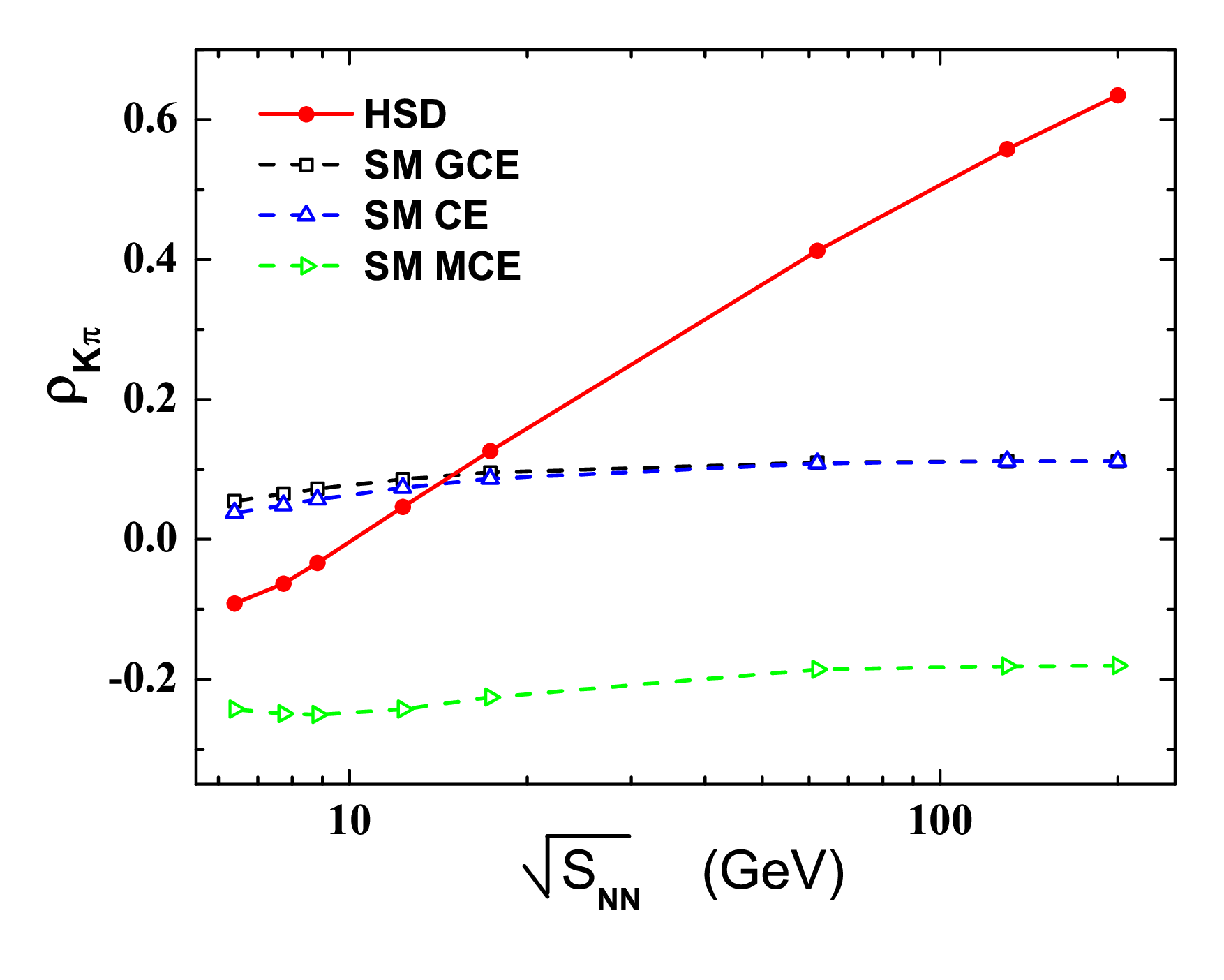,width=8.55cm}
\epsfig{file=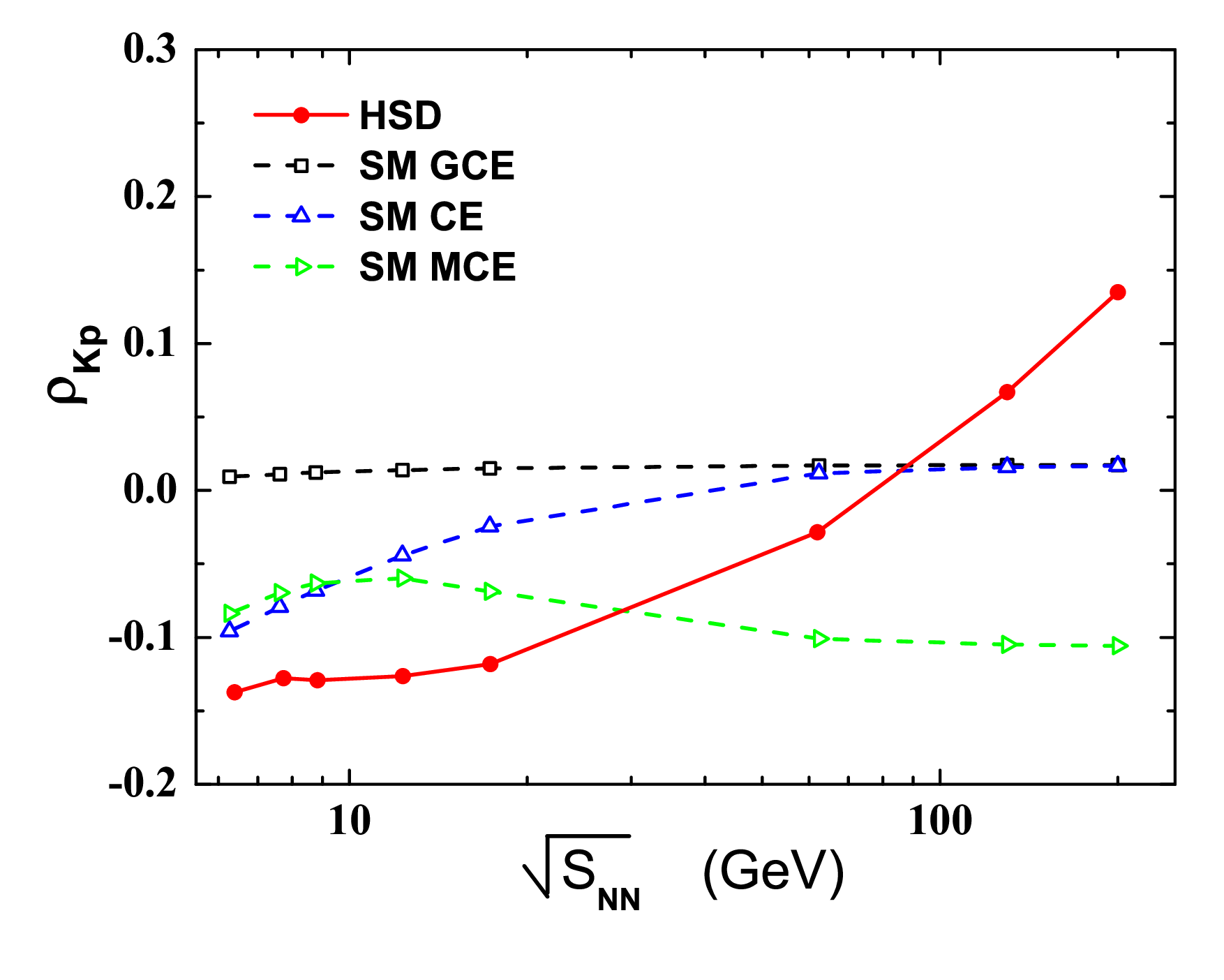,width=8.5cm}
\epsfig{file=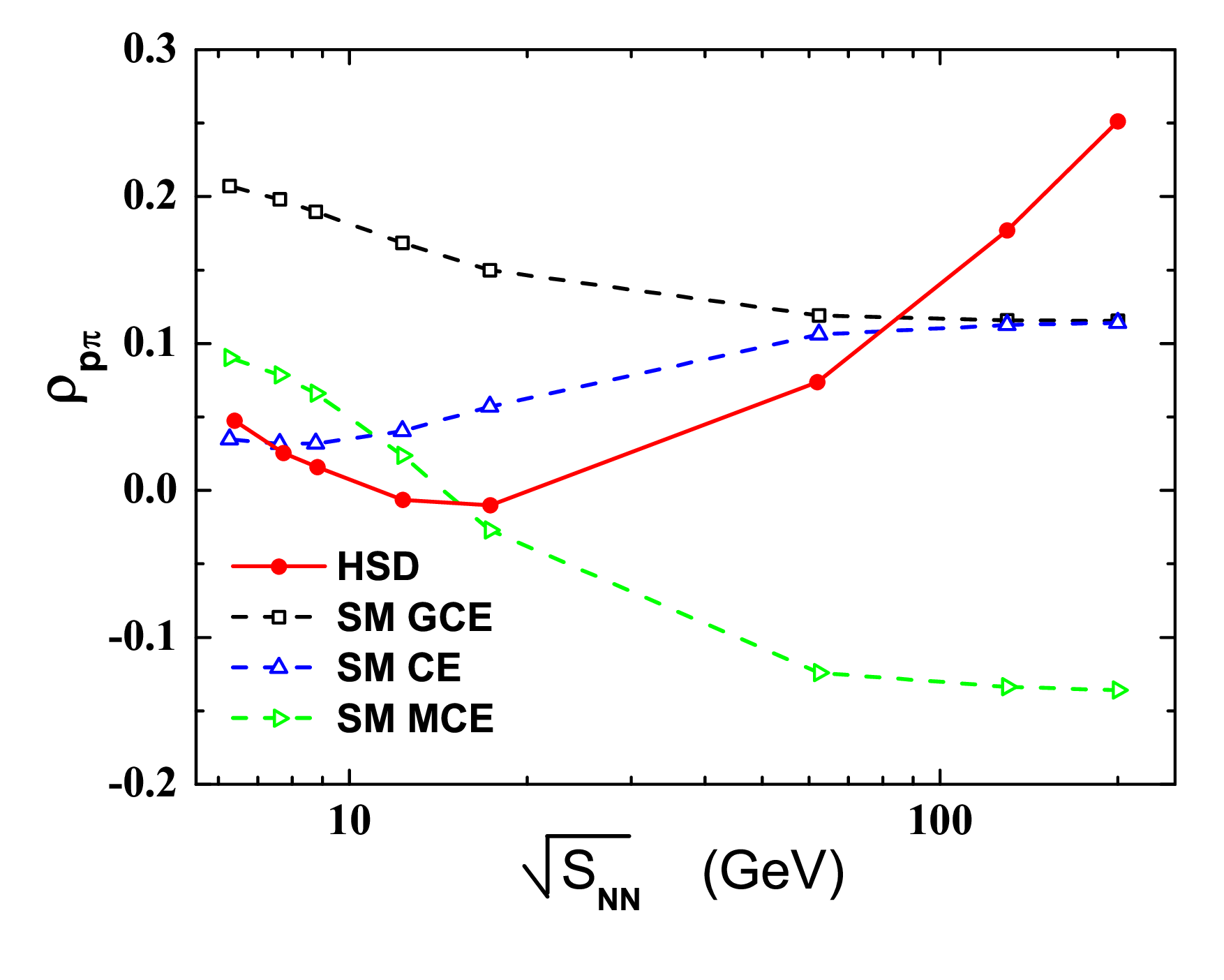,width=8.5cm}
\end{center}
\caption{ The SM results  in the GCE, CE, and MCE ensembles and
the HSD results (impact parameter $b=0$) are presented for the
correlation parameters $\rho_{K\pi}$, $\rho_{Kp}$, $\rho_{p\pi}$
for Pb+Pb (Au+Au) collisions at different c.m. energies
$\sqrt{s_{NN}}$} \label{fig-rhoAB}
\end{figure}

The values of $\omega_{\pi}$, $\omega_K$, $\omega_p$ and $\rho_{K\pi}$, $\rho_{K\pi}$, $\rho_{K\pi}$
for the HSD simulations of and statistical model (SM)
in Pb+Pb (Au+Au) central
collisions are presented in Figs.~\ref{fig-omegaAB} and \ref{fig-rhoAB}.

Let us first comment the SM (see more details in Refs.~\cite{GoHaKoBr09,KoHaGoBr09}).
In the SM the scaled variances $\omega_{A}$
and correlation parameter $\rho_{AB}$ approach finite values
in the thermodynamic limit of large volumes.
For central Pb+Pb and Au+Au collisions the corresponding volumes in the SM are
large enough.
Finite volume corrections are expected on the level of a few percent.
We discuss $K/\pi$ fluctuations in more details.
The $\pi$-$K$ correlations
$\rho_{K\pi}$ are due to resonances having simultaneously $K$ and $\pi$
mesons in their decay products.
In the hadron-resonance gas within the GCE ensemble, these
quantum statistics and resonance decay effects are responsible for
deviations of $\omega_K$ and $\omega_{\pi}$ from 1, and of
$\rho_{K\pi}$ from 0.
The most important effect of an exact charge
conservation in the CE ensemble is a suppression of the kaon number fluctuation.
This happens mainly due to exact strangeness conservation and is
reflected in smaller CE values of $\omega_K$ at low collision energies
in comparison to those from the GCE ensemble.
The MCE values of $\omega_K$ and
$\omega_{\pi}$ are further suppressed in comparison those from the CE ensemble.
This is due to exact energy conservation.
The effect is stronger for
pions than for kaons since pions carry a larger part of the total energy.
An important feature of the MCE is the anticorrelation between $N_{\pi}$
and $N_K$, i.e. negative values of $\rho_{K\pi}$.
This is also a
consequence of energy conservation for each microscopic state of the
system in the MCE \cite{Be07}.
The presented results demonstrate that global
conservation laws are rather important for the values of
$\omega_{\pi}$, $\omega_K$, and $\rho_{K\pi}$.
In particular, the
exact energy conservation strongly suppresses the fluctuations in the pion
and kaon numbers and leads to $\omega_K<1$ and $\omega_{\pi}<1$ in the
MCE ensemble, instead of $\omega_K>1$ and $\omega_{\pi}>1$ in the GCE and CE
ensembles.
The exact energy conservation changes also the $\pi$-$K$ correlation
into an anticorrelation: instead of $\rho_{K\pi}>0$ in the GCE and CE ensembles one
finds $\rho_{K\pi}<0$ in the MCE.
The effects of global conservation laws and resonance decays
are also seen for $\rho_{Kp}$, $\rho_{p\pi}$ and $\omega_p$.

The HSD results for $\omega_A$ and $\rho_{AB}$ shown by the solid lines
in Figs.~\ref{fig-omegaAB} and \ref{fig-rhoAB} are close to the CE and
MCE results for low SPS energies.  One may conclude  that  the
influence of conservation laws is more stringent at low collision
energies.  The HSD values for $\omega_A$ and $\rho_{AB}$ increase,
however, at high collision energies and a sizeable deviation of the HSD
results from those in the MCE SM is observed with increasing energies.

\begin{figure}[ht!]
\begin{center}
\epsfig{file=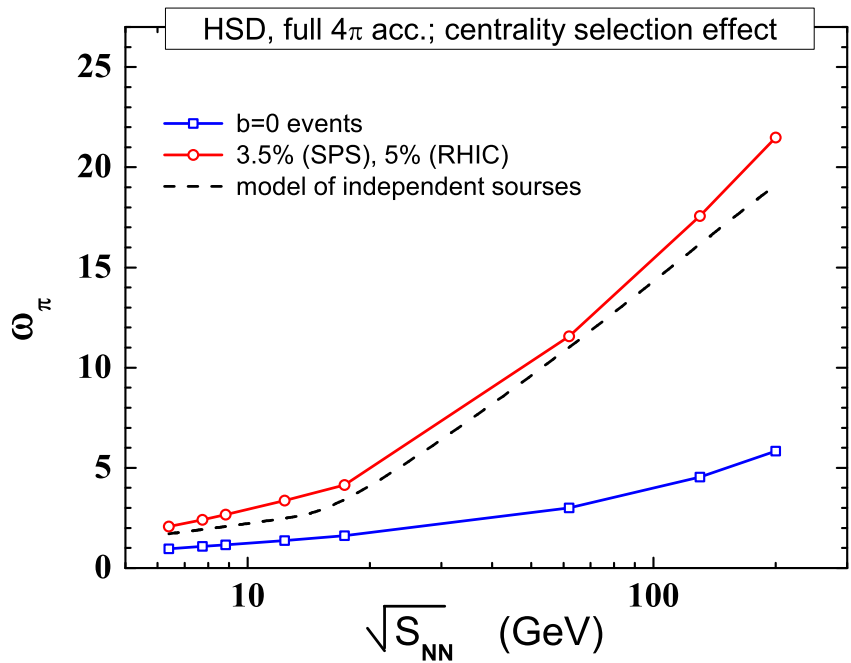,width=12cm}
\end{center}
\caption{(Color online)
The HSD results for $\omega_\pi$ for Pb+Pb (Au+Au) collisions
at different c.m. energies $\sqrt{s_{NN}}$ within the full $4\pi$-acceptance.
The lower solid line corresponds to zero impact parameter ($b=0$) and
the upper one to the experimentally selected samples of collision events.
The dashed line reflects the model of independent sources (\ref{omegai}).}
\label{fig:ratio_fig4}
\end{figure}

In Fig.~\ref{fig:ratio_fig4} the HSD scaled variance $\omega_\pi$ for the
full acceptance is shown in A+A collisions for $b=0$.
The samples of collision events selected experimentally are 3.5\% of most central  collision
events in Pb+Pb collisions at the SPS energies and 5\% in Au+Au collisions at RHIC energies.

Fig.~\ref{fig:ratio_fig4} presents the HSD results for $\omega_\pi$ in
these samples the comparison with the HSD results at $b=0$.  One finds
much larger values of $\omega_\pi$ in the centrality selected samples
than for $b=0$.  The effect is especially strong at RHIC energies.
Collisions with $b=0$ correspond to $\omega_P\cong 0$.  Thus,
$\omega_{\pi}^*$ in Eq.~(\ref{omegai}) can be approximately taken as
$\omega_{\pi}$ at $b=0$.  The HSD results correspond approximately to
$\omega_P\cong 0.5$ for the 3.5\% most central Pb+Pb collisions at SPS
energies and $\omega_P\cong 1$ for the 5\% most central Au+Au
collisions at RHIC energies.  The results of the MIS (\ref{omegai}) for
$\omega_\pi$ are  shown by the dashed line in Fig.~\ref{fig:ratio_fig4}
and they are close to the actual values of the HSD simulations for
$\omega_\pi$.

The fluctuation in the kaon to pion ratio is dominated by the
fluctuations of kaons alone since the average multiplicity of kaons is
about 10 times smaller than that of pions.
The model calculations of $\sigma (K\pi)$ require, in addition to
$\omega_K$, $\omega_{\pi}$, and $\rho_{K\pi}$ values, the
knowledge of the average multiplicities $\langle N_K\rangle$ and
$\langle N_{\pi}\rangle$.
To fix the average multiplicities in the SM one needs to choose
the system volume. For each collision energy the volume of the
statistical system has been fixed in a way to obtain the same kaon
average multiplicity in the SM as in the HSD calculations:
$\langle N_K\rangle_{stat}= \langle N_K\rangle_{HSD}$. Recall that
average multiplicities of kaons and pions are the same in all
statistical ensembles. The SM volume in central Pb+Pb (Au+Au)
collisions is large enough and all statistical ensembles are
thermodynamically equivalent for the average pion and kaon
multiplicities, since these multiplicities are much larger than 1.

\begin{figure}[ht!]
\begin{center}
\epsfig{file=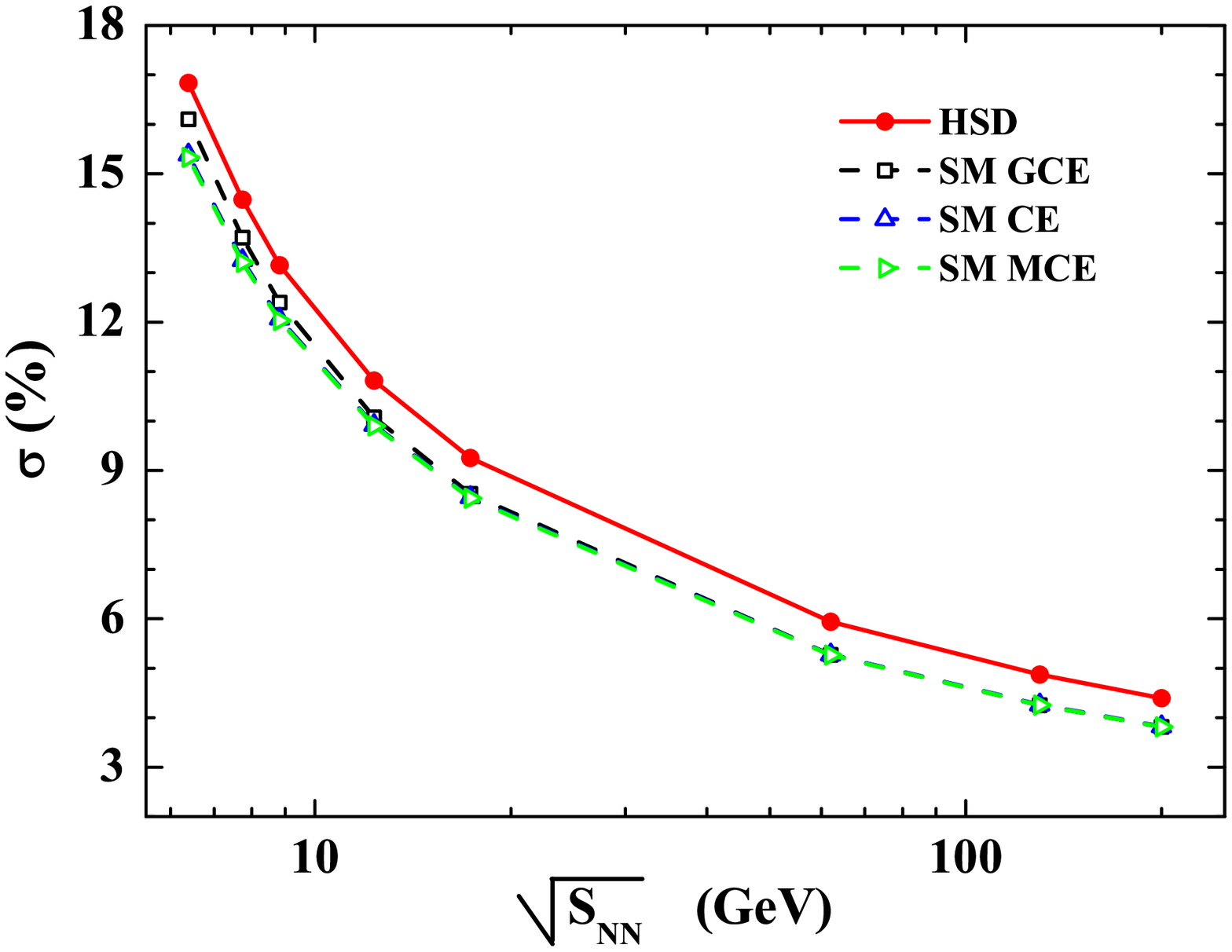,width=0.45\textwidth}
\epsfig{file=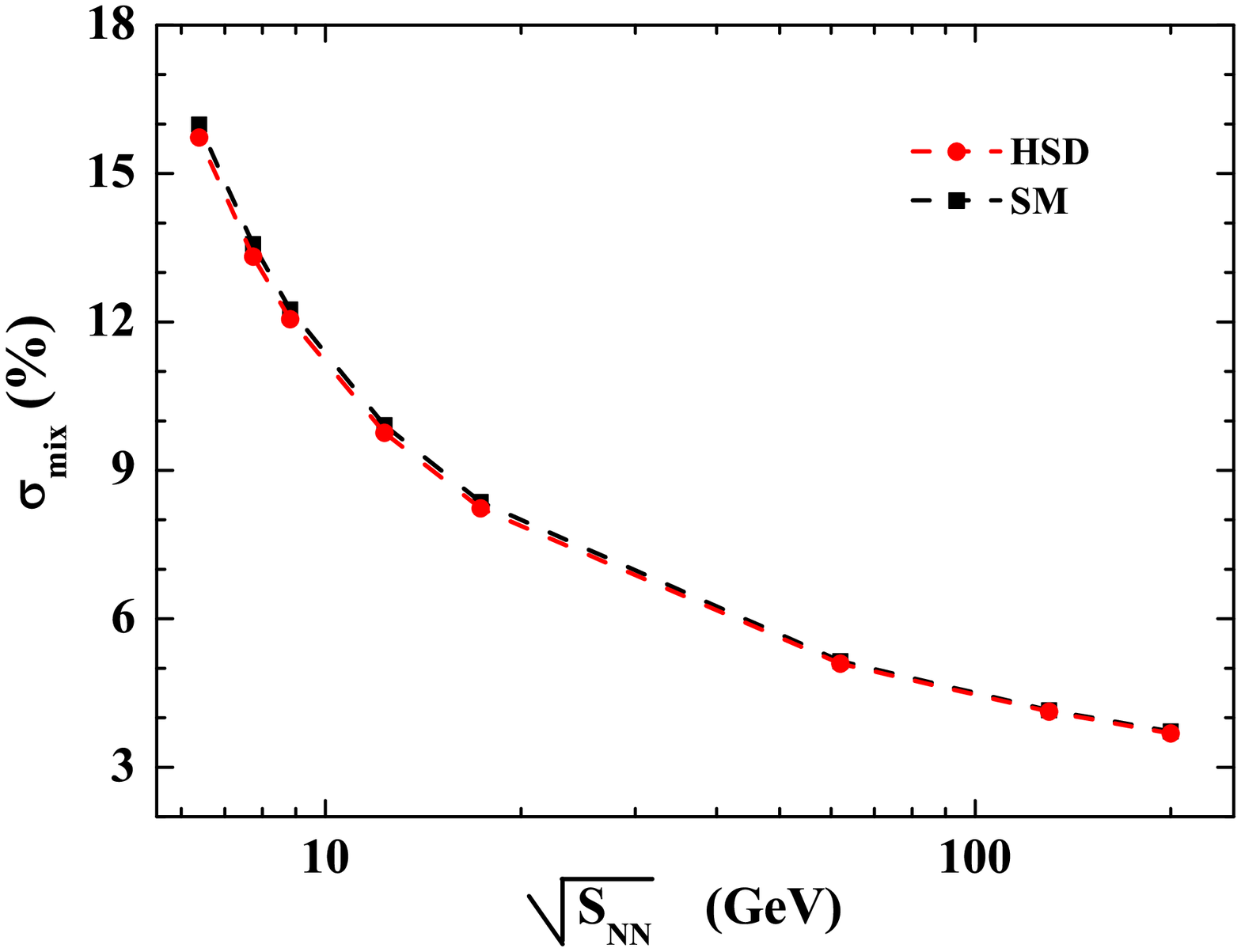,width=0.45\textwidth}
\end{center}
\caption{{\it Left:} The SM results in the GCE, CE, and MCE
ensembles as well as the HSD results (impact parameter $b=0$) are
presented for $\sigma \cdot$100\%
for $K/\pi$ ratio in Pb+Pb (Au+Au) collisions  at different c.m.
energies $\sqrt{s_{NN}}$. {\it Right:} The same as in the {\it
left} panel, but for $\sigma_{mix} \cdot$100\% in mixed events
defined by (\ref{Dmix})
} \label{fig:fig3}
\end{figure}

In Fig.~\ref{fig:fig3} the values of $\sigma$ and $\sigma_{mix}$
(in percents)
are presented, in the {\it left} and {\it right} panel,
respectively, for the SM in different ensembles as well as for the
HSD simulations. A first conclusion from Fig.~\ref{fig:fig3} ({\it
left}) is that all results for $\sigma$ in the different models
are rather similar. One observes a monotonic decrease of $\sigma$
with collision energy. This is just because of an increase of the
kaon and pion average multiplicities with collision energy. The
mixed event fluctuations $\sigma_{mix}$ in the model analysis are
fully defined by these average multiplicities according to
(\ref{Dmix}). The values of $\sigma_{mix}$ are therefore the same in
the different statistical ensembles. They are also very close to
the HSD values because the statistical system volume has been
defined to obtain the same kaon average multiplicities in the
statistical model as in HSD at each collision energy. As seen from
Fig.~\ref{fig:fig3} ({\it right}) the requirement of $\langle
N_K\rangle_{stat}= \langle N_K\rangle_{HSD}$ leads to almost equal
values of $\sigma_{mix}$ in both HSD and the SM.

\begin{figure}[ht!]
\begin{center}
\epsfig{file=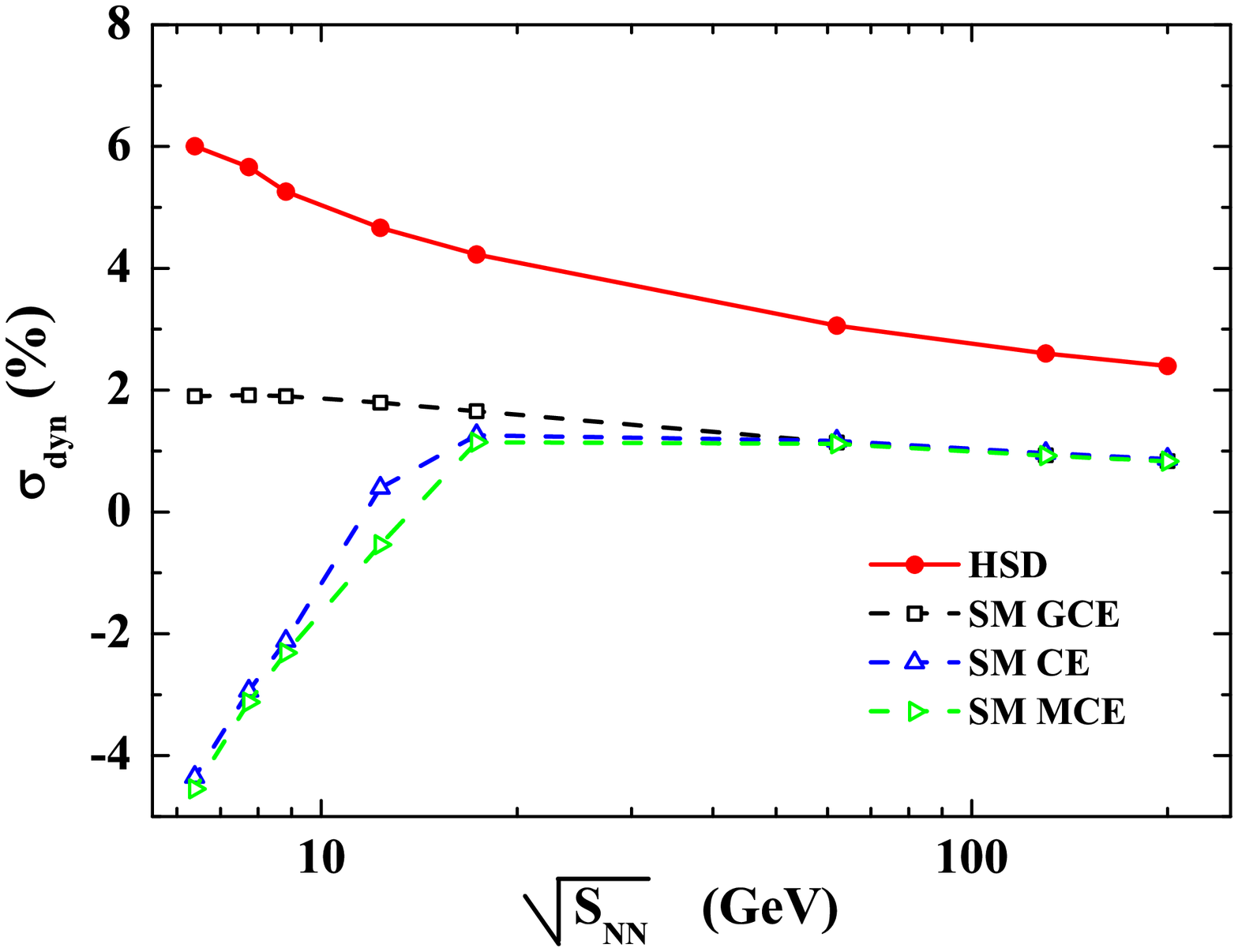,width=0.45\textwidth}
\epsfig{file=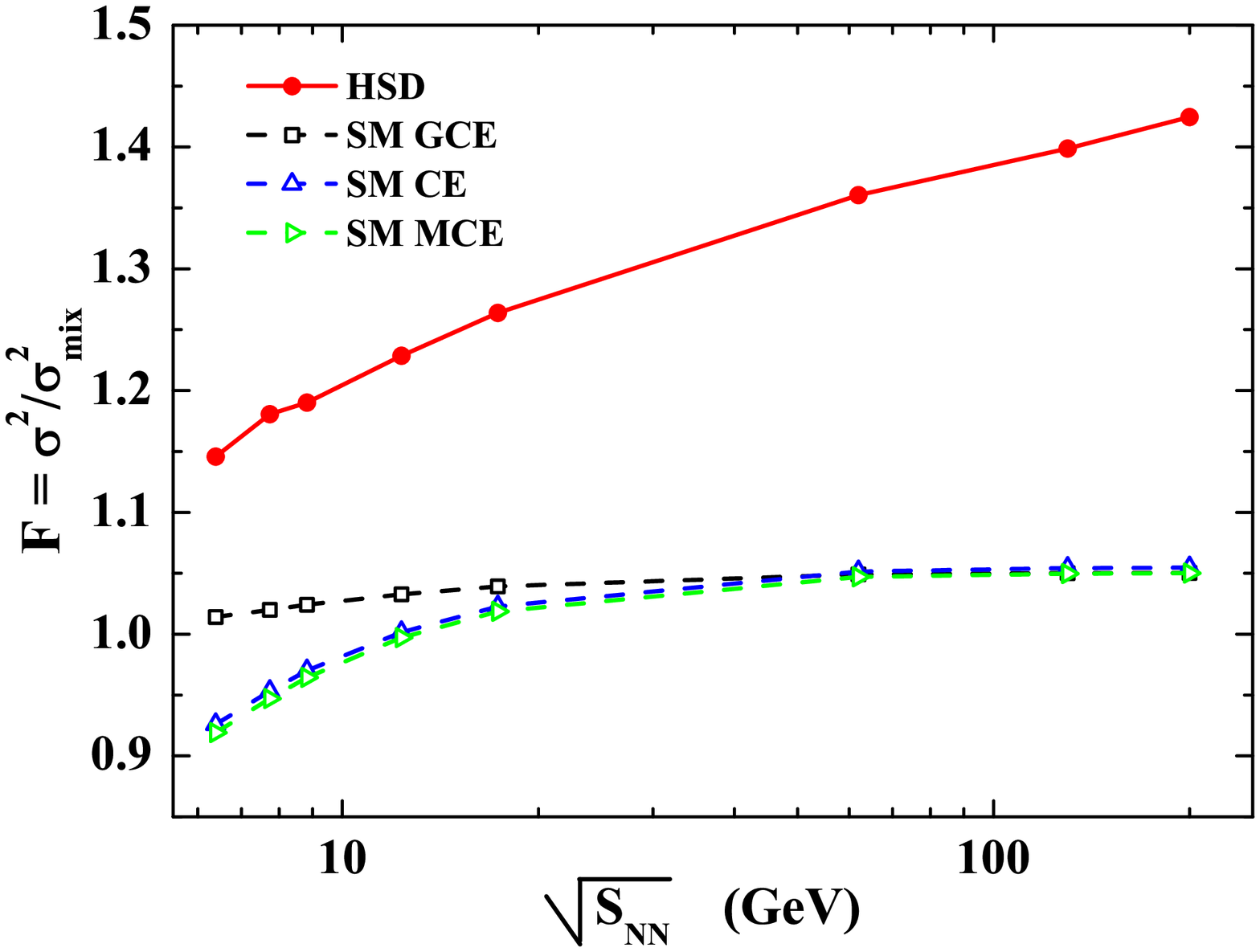,width=0.45\textwidth}
\end{center}
\caption{{\it Left}: The results for the $K/\pi$ fluctuations at
different c.m. energies $\sqrt{s_{NN}}$ in the GCE, CE, and MCE
ensembles as well as from HSD (impact parameter $b=0$) are
presented for $\sigma_{dyn}\cdot$100\% defined by (\ref{sigmadyn}).
{\it Right:} The same as in the {\it left} panel but for
$F=\sigma^2/\sigma_{mix}^2$.} \label{fig:fig4}
\end{figure}

Differences between the statistical ensembles as well as between
the statistical and HSD results become visible for other measures
of $K/\pi$ fluctuations, such as $\sigma_{dyn}$ defined by
(\ref{sigmadyn}) and $F=\sigma^2/\sigma_{mix}^2$. They are shown in
Fig.~\ref{fig:fig4}, {\it left} and {\it right}, respectively. At
small collision energies the CE and MCE results in
Fig.~\ref{fig:fig4} demonstrate negative values of $\sigma_{dyn}$,
respectively $F<1$. When the collision energy increases,
$\sigma_{dyn}$ in the CE and MCE ensembles becomes positive, i.e.
$F>1$. Moreover, the different statistical ensembles approach to
the same values of $\sigma_{dyn}$ and $F$ at high collision
energy. In the statistical model the values of $\sigma$ and
$\sigma_{mix}$ approach to zero at high collision energies due to
an increase of the average multiplicities. The same trivial limit
should be also valid for $\sigma_{dyn}$ in the SM. In contrast,
the measure $F$ shows a non-trivial behavior at high energies: the
SM gives $F\cong 1.05$ in high energy limit. The HSD result for
$F$ demonstrates a monotonic increase with collision energy. An
interesting feature of the SM is the approximately equal values of
$\sigma$ (and, thus, $\sigma_{dyn}$ and $F$) in the CE and MCE
ensembles. From
Figs.~\ref{fig-omegaAB} and \ref{fig-rhoAB} one observes that both
$\omega_K$, $\omega_{\pi}$ and $\rho_{K\pi}$ are rather different
in the CE and MCE. Thus, as discussed above, an exact energy
conservation influences the particle scaled variances and
correlations. These changes are, however, cancelled out in the
fluctuations of the kaon to pion ratio.

It has been mentioned in the literature (see, e.g., \cite{BaHe99})
that the fluctuations of particle number ratio $N_A/N_B$ are
independent of volume fluctuations since both multiplicities $N_A$
and $N+B$ are proportional to the volume. This is not correct. In
fact, the {\it average multiplicities} $\langle N_A\rangle$ and
$\langle N_{B}\rangle$, but not $N_A$ and $N_{B}$, are
proportional to the system volume. Let us consider the problem in
the SM assuming the presence of volume fluctuations with the
distribution function $W(V)$ at fixed values of $T$ and $\mu_B$.
This assumption corresponds approximately to the volume
fluctuations in A+A collisions because of different impact
parameters in each collision event. Under these assumptions the SM
values of $\omega_A$, $\omega_B$, and $\rho_{AB}$
remain the same for any volume $V$ (if only this volume is large
enough).
However, the average hadron multiplicities are proportional to the
volume $V$. Therefore, the SM result for $\sigma^2$ reads,
$\sigma^2= \sigma_0^2 V_0/V,$
where $V_0$ is the average system volume, and $\sigma^2_0$ is
calculated with the average multiplicities corresponding to this
average volume $V_0$. Expanding
$V_0/V=V_0/(V_0+\delta V)$ in a serious of $\delta V/V_0$,
one finds to the second order in $\delta V/V_0$,
\eq{\label{V3}
\sigma^2 ~\cong~ \sigma_0^2~\left[1~+~\frac{\langle \left(\delta V\right)^2\rangle}{V_0^2}\right]
~,
}
where
\eq{\label{V4}
\langle \left(\delta V\right)^2\rangle~=~\int dV~(V~-~V_0)^2~W(V)~
} corresponds to averaging over the volume distribution function
$W(V)$ which describes the volume fluctuations. As clearly seen
from Eq.~(\ref{V3}) the volume fluctuations influence, of course,
the $K/\pi$ particle number fluctuations and make them larger.
Comparing the $K/\pi$ particle number fluctuations in, e.g., 1\%
of most central nucleus-nucleus collisions with those in, e.g.,
10\% one should take into account two effects. First, in the 10\%
sample the average volume $V_0$ is smaller than that in 1\% sample
and, thus, $\sigma_0^2$ in Eq.~(\ref{V3}) is larger. Second, the
volume fluctuations (\ref{V4}) in the 10\% sample is larger, and
this gives an additional contribution to $\sigma^2$ according to
Eq.~(\ref{V3}).

One may also consider volume fluctuations at fixed energy and conserved
charges (see, e.g., Ref.~\cite{BeGaGo08}). In this case the connection
between the average multiplicity and the volume becomes more
complicated. The volume fluctuation within the MCE ensemble can
strongly affect the fluctuations in the particle number ratios. This
possibility will be discussed in more detail in a forthcoming study.

\subsection{Excitation Function for the Fluctuations of Ratios}
\label{sec:sigma_results}

A comparison of the SM results for fluctuations in different ensembles 
with the data looks problematic at present; the same is true for most 
other theoretical models.  This is because of difficulties in 
implementing the experimental acceptance and centrality selection 
which, however, can be taken into account in the transport approach.  
In order to compare the HSD calculations with the measured data, the 
experimental cuts are applied for the simulated set of HSD events.

\begin{figure}[htp]
\begin{center}
\epsfig{file=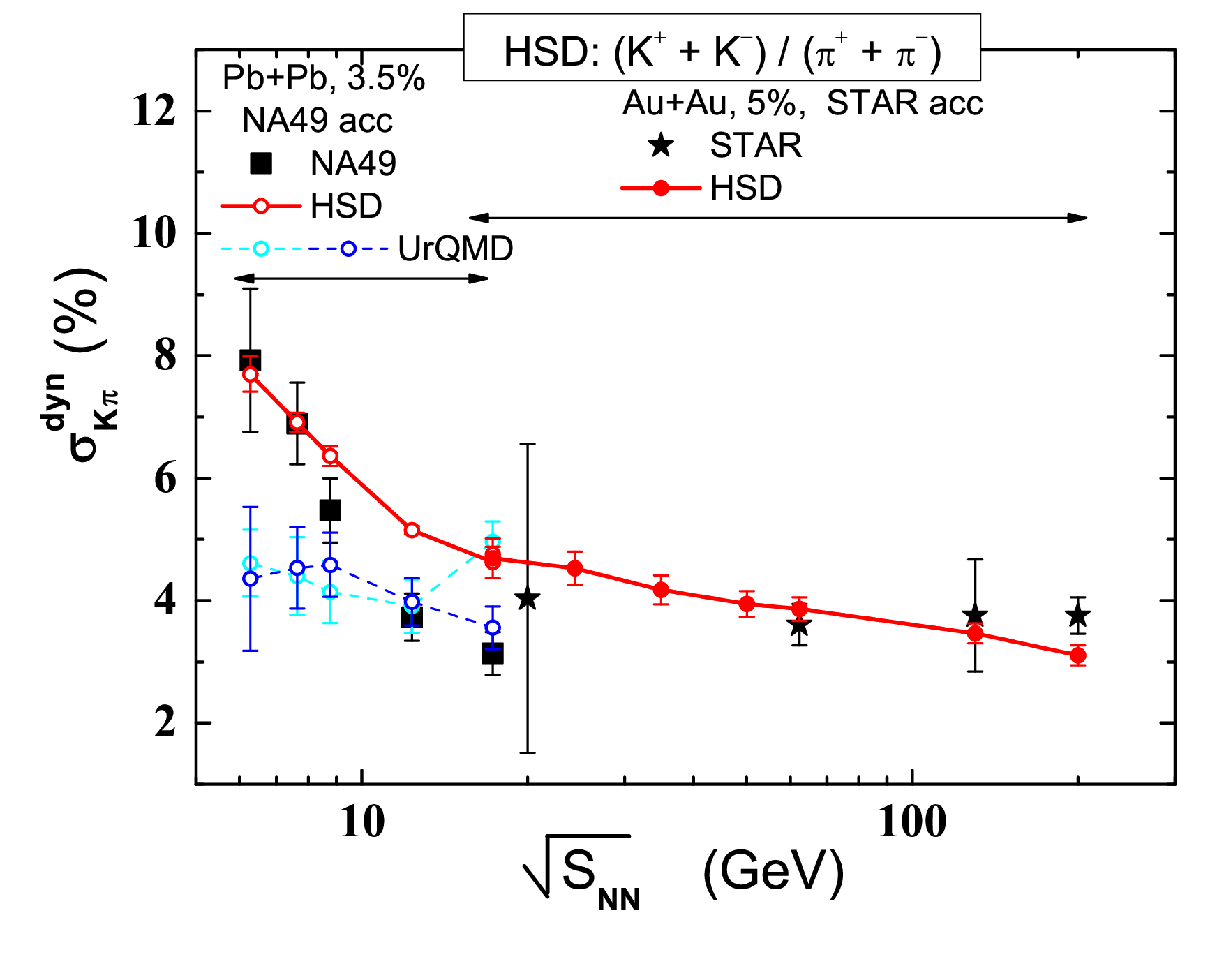,width=8.5cm}
\epsfig{file=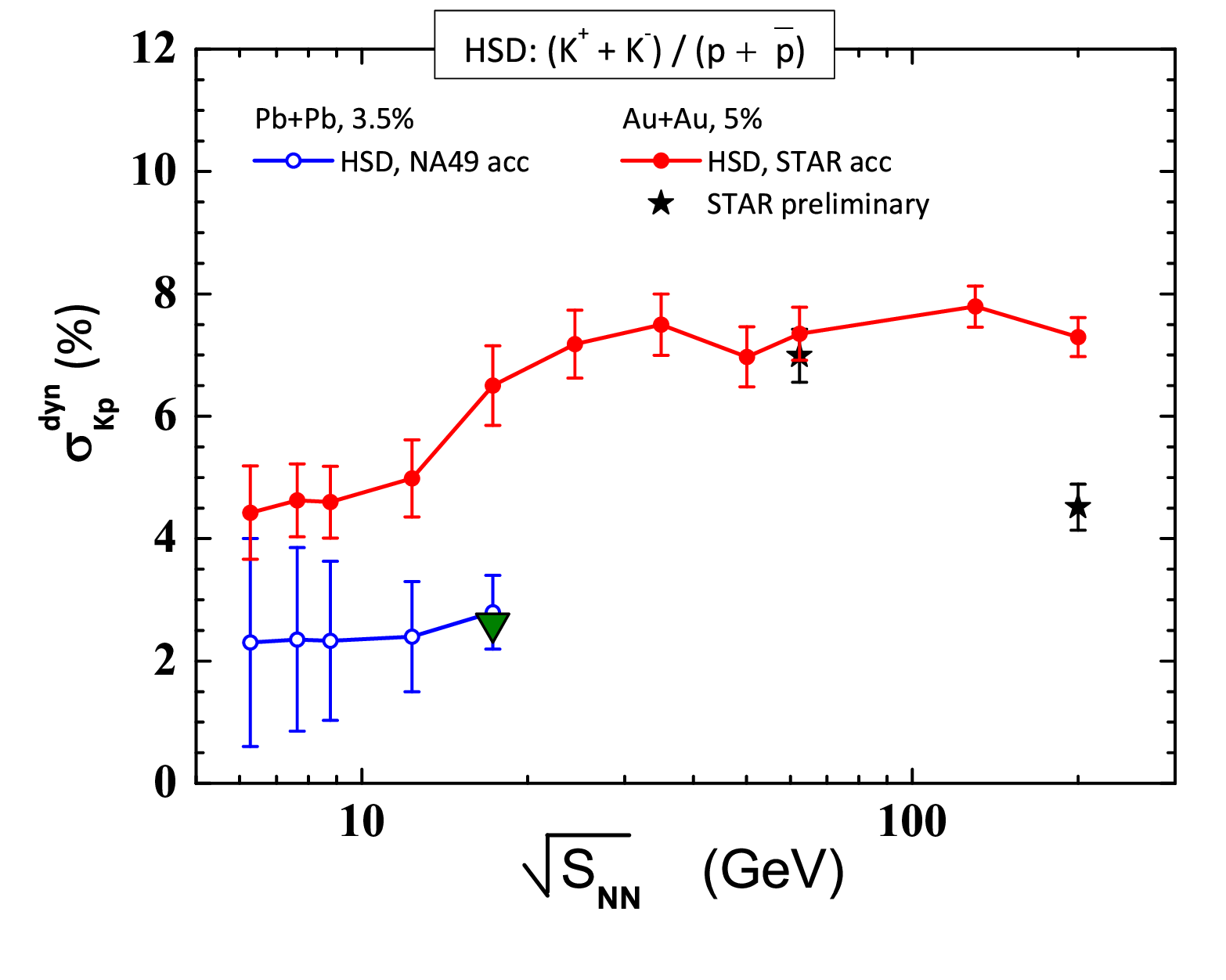,width=8.5cm}
\epsfig{file=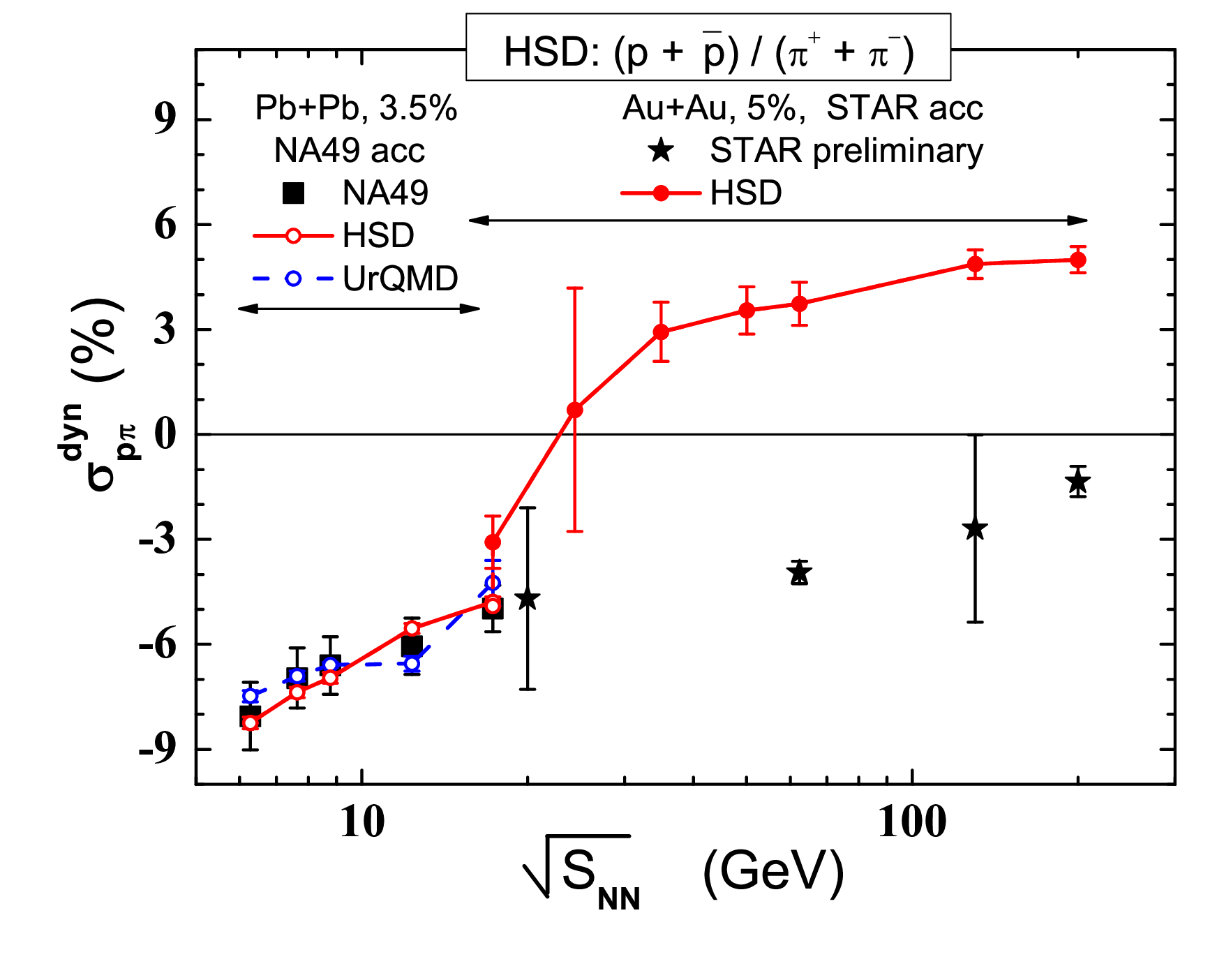,width=8.5cm}
\end{center}
\caption{The HSD results for the excitation function in
$\sigma_{dyn}$ for the $K/\pi$, $K/p$, $p/\pi$ ratios within the
experimental acceptance (solid line) in comparison to the
experimental data measured by the NA49 Collaboration at
SPS~\cite{Al08} and by the STAR Collaboration at
RHIC~\cite{Da06,Ab09,WeQM09,TiQM09}. The UrQMD calculations are
shown by dotted lines. } \label{fig:fig5}
\end{figure}

In Fig.~\ref{fig:fig5} the HSD results of $\sigma_{dyn}$ for the 
$K/\pi$, $p/\pi$ and $K/p$ ratios are shown in comparison with the 
experimental data by the NA49 Collaboration at the SPS~\cite{Al08} and 
the preliminary data of the STAR Collaboration at 
RHIC~\cite{Da06,Ab09,WeQM09,TiQM09}.  The available results of UrQMD 
calculations (from Refs.~\cite{Al08,Ro04,KrFr06}) are also shown by the 
dashed lines.

For the SPS energies we use the NA49 acceptance tables from
Ref.~\cite{Al08}. For the RHIC  energies we employ the following
cuts: in pseudorapidity, $|\eta|<1$, and in transverse momentum,
$0.2<p_T<0.6$~GeV/c for kaons and pions and $0.4<p_T<1$~GeV/c for
protons~\cite{Da06,Ab09,WeQM09,TiQM09}. We note also, that the HSD
results presented in Fig.~\ref{fig:fig5} correspond to the
same centrality selection as in the experiment: the NA49 data
correspond to the 3.5\% most central collisions selected via the veto
calorimeter, whereas in the STAR experiment the 5\% most central
events with the highest multiplicities in the pseudorapidity range
$|\eta|<0.5$ have been selected.

One sees that the UrQMD model gives practically a constant
$\sigma_{dyn}^{K\pi}$, which is by about $40\%$ smaller than the
results from HSD at the lowest SPS energy. This difference between
the two transport models may be attributed to different
realizations of the string and resonance dynamics in HSD and
UrQMD: in UrQMD the strings decay first to heavy baryonic and
mesonic resonances which only later on decay to `light' hadrons
such as kaons and pions. In HSD the strings dominantly decay
directly to `light' hadrons (from the pseudoscalar meson octet) or
the vector mesons $\rho$, $\omega$ and $K^*$ (or the baryon octet
and decouplet in case of baryon number $\pm 1$). Such a
`non-equilibrated' string dynamics leads to stronger
fluctuations of the $K/\pi$ ratio. Note that all differences
between SM and transport models, as well as between different
versions of the transport models, become clearly seen at the
lower bombarding energies. This is only because of using
$\sigma_{dyn}$ as a measure of the $K/\pi$ ratio fluctuations. If
one uses $F=\sigma^2/\sigma_{mix}^2$ as a measure of the $K/\pi$
fluctuations the conclusion will be opposite: as
Fig.~\ref{fig:fig4} ({\it right}) demonstrates the difference
between the SM and HSD predictions measured in $F$ would increase
with collision energy.

The excitation function of $\sigma_{dyn}^{Kp}$ for the $K/p$ ratio
is presented in the middle panel of Fig.~\ref{fig:fig5}.
The 'stars' correspond to the preliminary STAR data in Au+Au collisions
\cite{TiQM09}.
The HSD results for $\sigma_{dyn}^{Kp}$ show a weak energy dependence
in both SPS and RHIC energy regions.
However, as follows from our analysis this observable is very
sensitive to the experimental acceptance. In order to illustrate it, we
extended the HSD calculations for Au+Au within the STAR acceptance for SPS
energies also - cf. the upper red line with solid dots, which
it is higher then the HSD result for Pb+Pb obtained for the 
NA49 acceptance - lower blue line with open dots.
The influence of the experimental acceptance is clearly seen at 158
AGeV where a switch from the NA49 to the STAR acceptance leads to a
jump in $\sigma_{dyn}^{Kp}$ by more than 3\%.
On the other hand, our calculations for Pb+Pb (3.5\% central) and for
Au+Au (5\% central) collisions - performed within the NA49 acceptance
for  both cases at 160 AGeV - shows a very week sensitivity of
$\sigma_{dyn}^{Kp}$ on the actual choice of the collision system and
centrality -- cf. the coincident open circle and triangle at 158 AGeV
in the middle panel of Fig.~\ref{fig:fig5}.
Indeed, new data from NA49 and STAR should clarify the situation.

At SPS energies the HSD simulations lead to negative values of
$\sigma_{dyn}$ for the proton to pion ratio -- cf. Fig. \ref{fig:fig5},
lower panel.  This is in agreement with the NA49 data in Pb+Pb
collisions. On the other hand HSD gives large positive values of
$\sigma_{dyn}^{p\pi}$ at RHIC energies which strongly overestimate the
preliminary STAR data for Au+Au collisions \cite{WeQM09}.

\section{Summary}
\label{ch:summ}

The present review has been devoted to a systematic study of
fluctuations and correlations in heavy-ion collisions within the
HSD and UrQMD transport models. This provides a powerful tool to
simulate experimental collisions on an event-by-event basis.

In Section \ref{ch:mult} the centrality dependence of multiplicity
fluctuations has been studied in fixed target A+A experiments.  The
centrality selection in Pb+Pb collisions at 158~AGeV has been performed
in full correspondence to the NA49 experimental procedure:  only the
samples of events with fixed numbers of the projectile participants,
$N_P^{proj}$, have been considered.  A decrease of the scaled variance
$\omega_P$  for the participant number fluctuations with $N_P^{proj}$
in central A+A collisions has been found.  The transport model results
for the scaled variances $\omega_i$ of negative, positive, and all
charged hadrons in Pb+Pb at 158 AGeV have been presented and
interpreted.  The samples with $N_P^{proj}=20\div 30$ show a maximum of
fluctuations of the number of target nucleon participants $N_P^{targ}$.
The final hadron multiplicity fluctuations exhibit an analogous
behavior. This explains the large values of the HSD and UrQMD scaled
variances $\omega_i$ for positive, negative, and all charged hadrons in
the target hemispheres and in the full $4\pi$ acceptance.  The
asymmetry between the projectile and target participants -- introduced in
the data samples by the trigger condition of fixed $N_P^{proj}$ --  can be
used to explore different dynamical scenarios for A+A collisions by
measuring the final multiplicity fluctuations as a function of rapidity
(cf. Fig.~\ref{fig:w-yNproj}). The analysis reveals that the NA49 data
indicate a rather strong mixing of the longitudinal flows of the
projectile and target hadron production sources. This sheds  new light
on the A+A reaction dynamics at top SPS energies. The data demonstrate
a significantly larger amount of mixing than generated in
hadron-string transport approaches. This probably points towards the
presence of strongly interacting quark-gluon phase.

In Section \ref{ch:ppAA} the particle number fluctuations in
central A+A collisions from 10 to 21300 AGeV have been studied
within the HSD transport model. HSD predicts that the scaled
variances $\omega_i$ in central A+A collisions remain close to the
corresponding values in $p+p$ collisions and  increase with
collision energy with the multiplicity per participating nucleon,
$\omega_i\propto n_i$. The scaled variances $\omega_i$ calculated
within the statistical hadron-resonance gas (HG) model along the
chemical freeze-out line show a rather different behavior:
the $\omega_i$ approach finite values at high collision energy. At the
top RHIC energy $\sqrt{s_{NN}}=200$~GeV the HSD results for
$\omega_{i}$(HSD) are already about 10 times larger than the
corresponding micro-canonical (MCE) HG values for $\omega_i$(MCE).
Thus, the HSD and HG scaled variances $\omega_i$ show a different
energy dependence and are very different numerically at high
energies. However, a comparison with the NA49 data of very central,
$\leq 1 \%$, Pb+Pb collisions in the SPS energy range does not
distinguish between the HSD and MCE HG results. This occurs
because of two reasons: First, the MCE HG and HSD results for
$\omega_{i}$ at SPS energies are not too much different from each
other at SPS energies as well as from the $\omega_{i}$ values
in p+p collisions. Second, small
experimental values of the acceptance, $q=0.04\div 0.16$, and
about 10\% possible ambiguities coming from the acceptance scaling
relation (\ref{acc}) make the difference between the HSD and MCE
HG results almost invisible. New measurements of $\omega_i$ for
samples of very central A+A collisions with large acceptance at
both SPS and RHIC energies are needed to allow for a proper
determination of the underlying dynamics.

In Section \ref{ch:NA61} the event-by-event multiplicity
fluctuations in A+A collisions have been studied for different
energies and system sizes within the transport approach.
This study is in full correspondence to the future experimental
program of the NA61 Collaboration at the SPS. The central C+C,
S+S, In+In, and Pb+Pb nuclear collisions from $E_{lab}$= 10,~ 20,~
30,~ 40,~ 80, 158~AGeV have been investigated. The influence of
participant number fluctuations on hadron multiplicity
fluctuations has been emphasized and studied in detail. To make
these `trivial' fluctuations smaller, one has to consider the most
central collisions. Indeed, one needs to make a very rigid
selection -- 1\% or smaller -- of the `most central' collision
events. In addition, one wants to compare the event-by-event
fluctuations in these `most central' collisions for heavy and for
light nuclei. Under these new requirements different centrality
selections are not equivalent to each other. As a consequence,
there is no universal geometrical selection by the impact
parameter. This is a new and serious problem for theoretical
models (e.g. for hydrodynamical models) in a precision
description of the event-by-event fluctuation data. The above
statements have been illustrated by the $b=0$ selection criterium.
For light nuclei even these `absolutely
central' geometrical collisions lead to rather large fluctuations
of the number of participants, essentially larger than in the  1\%
most central collisions selected by the largest values of the
projectile participants $N_P^{proj}$.
The multiplicity fluctuations calculated in these samples show a
much weaker dependence on the atomic mass  number A than for the
criterium of zero impact parameter $b=0$. A monotonic energy
dependence for the multiplicity fluctuations is obtained in both
the HSD and UrQMD transport models. The two models demonstrate a
similar qualitative behavior of the particle number fluctuations.
Our study has also demonstrated  a sensitivity of the
multiplicity fluctuations to some specific details of the
transport models.
Thus, the present HSD and UrQMD results for the scaled variances
provide a general trend of their dependencies on A and $E_{lab}$
and indicate quantitatively the systematic uncertainties.
It has to be stressed again, that HSD and UrQMD - in their present
formulations - do not include explicitly a phase transition to the
QGP. The expected enhanced fluctuations -- attributed to the
critical point and phase transition -- can be observed
experimentally on top of a monotonic and smooth `hadronic
background'.  The most promising signature of the QCD critical
point would be an observation of a non-monotonic dependence of the
scaled variances with bombarding energy $E_{lab}$ for central A+A
collisions with fixed atomic mass number. In the fixed target SPS
experiments the measurements of $\omega_-$, $\omega_+$, and
$\omega_{ch}$ are preferable in the forward hemispheres. In this case
the remaining small fluctuations of the number of target participants
$N_P^{targ}$ in the 1\% most central collisions -- defined by the
number of the projectile participants $N_P^{proj}$ -- become even less
important, as they contribute mainly to the particle fluctuations in
the backward hemisphere. These findings are expected to be helpful for
the optimal choice of collision systems and collision energies for the
experimental search of the QCD critical point.

In Section \ref{ch:corr} the HSD transport model has been used to
calculate the scaled variance of participant number fluctuations,
$\omega_P$, and the number of $i$'th hadrons per nucleon accepted by
the mid-rapidity PHENIX detector, $q_in_i$, in different Beam-Beam
Counter centrality classes. The HSD model for N+N collisions has been
also used to estimate the fluctuations from a single source,
$\omega_i^*$. It has been found that the model of independent sources
based on these HSD results leads to a good agreement with the PHENIX
data \cite{Ad05a,Mi05}.  In different 5\% centrality classes,
$\omega_P$ goes down and $q_i n_i$ goes up  with increasing of $\langle
N_P\rangle$. This results in  a non-monotonic dependence of
$\omega_i^{acc}$ on $\langle N_P\rangle$ as seen in the PHENIX data.
Thus, one may conclude that the centrality dependence of the
fluctuations seen in the present PHENIX data are the consequences of
participant number fluctuations. To avoid a dominance of the
participant number fluctuations we stress again that one needs to
analyze  most central collisions with a much more rigid ($\le 1\%$)
centrality selection.

The system size event-by-event fluctuations induce also
rapidity forward-backward correlations in relativistic A+A
collisions. The analysis has been based on two independent models
-- the microscopic HSD transport approach and a `toy' wounded
nucleon model realized as a Glauber Monte Carlo event generator.
It has been shown that strong forward-backward correlations arise
due to an averaging over many different events that belong to one
10\% centrality bin. In contrast to average multiplicities, the
resulting fluctuations and correlations depend strongly on the
specific centrality trigger. For example, the centrality selection
via impact parameter $b$ (used in most theoretical calculations) and
via $N_{ch}^{ref}$ (used experimentally) lead to rather different
values of participant number fluctuations $\omega_P$ and
forward-backward correlation parameters $\rho$. These different
centrality selections give also quite different dependencies of
$\omega_P$ and $\rho$ on centrality.
The HSD simulations reveal strong forward-backward correlations
and reproduce the main qualitative features of the STAR data in
Au+Au collisions at RHIC energies \cite{ArMcPa07,LaMc06}.
The forward-backward correlations can be studied experimentally
for smaller size centrality bins defined by $N_{ch}^{ref}$. When
the size of the bins decreases, the contribution of `geometrical'
fluctuations should lead to weaker forward-backward correlations
and to a less pronounced centrality dependence. Note, that the
`geometrical' fluctuations discussed here are in fact present in
all dynamical models of A+A collisions.  Thus, they should be
carefully accounted before any discussion of new physical effects
is addressed. A future experimental analysis -- in the direction
examined here -- should clarify whether the observed correlations
by the STAR Collaboration at RHIC contain really additional
contributions from `new physics'.

In Section \ref{ch:QB} the transport approach has been used to
investigate the event-by-event fluctuations of baryon number and
electric charge in A+A collisions.  The study has been based on the
microscopic HSD transport model which allows to analyze on the
event-by-event basis the influence of two main sources of the conserved
charge fluctuations -- the fluctuations of the number of participants
and the finite experimental acceptance.  It has been found that the
fluctuations in the number of target participants strongly influences
the baryon number and electric charge fluctuations.  The consequences
depend crucially on the dynamics of the initial flows of the conserved
charges.  For a better understanding of the HSD results three limiting
groups of models for A+A collisions have been discussed:  transparency,
mixing and reflection. These ``pedagogical'' considerations indicate
that the HSD model shows only a small mixing on initial baryon flow and
is closer to the transparency models. This makes important the findings
of Refs.~\cite{WeBrCaSt03,Br04,Br04a,BrCaSt03} about the influence of
the partonic degrees of freedom on the initial phase dynamics.  This
phase might increase the mixing of conserved charges by additional
strong parton-parton interactions.

In Section \ref{ch:ratio} the event-by-event multiplicity fluctuations
of pions, kaons, protons as well as their correlations and ratio
fluctuations in central Au+Au (or Pb+Pb) collisions from low SPS up to
top RHIC energies have been studied within the HSD transport approach
and in the statistical hadron-resonance gas model for different
statistical ensembles -- the grand canonical ensemble (GCE), canonical
ensemble (CE), and micro-canonical ensemble (MCE).  The HSD results at
SPS energies are close to those in the CE and MCE statistical models.
This indicates a dominant role of resonance decays and global
conservation laws for low energy A+A collisions. On the other hand,
substantial differences in HSD and statistical model results have been
observed at RHIC energies which can be attributed to non-equilibrium
dynamical effects in the HSD simulations.  Thus, the scaled variances
of different particle species may serve as good observables to probe
the amount of equilibration achieved in central A+A collisions at RHIC
energies.  On the other hand, it has been found that the observable
$\sigma_{dyn}$, which characterizes ratio fluctuations, appears to be
rather sensitive to the details of the model at low collision energies.
It has been found that HSD can qualitatively reproduce the measured
excitation function for the $K/\pi$ ratio fluctuations in central Au+Au
(or Pb+Pb) collisions from low SPS up to top RHIC energies.  The HSD
results for $\sigma_{dyn}^{p\pi}$ appear to be close to the NA49 data
at the SPS. The data for $\sigma_{dyn}^{Kp}$ in Pb+Pb collisions at the
SPS energies will be available soon and allow for further insight. A
comparison of the HSD results with preliminary STAR data in Au+Au
collisions at RHIC energies are not fully conclusive: $\sigma_{dyn}$
from the HSD calculations is approximately in agreement with data
\cite{TiQM09} for the kaon to proton ratio, but overestimates the
experimental results \cite{WeQM09} for the proton to pion ratio. New
data on event-by-event fluctuations in Au+Au at RHIC energies will help
to clarify the situation.

It has been argued in the Introduction  that event-by-event
fluctuations in nucleus-nucleus collisions can be considered as a
probe for the phase transition and the QCD critical point.
The transport models have been employed to study the fluctuations
and correlations including the influence of experimental
acceptance as well as the centrality selection, the dependence on
the system size and collision energy.

So far, the transport models employed do not include an explicit phase
transition to the QGP as well as a dynamical description of hadronization.
Recent steps have been taken in this direction within
Parton-Hadron-String-Dynamics (PHSD) approach \cite{CaBr08,CaBr09}
that incorporates the partonic dynamics in line with an equation
of state from lattice QCD. Nevertheless, the comparison of results from
microscopic transport models with experimental data will allow to
separate the effects indicated by a phase transition to the QGP and the
QCD critical point.

\section*{Acknowledgments}
\addcontentsline{toc}{section}{Acknowledgments} We like to thank
V.V. Begun, M.~Bleicher, H. B\"usching, W.~Cassing, M.~Ga\'zdzicki, M.~Hauer,
C.~H\"ohne, D.~Kresan, O. Linnyk, B.~Lungwitz, I.~N.~Mishustin,
J. T. Mitchell, M.~Mitrovski, L.~M.~Satarov, T.~Schuster, Yu.~M.~Sinyukov,
R.~Stock, H.~St\"ocker, H.~Str\"obele, T. Tarnowsky, G.~Torrieri, G. Westfall
and O.~S.~Zozulya for useful discussions.
V.K.'s work was supported by the Helmholtz International Center
for FAIR within the framework of the LOEWE program
(Landesoffensive zur Entwicklung Wissenschaftlich-\"Okonomischer
Exzellenz) launched by the State of Hesse.

\section*{References}
\addcontentsline{toc}{section}{References}
\markboth{References}{}
\bibliographystyle{h-physrev3} 
\bibliography{my}

\end{document}